\documentclass[reprint,superscriptaddress,amsmath,amssymb,aps,prd,notitlepage,longbibliography,floatfix,nofootinbib,onecolumn]{revtex4-1}

\usepackage{tensor}     
\usepackage{graphicx}   
\usepackage[
colorlinks=true,        
citecolor=blue,         
linkcolor=blue,         
urlcolor=blue           
]{hyperref}             
\usepackage{bm}         
\usepackage{tabulary}
\usepackage{color}      
\usepackage[utf8]{inputenc} 
\usepackage[section]{placeins} 
\usepackage{orcidlink}
\newcommand{\nc}{\newcommand*} 
\nc{\mH}{\mathcal{H}}
\def\({\left(}
\def\){\right)}
\def\[{\left[}
\def\]{\right]}
\def\e{\begin{equation}}
\def\q{\end{equation}}
\def\m{\begin{eqnarray}}
\def\n{\end{eqnarray}}

\nc{\Eq}[1]{Eq.~\eqref{#1}}     
\nc{\Fig}[1]{Fig.~\ref{#1}}     
\nc{\Table}[1]{Table~\ref{#1}}  
\nc{\Sec}[1]{Sec.~\ref{#1}}     

\nc{\ogw}{\Omega_{\mathrm{GW}}}
\nc{\eg}{\textit{e.g.~}}
\nc{\app}{\approx}
\nc{\hf}{\frac{1}{2}}
\nc{\red}[1]{\textcolor{red}{#1}}

\begin{document}
\title{Gauge Dependence of Gravitational Waves Induced by Primordial Isocurvature Fluctuations} 

\author{Chen Yuan\orcidlink{0000-0001-8560-5487}}
\email{chenyuan@tecnico.ulisboa.pt}
\affiliation{CENTRA, Departamento de Física, Instituto Superior Técnico – IST, Universidade de Lisboa – UL, Avenida Rovisco Pais 1, 1049–001 Lisboa, Portugal}
\author{Zu-Cheng Chen\orcidlink{0000-0001-7016-9934}}
\email{Corresponding author: zuchengchen@hunnu.edu.cn}
\affiliation{Department of Physics and Synergetic Innovation Center for Quantum Effects and Applications, Hunan Normal University, Changsha, Hunan 410081, China}
\affiliation{Institute of Interdisciplinary Studies, Hunan Normal University, Changsha, Hunan 410081, China}
\author{Lang~Liu\orcidlink{0000-0002-0297-9633}}
\email{Corresponding author: liulang@bnu.edu.cn}	
\affiliation{Faculty of Arts and Sciences, Beijing Normal University, Zhuhai 519087, China}
	
\begin{abstract}
Primordial isocurvature perturbations, which can arise from various sources in the early Universe, have the potential to leave observable imprints on the gravitational-wave background and provide insights into the nature of primordial fluctuations. In this study, we investigate the gauge dependence of induced gravitational waves (IGWs) sourced by these isocurvature perturbations {during radiation dominated era and the kination period in the early universe}. We analyze the energy density spectra of IGWs in three different gauges: synchronous, Newtonian, and uniform curvature gauges. To facilitate this analysis, we derive analytical solutions for the perturbations that contribute to the IGW spectra {and a general gauge transformation from Newtonian gauge to an arbitrary gauge}. Our results reveal significant differences in the energy spectra across these gauges. We find that the energy density of IGWs {during radiation domination} increases with conformal time as $\eta^8$ and $\eta^4$ for synchronous and uniform curvature gauges, respectively, while it converges in the Newtonian gauge. These findings highlight the importance of gauge choice in calculating IGWs and have implications for the interpretation of future observations of the gravitational-wave background.
\end{abstract}

\maketitle
\section{Introduction} 	
In the early Universe, primordial density perturbations can arise as two distinct types - adiabatic fluctuations and isocurvature fluctuations~\cite{Kodama:1984ziu,Bucher:1999re}. Adiabatic fluctuations, which represent perturbations in the total energy density, are the predominant source of the inhomogeneities we observe today in the cosmic microwave background (CMB) and large-scale structure~\cite{Planck:2018jri}. On the other hand, isocurvature fluctuations, which correspond to spatially varying differences in the relative number densities of different particle species, are predicted in many inflationary models with multiple scalar fields~\cite{Linde:1985yf,Polarski:1994rz}.

On large scales, observations have revealed primordial fluctuations to be remarkably small in amplitude, nearly scale-invariant, predominantly adiabatic, and almost Gaussian~\cite{Planck:2018jri,Planck:2019kim}. However, our knowledge of the state of the Universe on small scales remains far more limited. In fact, the primordial fluctuations on small scales may be substantially larger than those on cosmic scales. Since curvature perturbations couple to tensor perturbations at second order, such enhanced small-scale fluctuations can produce induced gravitational waves (IGWs) during the radiation-dominated era~\cite{Ananda:2006af,Baumann:2007zm,Garcia-Bellido:2016dkw,Inomata:2016rbd,Garcia-Bellido:2017aan,Kohri:2018awv,Cai:2018dig}. IGWs have emerged as a promising new probe of primordial black holes (PBHs)~\cite{Zeldovich:1967lct,Hawking:1971ei,Carr:1974nx}, which have garnered significant interest in recent years~\cite{Saito:2008jc,Belotsky:2014kca,Carr:2016drx,Garcia-Bellido:2017mdw,Carr:2017jsz,Germani:2017bcs,Chen:2018czv,Liu:2018ess,Liu:2019rnx,Chen:2018rzo,Chen:2019irf,Wang:2019kaf,Cai:2019bmk,Liu:2020cds,Wu:2020drm,DeLuca:2020sae,Vaskonen:2020lbd,Chen:2021nxo,DeLuca:2020agl,Domenech:2020ers,Liu:2020vsy,Liu:2020bag,Cai:2021wzd,Yuan:2021qgz,Liu:2021jnw,Liu:2022wtq,Inomata:2022yte,Meng:2022low,Chen:2022qvg,Liu:2022iuf,Zheng:2022wqo,Choudhury:2023jlt,Chen:2022fda,Choudhury:2023vuj,Hai-LongHuang:2023atg,Choudhury:2023kdb,Choudhury:2023hvf,Choudhury:2023rks,Wang:2024vfv,Choudhury:2024jlz,Yuan:2024yyo,Chen:2024dxh,Chen:2024joj,Choudhury:2024kjj,Huang:2024wse,Ding:2024mro,Choudhury:2024dei,Calza:2024fzo,Calza:2024xdh,Sasaki:2018dmp,Carr:2020gox,Carr:2020xqk,Choudhury:2024aji} due to their potential to explain dark matter~\cite{Sasaki:2018dmp,Carr:2020gox,Carr:2020xqk} and serve as sources for the gravitational wave events detected by the LIGO-Virgo-KAGRA collaboration~\cite{Bird:2016dcv,Sasaki:2016jop}.

The topic of IGW from large amplitude curvature perturbations has attracted significant interest recently~\cite{Alabidi:2012ex,Alabidi:2013lya,Choudhury:2013woa,Espinosa:2018eve,Bartolo:2018rku,Inomata:2018epa,Yuan:2019udt,Inomata:2019zqy,Inomata:2019ivs,Chen:2019xse,Yuan:2019wwo,Domenech:2017ems,Domenech:2019quo,Ota:2020vfn,Cai:2019jah,Cai:2019elf,Cai:2019amo,Bhattacharya:2019bvk,Pi:2020otn,Choudhury:2023hfm,Choudhury:2023kam,Bhattacharya:2023ysp,Liu:2023ymk,Jin:2023wri,Liu:2023pau,Yi:2023npi,Liu:2023hpw,Choudhury:2023fwk,Choudhury:2023fjs,Chen:2024fir,Choudhury:2024dzw,Chen:2024twp,Choudhury:2024one}. However, an important issue arises when considering the gauge invariance of these IGWs. While linear gravitational waves are gauge invariant, this invariance breaks down at second order~\cite{Noh:2003yg}. Consequently, the energy density spectrum of IGWs may depend on the choice of gauge. This gauge dependence issue has been extensively studied for IGWs sourced by primordial adiabatic fluctuations~\cite{Hwang:2017oxa,Gong:2019mui,Tomikawa:2019tvi,Wang:2019zhj,DeLuca:2019ufz,Inomata:2019yww,Yuan:2019fwv,Nakamura:2020pre,Giovannini:2020qta,Lu:2020diy,Chang:2020tji,Ali:2020sfw,Chang:2020iji,Chang:2020mky,Giovannini:2020soq,Domenech:2020xin,Cai:2021ndu,Cai:2021jbi}. However, the question of how gauge choice affects IGWs sourced by primordial isocurvature fluctuations has not yet been addressed, leaving a significant gap in our understanding of these cosmological signals.

In this paper, we aim to extend the analysis of gauge dependence to IGWs sourced by primordial isocurvature fluctuations. Specifically, we investigate how the energy spectrum of these IGWs varies in different gauges, including the synchronous, Newtonian, and uniform curvature gauges. Interestingly, we find that the spectra of IGW from primordial isocurvature fluctuations are totally different in these three gauges. 
The rest of this paper is organized as follows. In Section \ref{igw}, we briefly review the energy density of IGWs and derive the equations of motion for the linear perturbations and the second-order source term in a general gauge. Meanwhile, we consider three specific gauges and analyze the evolution of perturbations and the source term in each of them: the synchronous gauge in Section \ref{s-gauge}, the uniform curvature gauge in Section \ref{u-gauge}, and the Newtonian gauge in Section \ref{n-gauge}.
We also derive a general gauge transformation from Newtonian gauge to any gauge in Section \ref{gauge-trans} and the IGWs during kination period is studied in Section \ref{w=1}.
Finally, we summarize our findings and discuss their implications in Section~\ref{summary}.

\section{Gravitational waves induced by isocurvature perturbations}\label{igw}

In this section, we will explore the energy density spectra of gravitational waves induced by primordial isocurvature perturbations. During the early stages of the Universe, second-order tensor perturbations can be generated by the quadratic terms of linear scalar perturbations~\cite{Tomita:1967wkp,Matarrese:1992rp,Matarrese:1993zf,Matarrese:1997ay,Noh:2004bc,Carbone:2004iv,Nakamura:2004rm}. These IGWs offer a unique window into the physics of the early Universe on scales significantly smaller than those probed by CMB observations. For comprehensive reviews of IGWs, we refer the reader to Ref.~\cite{Domenech:2023jve} for the isocurvature case and Refs.~\cite{Yuan:2021qgz,Domenech:2021ztg} for the adiabatic case.

Let us start by considering the perturbed metric in its most generic form, which includes linear scalar perturbations
\begin{equation}
    d s^{2}=a^{2}\left[-(1+2 \phi) \mathrm{d} \eta^{2} + 2(B_i+B_{,i}) \mathrm{d} x^i \mathrm{d} \eta +\left((1-2 \psi) \delta_{i j} + (E_{i,j}+E_{j,i})+2E_{,ij} 
    + 2 h_{i j}^{(1)}
    +\frac{1}{2} h_{i j}^{(2)}\right) \mathrm{d} x^{i} \mathrm{~d} x^{j}\right],
\end{equation}
where $\phi$, $\psi$, $B$, and $E$ represent different linear scalar perturbations, while $E_i$ and $B_i$ denote vector perturbations. A comma indicates spatial derivatives. The tensor perturbations $h_{ij}^{(1)}$ and $h_{ij}^{(2)}$ correspond to the transverse and traceless modes at first and second order, respectively. In the following analysis, we will neglect the vector modes and the first-order tensor modes, as they are weak compared to the linear-order scalar modes. For simplicity, we will use the notation $h_{ij}$ to represent the second-order tensor mode $h_{ij}^{(2)}$. Here, $\eta$ represents the conformal time, and $a$ denotes the scale factor.

We concentrate on the radiation-dominated (RD) era where the stress-energy tensor has two part. The first part is $T_{m \mu \nu}  =\rho_m u_{m \mu} u_{m \nu}$ representing the stress-energy tensor for matter and $T_{r \mu \nu}  =\left(\rho_r+p_r\right) u_{r \mu} u_{r \nu}+p_r g_{\mu \nu}$ for radiation.
In these expressions, $u_\mu = (u_0 , a\,v_{,i})$ represents the four-velocity, while $\rho$ and $p$ denote the energy density and pressure, respectively. During RD, we have $p_r=1/3\rho_r$. In the following analysis, we will use the notations $\delta \rho$ and $\delta p$ to represent the perturbations in the energy density and pressure respectively.

By performing a first order transformation, $\tilde{\eta}=\eta+T$ and $\tilde{x^i}=x^i+L^{,i}$, the metric perturbations follow the transformation rules:
\begin{equation}\label{transf}
    \begin{aligned}
    \tilde{\phi} & = \phi + \mH T +T',\\
    \tilde{\psi} & = \psi -\mH T,\\
    \tilde{B} & = B- T + L',\\
    \tilde{E} & = E + L,\\
    \end{aligned}
\end{equation}
where a prime represents the derivative of conformal time and $\mH \equiv a'/a$ is the conformal Hubble parameter. By choosing specific $T$ and $L$, one can eliminate two degrees of freedom from the four scalar modes.
Furthermore, the first order $ij$ component of the Einstein equation eliminates another degree of freedom, leaving only one scalar mode, governed by
\begin{equation}
\begin{aligned}\label{eom1}
    &\psi''+\mH\left[ \phi'+(2+3c_s^2)\psi')\right]
    +\left[ (1+3c_s^2)\mH^2 +2\mH' \right] \phi
    +c_s^2\left[\mH \Delta(B-E')-\Delta\psi\right]=4\pi a^2
\tau\delta s,\\
    &\phi-\psi+(B-E')'+2\mH(B-E')=0.
\end{aligned}
\end{equation}
The first line represents the equation of motion for the scalar modes, where $c_s^2$ and $\delta s$ are the sound speed and entropy perturbation, respectively, arising from $\delta p = c_s^2 \delta \rho +\tau \delta s$, which takes the form
\begin{equation}
    c_s^2 = {1\over 3}\left(1+{3\over 4}{\rho_m\over \rho_r}\right)^{-1}, \qquad \tau={c_s^2 \rho_m \over s}, \qquad S\equiv {\delta s \over s }={3\over 4}{\delta\rho_r\over \rho_r}-{\delta\rho_m\over \rho_m}.
\end{equation}
Here, $S$ is a gauge-invariant quantity by definition. Moreover, by defining the relative velocity as $v_{\mathrm{rel}}\equiv v_m-v_r$, one can obtain the relation $ S' = \Delta v_{\mathrm{rel}}$. Combining the Einstein equation and energy conservation, $\nabla_{\mu} T^{\mu\nu}=0$, up to first order, one obtains the equation of motion for the entropy (see Appendix~\ref{appendix} for details)
\begin{equation}\label{eom2}
    S''+3\mH c_s^2 S'+{3\rho_m \over 4\rho_r}c_s^2 k^2 S + {3\over 16\pi a^2 \rho_r}c_s^2 k^4 [\mH (B-E')-\psi] =0.
\end{equation}
In the adiabatic case, the initial values of the scalar modes are determined by the inflation field, while in the isocurvature case, the metric perturbations are sourced by the entropy. We extract the initial value from $S$ in Fourier space, such that
\begin{equation}
    S = S_{\bm{k}} T_{S}(k\eta),
\end{equation}
where $T_S$ is the transfer function of the entropy, representing its time evolution, normalized as $T_S(0)=1$. The initial value $S_{\bm{k}}$ is related to the dimensionless primordial entropy spectrum as
\begin{equation}
    \left\langle S_{\bm{k}}S_{\bm{k}}' \right\rangle={2\pi^2\over k^3}\mathcal{P}_S(k) \delta^{(3)}(\bm{k}+\bm{k'}),
\end{equation}
where $\delta^{(3)}$ is the three-dimensional Dirac delta function.

Following the notation in Ref.~\cite{Domenech:2021and}, we introduce two dimensionless quantities, $x\equiv k\eta$ and $\kappa = k/k_{\mathrm{eq}}$, to simplify \Eq{eom1} and \Eq{eom2}. During the RD era, we have $\kappa \gg 1$ and $x/\kappa \ll 1$ \cite{Domenech:2021and}. This allows us to expand the equations of motion to the order of $\mathcal{O}(\kappa ^{-1})$, yielding
\begin{equation}
    \begin{aligned}
        &{d^2T_S \over dx^2}+\left({1\over x}-{1\over 2\sqrt{2}\kappa} \right){dT_S \over dx}+{x \over 4\sqrt{2}\kappa }T_S+{x\over 6} (T_B-T_{E'}-xT_{\psi})\simeq  0,\\
        &{d^2 T_\psi \over dx^2}+{3\over x}{d T_{\psi}\over dx}+\left({1\over x} + {1\over 4\sqrt{2}\kappa} \right){d T_\phi \over dx} 
        -\left( {1\over 3x}-{1\over 6\sqrt{2}\kappa } \right)(T_B-T_{E'})
        +{1\over 4\sqrt{2}x \kappa}T_{\phi}
        +\left( {1\over 3} - {x\over 4\sqrt{2}\kappa} \right)T_{\psi}
        -{1\over 2\sqrt{2}x\kappa}T_S\simeq0,\\
        &T_{\phi}-T_{\psi}+{d\over dx}\left( T_B -T_{E'}  \right)+\left( {2\over x}+{1\over 2\sqrt{2}\kappa} \right)\left( T_B-T_{E'}\right)   \simeq0.
    \end{aligned}
\end{equation}
Here, we define the transfer functions for the scalar modes as
\begin{equation}
    T_{X}(k\eta) = { X \over  S_{\bm{k}} },\qquad \text{where } X = \phi, \psi, B, E.
\end{equation}

At second order, linear scalar perturbations will source $h_{ij}$ in the form of quadratic terms, generating the so-called scalar-IGW. From the second-order Einstein equation, one obtains
\begin{equation}\label{hij2}
    h_{ij}''+2\mH h_{ij}' + k^2 h_{ij} = -4\mathcal{T}_{ij}^{\ell m}\mathcal{S}_{\ell m},
\end{equation}
where $\mathcal{T}_{i j}^{\ell m} = e_{ij}^{(+)}(\bm{k})e^{(+)\ell m}(\bm{k})+e_{ij}^{(\times)}(\bm{k})e^{(\times)\ell m}(\bm{k}) $ is the transverse and traceless projection operator. The polarization tensors of $+$ and $\times$ modes are given by $e_{ij}^{(+)} = (e_i e_j - \bar{e}_i \bar{e}_j)/\sqrt{2}$ and $e_{ij}^{(\times)} = (e_i \bar{e}_j + \bar{e}_i e_j)/\sqrt{2}$ respectively. Here, $e$ and $\bar{e}$ are two linearly independent unit vectors that are both perpendicular to $\bm{k}=(0,0,k)$. For convenience, we chose  $e=(1,0,0)$ and $\bar{e}=(0,1,0)$. During RD era, the source term in the most generic gauge reads
\begin{equation}
    \begin{aligned}
       \mathcal{S}_{ij} & = \Delta B (B-E')_{,ij}-B_{,ib}B_{,jb}+E_{,ibc}E_{,jbc}-E_{,ijb}(\Delta E-\phi-\psi)_{,b}-2\psi_{,ij}\Delta E
        + 2(E_{,ib}\psi_{,jb}+E_{jb}\psi_{ib})+2\mH B_{,b}E_{bij}\\
        & +2\mH \psi B_{,ij}+2(\psi B_{,ij})'
        + E_{,ijb}B_{,b}'+(B_{,jb}E_{ib}'+B_{,ib}E_{jb}') - 2E_{,ib}'E_{jb}'-[(B-E')_{,ij}(\Delta E-\phi)']+\psi'(B+E')_{,ij}\\
        & + {2\over 3}E_{,ij}[
        \Delta(3\phi-8\psi + 3(B-E')'+8\mH(B-E') )+6\mH (\phi+4\psi)'+9\psi''
        ]+2\phi \psi_{,ij}+(\phi-\psi)_{,i}(\phi-\psi)_{,j}\\
        & - \left( \phi + {\psi'\over \mH} \right)_{,i}
        \left( \phi + {\psi'\over \mH} \right)_{,j}.
    \end{aligned}
\end{equation}
The solution to \Eq{hij2} can be written as
\begin{equation}\label{h_ft}
h_{ij}(\eta,\bm{k})= h^{(+)}(\eta,\bm{k})e_{ij}^{(+)}(\bm{k}) + 
h^{(\times)}(\eta,\bm{k})e_{ij}^{(\times)}(\bm{k}),
\end{equation}
where $h(\eta,\bf{k})$ for either $+$ or $\times$ mode can be solved using Green's function method in Fourier space
\begin{equation}\label{solh}
    h(\eta,\bm{k})={1\over a(\eta)}\int_0^\eta g_k(\eta;\tilde{\eta})a(\tilde{\eta})\mathcal{S}(\tilde{\eta},\bm{k})\mathrm{d}\tilde{\eta}.
\end{equation}
Here, $g_k(\eta ; \tilde{\eta}) =\sin(k\eta-k\tilde{\eta} )/k$ is the Green's function, and the source term in Fourier space is
\begin{equation}
    \mathcal{S}(\eta,\bm{k})=-4\int\frac{\mathrm{d}^3p}{(2\pi)^{3}} \Big(e^{ij}p_ip_j\Big)S_p S_{|\bm{p}-\bm{k}|} F(p,|\bm{k}-\bm{p}|,\eta),
\end{equation}
where $F$ is the transfer function of the source term, defined by extracting $p_i$, $p_j$, $S_p$ and $S_{|\bm{p}-\bm{k}|}$ from the Fourier transform of $\mathcal{S}_{ij}$. The expression for $F$ without gauge fixing is lengthy and we will only present the expression for $F$ in certain gauges in subsequent sections.

An important observational quantity that characterizes IGW is the energy density parameter, $\Omega_{\mathrm{GW}}(f)$, defined as the energy density of gravitational waves per logarithm frequency (or per logarithm wavelength if using $k=2\pi f$) divide by the critical energy density of the Universe. It can be evaluated as
\begin{equation}\label{ogwd}
\ogw(k,\eta)\equiv{1 \over \rho_c}{\mathrm{d}\rho_{\mathrm{GW}} \over \mathrm{d} \ln k} = {1\over 24}\left(k\over \mH \right)^2\overline{\mathcal{P}_h(k,\eta)},
\end{equation}
where an overline stands for the oscillation average such that $\sin^2x=\cos^2 x\to 1/2$. We sum over the two polarization modes in Eq.~(\ref{ogwd}) and the dimensionless power spectrum for the second order tensor mode is defined as 
\begin{equation}
    \left\langle h(\eta,\bm{k}) h(\eta,\bm{k}')  \right\rangle = \frac{2\pi^2}{k^3} \mathcal{P}_h(k,\eta) \delta^{(3)}(\bm{k}+\bm{k}').
\end{equation}
When computing IGW during RD, one should evaluate $\ogw(k,\eta)$ in the sub-horizon limit $\eta\to\eta_{c}$, to ensure that the source term decays to negligible at $\eta_c$, indicating that the IGW signal has stabilized. Then, at matter-radiation equality, the energy density parameter is given by $\ogw(k) \equiv \ogw(k,x_c)$ for $x_c\gg 1$. Combining the above equations, one can express $\ogw(k)$ in terms of the primordial power spectrum as 
\begin{equation}\label{Omegagw}
    \ogw(k) = {1\over 6}\int_0^\infty\mathrm{d}u\int_{|1-u|}^{1+u}\mathrm{d}v~{v^2\over u^2}\Big[1-\left({1+v^2-u^2\over 2v}\right)^2\Big]^2 \mathcal{P}_S(uk) \mathcal{P}_S(vk)\overline{I^2(u,v,x\to \infty)}.
\end{equation}
For convenience, we introduce two dimensionless variables: $u\equiv p/k$ and $v\equiv |\vec{p}-\vec{k}|/k$. The kernel function is given by
\begin{equation}\label{iuv}
    I(u,v,x)\equiv\int_0^x\mathrm{d}\tilde{x}~\tilde{x}\sin(x-\tilde{x})\frac{1}{2}[F(u,v,\tilde{x})+F(v,u,\tilde{x})].
\end{equation}

In the following subsections, we will explore three specific gauge choices and present the corresponding analytical results for each case. 

\subsection{IGWs in synchronous gauge} \label{s-gauge}
The metric perturbations in the synchronous gauge satisfy $\phi=B=0$ so that only the $ij$ components in the perturbated metric survive. During RD, the solutions for the transfer functions are given by
\begin{equation}
    \begin{aligned}
        T_\psi(x) &= \frac{3}{\sqrt{2}x^2\kappa} \left(x-\sqrt{3} \sin{x\over \sqrt{3}}\right)+\mathcal{O}\left({x\over \kappa}\right)^2,\\
        T_{E'}(x)& = -\frac{3}{2\sqrt{2}x^2 \kappa}\left(-6+x^2 + 6\cos{x\over \sqrt{3}}\right)+\mathcal{O}\left({x\over \kappa}\right)^2,\\
        T_S(x)&= 1+{3\over 2\sqrt{2}\kappa} \left[
    x+\sqrt{3}\sin\left({x\over \sqrt{3}}\right)-2\sqrt{3} \mathrm{Si}\left({x\over \sqrt{3}}\right)
    \right]+\mathcal{O}\left({x\over \kappa}\right)^2.
    \end{aligned}
\end{equation}
From \Eq{transf}, it can be seen that $\phi=0$ can be achieved by choosing treads (i.e., choosing $T$) such that an observer moving along a thread measures the coordinate time. The condition $B=0$ indicates that the threads are orthogonal to the time slices. Furthermore, $B=0$ only fixes $L'$ in \Eq{transf}, leaving the choice of initial time slice as a remaining degree of freedom. This remaining gauge freedom can be removed by fixing the integration constant in the transfer function of $E$ as follows
\begin{equation}
    T_E(x) = \int T_{E'}(x) \mathrm{d}x = -\frac{3}{2\sqrt{2}x\kappa}
    \left[
    6+x^2-6\cos{x\over\sqrt{3}}-2\sqrt{3} x \mathrm{Si}\left({x\over\sqrt{3}} \right)
    \right] + C,
\end{equation}
where we set $C=0$ in the following computation. The transfer function, $F$, can be written as
\begin{equation}\label{FuvS}
    \begin{aligned}
        F^S(u,v,x) &= \frac{(-1+u^2+v^2)(-1+3u^2+v^2)}{4u^2v^2}T_E(u x)T_E(vx) -{16 u \over 3 v^2 x}T_{E'}(ux)T_E(vx) + {1+u^2-v^2 \over u v}T_{E'}(ux)T_{E'}(vx) \\\
        &+ \frac{-3-13u^2+3v^2}{3v^2}T_\psi(ux)T_E(vx) - \frac{3+u^2-3v^2}{2u^2}T_E(ux)T_\psi(vx) + T_\psi(ux)T_\psi(vx)
        - {2u^2\over v^2}T_{E''}(ux)T_E(vx)\\
        &-{16u\over v^2x}T_{\psi'}(ux)T_E(vx)-{u\over v}T_{\psi'}(ux)T_{E'}(vx)-{6u^2\over v^2}T_{\psi''}(ux)T_E(vx).
    \end{aligned}
\end{equation}
Here we introduce a notation for the transfer function of the derivative of the perturbations, such that $T_{\psi'}(x) \equiv d T_{\psi}(x)/ dx$. In the sub-horizon limit, where $x\gg1$, the metric perturbation $\psi$ scales as $1/x$. However, $E'$ approaches a constant, and $E$ diverges as $E\sim x$ in this limit. Therefore, the transfer function of the source term in the sub-horizon limit is dominated by the first term in Eq.~(\ref{FuvS}), scaling as $F^S(u,v,x\gg1)\sim x^2$. 
According to Eq.~(\ref{iuv}), $I(u,v)$ will increase as $x^4$, and hence $\Omega_{\mathrm{GW}}$ will increase as $x^8$.
As a result, IGWs in synchronous gauge will diverge, as the perturbations will  continuously induce the gravitational waves.

\subsection{IGWs in uniform curvature gauge} \label{u-gauge}
In the uniform curvature gauge, the metric perturbations satisfy $\psi=E=0$ and the transfer functions of the remaining perturbations read
\begin{equation}
\begin{aligned}
    T_B(x)&= -{3\over 2\sqrt{2}\kappa x^2 }\left[
    6+x^2-2\sqrt{3}x\sin\left({x\over \sqrt{3}}\right)
    -6\cos\left({x\over \sqrt{3}}\right)\right]
    +\mathcal{O}\left({x\over \kappa}\right)^2,\\
    T_\phi(x)&=-{3\over\sqrt{2}x \kappa} \left[1-\cos\left({x\over\sqrt{3}}\right)\right]+\mathcal{O}\left({x\over \kappa}\right)^2.
\end{aligned}
\end{equation}
Since $S$ is a gauge-invariant quantity, the expression for $T_S(x)$ remains the same as in the synchronous gauge.
Furthermore, the source term in the uniform curvature gauge can be simplified to
\begin{equation}
        S_{ij}=B_{,ij}\partial^2B -B_{,bi}B_{,bj}+\phi'B_{,ij},
\end{equation}
and its transfer function can be expressed as
\begin{equation}
    F^U(x,u,v) = \frac{1+u^2-v^2}{2uv}T_B(ux)T_B(vx) -{u\over v}T_\phi'(ux)T_B(vx).
\end{equation}
Although $\phi$ decays as $1/x$ in the sub-horizon limit, $B$ approaches a constant value. Consequently, the transfer function of the source term in the sub-horizon limit is $F^U(u,v,x\gg1)\sim \mathcal{O}(1)$.
This leads to $I(u,v,x\gg 1) \sim x^2$ and $\Omega_{\mathrm{GW}} \sim x^4$ in the sub-horizon limit.

\subsection{IGWs in Newtonian gauge} \label{n-gauge}

IGWs generated by isocurvature perturbations were first studied in the Newtonian gauge~\cite{Domenech:2021and}, which is defined as $B=E=0$. In this gauge, one has ${\psi}={\phi}$, and the solution is given by
\begin{equation}
    T_\phi(x) = {3\over 2\sqrt{2}\kappa x^3 }\left[
    6+x^2-2\sqrt{3}x\sin\left({x\over \sqrt{3}}\right)
    -6\cos\left({x\over \sqrt{3}}\right)\right]
    +\mathcal{O}\left({x\over \kappa}\right)^2.
\end{equation}
The source term in Newtonian gauge can be simplified as
\begin{equation}
        S_{ij} =2\phi \phi_{,ij} - \left( \phi + {\phi'\over \mH} \right)_{,i}
        \left( \phi + {\phi'\over \mH} \right)_{,j},
\end{equation}
with its transfer function to be
\begin{equation}
    F^N(u,v,x) = -2 T_{\phi}(ux)T_{\phi}(vx)-\left[T_{\phi}(ux)+ux T_{\phi'}(ux) \right]
    \left[T_{\phi}(vx)+vx T_{\phi'}(vx) \right].
\end{equation}
Note that $\phi$ decays as $1/x$ in the sub-horizon limit, and we have $F^{N}(u,v,x\gg1)\sim 1/x^2$. As a result, IGWs in Newtonian gauge converge, and $\overline{I^2(u,v,x\to\infty)}$ can be simplified as
\begin{equation}
    \overline{I^2(u,v,x\to\infty)}=\frac{1}{2}  \left(\int_0^\infty \tilde{x} \cos \tilde{x} \tilde{F}(u,v,\tilde{x}) \mathrm{d} \tilde{x}\right)^2 
    +\frac{1}{2}  \left(\int_0^\infty \tilde{x} \sin \tilde{x} 
    \tilde{F}(u,v,\tilde{x}) \mathrm{d} \tilde{x}\right)^2\equiv \frac{1}{2}\left[I_c(u,v)^2+I_s(u,v)^2\right],
\end{equation}
where we have defined $\tilde{F}(u,v,x)=1/2 \left[F(u,v,x)+F(v,u,x)\right]$. The expressions for $I_c$ and $I_s$ are
\begin{equation}
    \begin{aligned}
        I_c(u,v)&=\frac{9}{64 u^{3} v^{3}\kappa^{2} }\Bigg [-3 u^{2} v^{2}+\left(-3+u^{2}\right)\left(-3+u^{2}+2 v^{2}\right)\log\Bigg|1-\frac{u^{2}}{3}\Bigg|+\left(-3+v^{2}\right)-3+v^{2}+2 u^{2}\log\Bigg|1-\frac{v^{2}}{3}\Bigg|\\
        &-\frac{1}{2}\left(-3+v^{2}+u^{2}\right)^{2}\log\Bigg|\left(1-\frac{(u+v)^{2}}{3}\right)\left(1-\frac{(u-v)^{2}}{3}\right)\Bigg|\Bigg],
    \end{aligned}
\end{equation}
and
\begin{equation}
\begin{aligned}
I_{s}(u, v) & =\frac{9 \pi}{32 u^{3} v^{3} \kappa^{2} }\Bigg\{9-6 v^{2}-6 u^{2}+2 u^{2} v^{2} +\left(3-u^{2}\right)\left(-3+u^{2}+2 v^{2}\right) \Theta\left(1-\frac{u}{\sqrt{3}}\right) \\
& +\left(3-v^{2}\right)\left(-3+v^{2}+2 u^{2}\right) \Theta\left(1-\frac{v}{\sqrt{3}}\right) +\frac{1}{2}\left(-3+v^{2}+u^{2}\right)^{2}\left[\Theta\left(1-\frac{u+v}{\sqrt{3}}\right)+\Theta\left(1+\frac{u-v}{\sqrt{3}}\right)\right]\Bigg\}.
\end{aligned}
\end{equation}
With the expressions for kernel functions derived, one can calculate the energy density of IGWs in the Newtonian gauge using \Eq{Omegagw}, which is identical to the results obtained in Ref.~\cite{Domenech:2021and}.

\begin{figure*}[htbp!]
\centering
\includegraphics[width=0.48\textwidth]{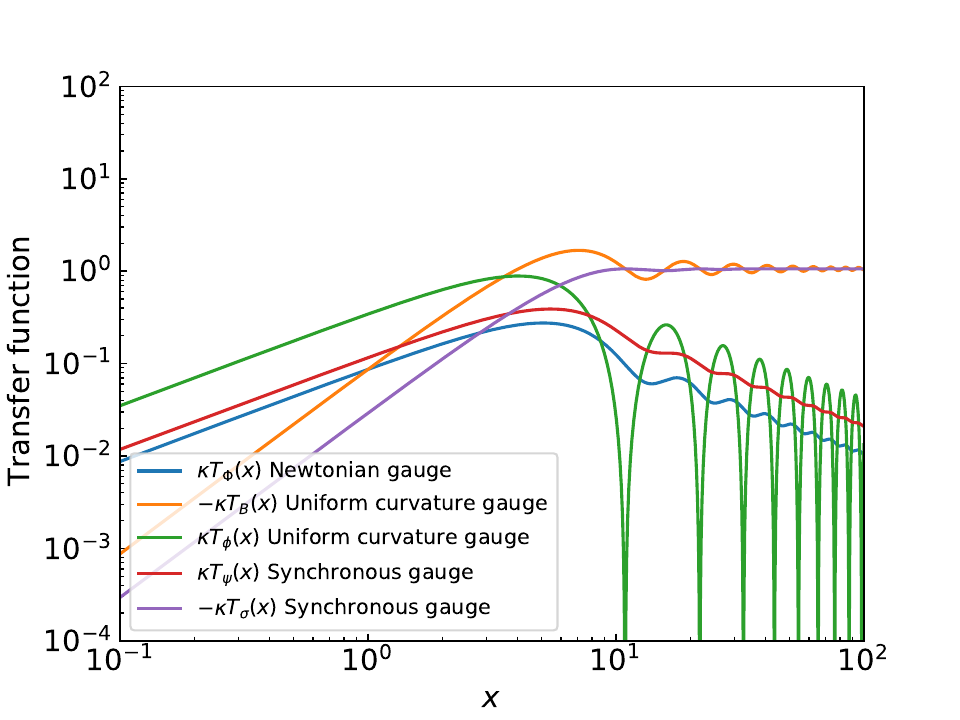}
\includegraphics[width =0.48\textwidth]{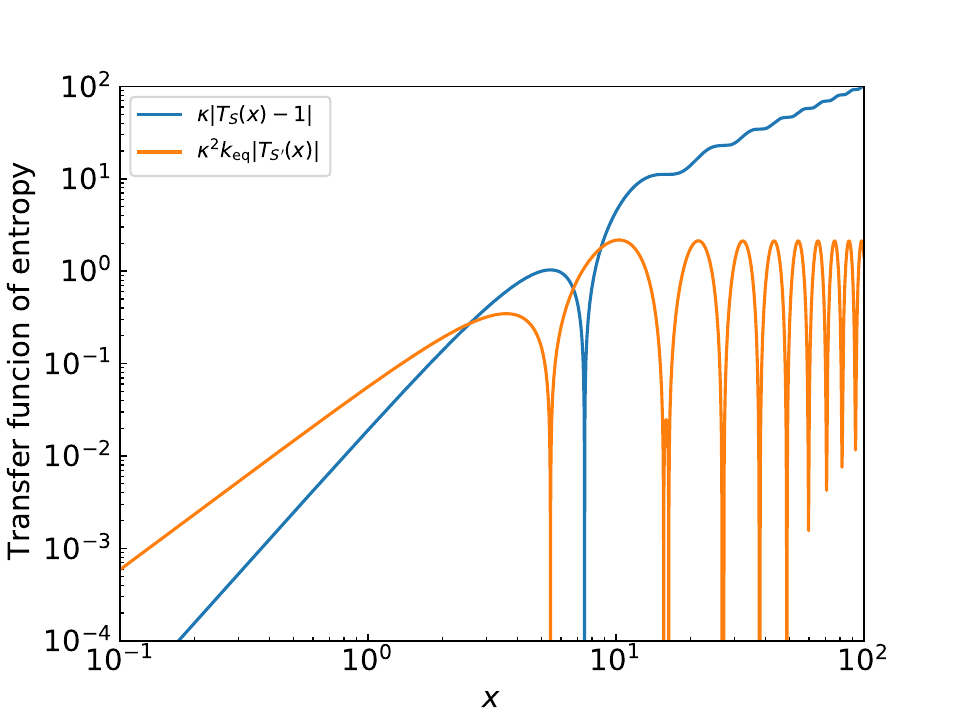}
\caption{\label{fig:T} Transfer functions of first-order perturbations in synchronous, uniform curvature and Newtonian gauges as a function of $x=k \eta$.
\textit{Left panel}: The transfer function for the metric perturbations.
\textit{Right panel}: The transfer function for the gauge-invariant entropy perturbation and its derivative.
}
\end{figure*}

\begin{figure}[htbp]
    \centering
    \includegraphics[width=0.48\columnwidth]{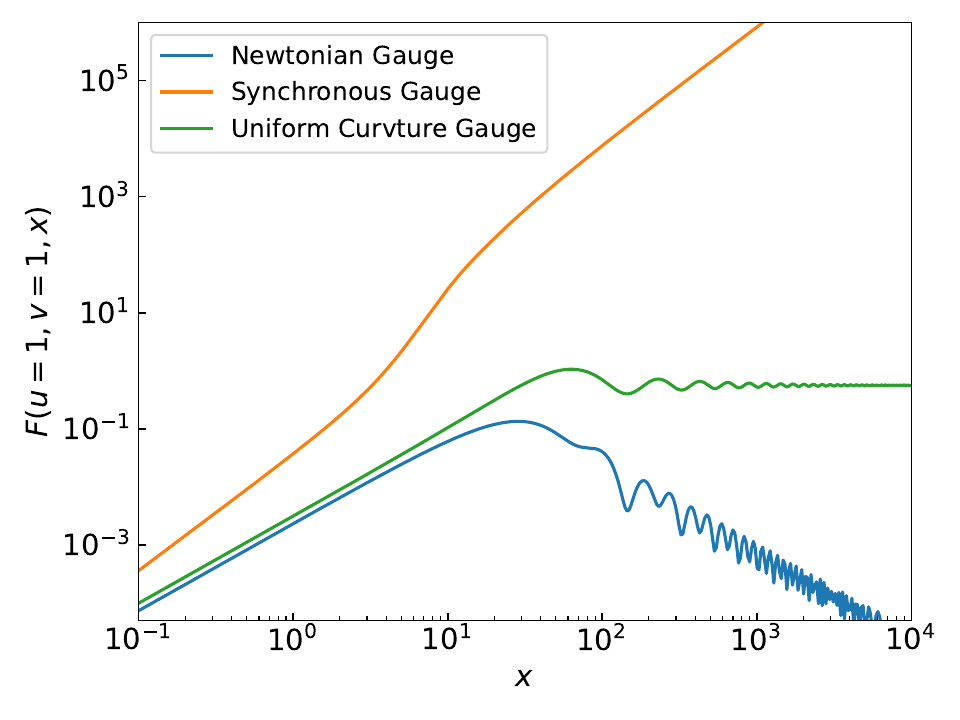}
    \includegraphics[width=0.48\columnwidth]{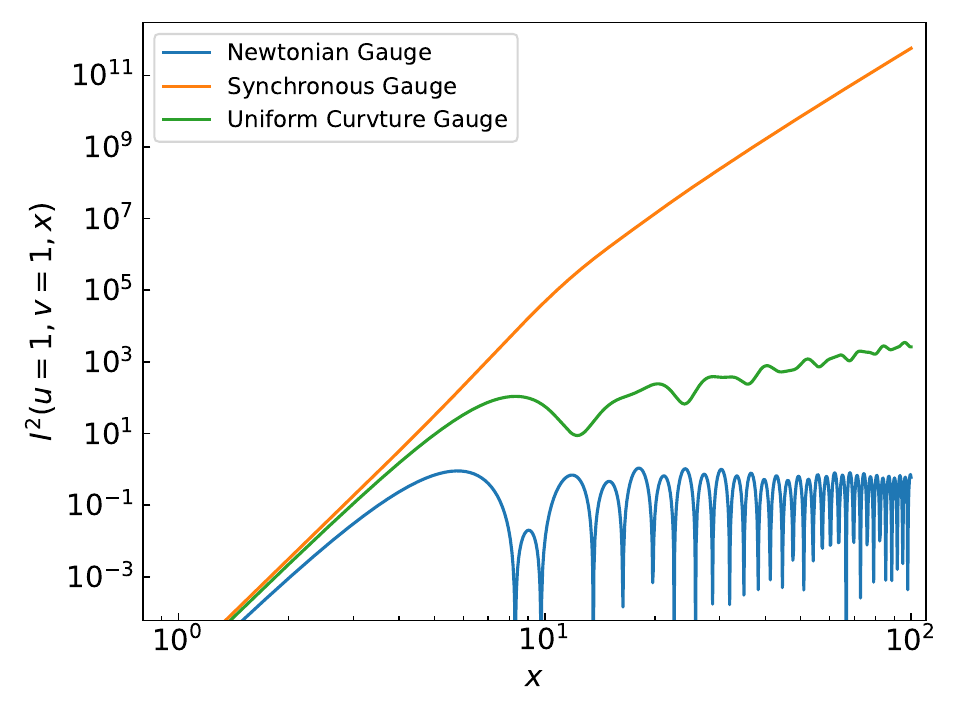}
    \caption{\textit{Left panel}: Evolution of the second-order source term transfer functions in Newtonian, synchronous, and uniform curvature gauges as a function of the dimensionless time variable $x=k\eta$, evaluated at $u=v=1$. \textit{Right panel}: Time evolution of the corresponding  $I^2(u,v,x)$ for IGWs in these three gauges, demonstrating their distinct asymptotic behaviors.}
    \label{fig:Fuv}
\end{figure}

In summary, we find that the results of IGW in synchronous, uniform curvature and Newtonian gauges are different, rendering the gauge choice in calculating IGW is important. 
Fig.~\ref{fig:T} presents a comparison of the transfer functions for linear perturbations in the synchronous, uniform curvature and Newtonian gauges, while Fig.~\ref{fig:Fuv} illustrates the transfer function of the source term for each gauge.

\section{Gauge Transformation}\label{gauge-trans}
In this section, we provide the gauge transformation from Newtonian gauge to an arbitrary gauge such that $x^\mu \to \tilde{x}^\mu =x^\mu +\xi^\mu$, where $\xi^\mu=(\alpha,\partial^{i}L)$. Under this transformation, the second-order tensor perturbation transforms as $h_{ij}\to h_{ij}+\chi_{ij}$, where the explicit expression for $\chi_{ij}$ is given in Ref.~\cite{DeLuca:2019ufz}. Under the gauge transformation, the modification to the IGWs is equivalent to transform the kernel function, namely
\begin{equation}
    I(u,v,x)\to I(u,v,x) +I_{\chi}(u,v,x).
\end{equation}For a general cosmological background characterized by equation of state $w$, the gauge transformation term takes the form (see e.g. Ref.~\cite{Lu:2020diy})
\begin{equation}
\begin{aligned}
    I_{\chi} = &-\frac{x^{2/(1+3w)}}{4uv} \Bigg[2 T_\alpha(ux)T_\alpha(vx)+ \frac{1 - u^2 - v^2}{uv} T_L(ux)T_L(vx) \\
    & -4\left( \frac{u}{v} T_{\phi}(ux)T_L(vx) + \frac{v}{u} T_{\phi}(vx)T_L(ux) \right) + \frac{8}{x(1+3w)}\left( \frac{1}{v} T_\alpha(ux)T_L(vx) + \frac{1}{u} T_L(ux)T_\alpha(vx) \right) 
\Bigg],
\end{aligned}
\end{equation}where $T_{\phi}$ is the transfer function of $\phi$ in Newtonian gauge and $T_\alpha$, $T_L$ are the transfer function of $\alpha$ and $L$ respectively, defined as
\begin{equation}
    T_\alpha(k\eta) = \frac{\alpha}{k S_{\boldsymbol{k}}}, \quad T_L(k\eta) = \frac{L}{k^2 S_{\boldsymbol{k}}}.
\end{equation}
The complete expression for the Newtonian gauge kernel function $I(u,v,x)$ with general $w$ can be found in Ref.~\cite{Domenech:2024wao}. With these expressions, one can compute the IGWs generated by isocurvature perturbations in any gauge for a general equation of state $w$.
{A special case arises for $w=0$, where both curvature and entropy perturbations remain constant at leading order and thus the generation of IGWs is highly suppressed in this case \cite{Domenech:2024wao}.}

\section{IGWs during the kination period}\label{w=1}
\begin{figure}[htbp]
    \centering
    \includegraphics[width=0.48\columnwidth]{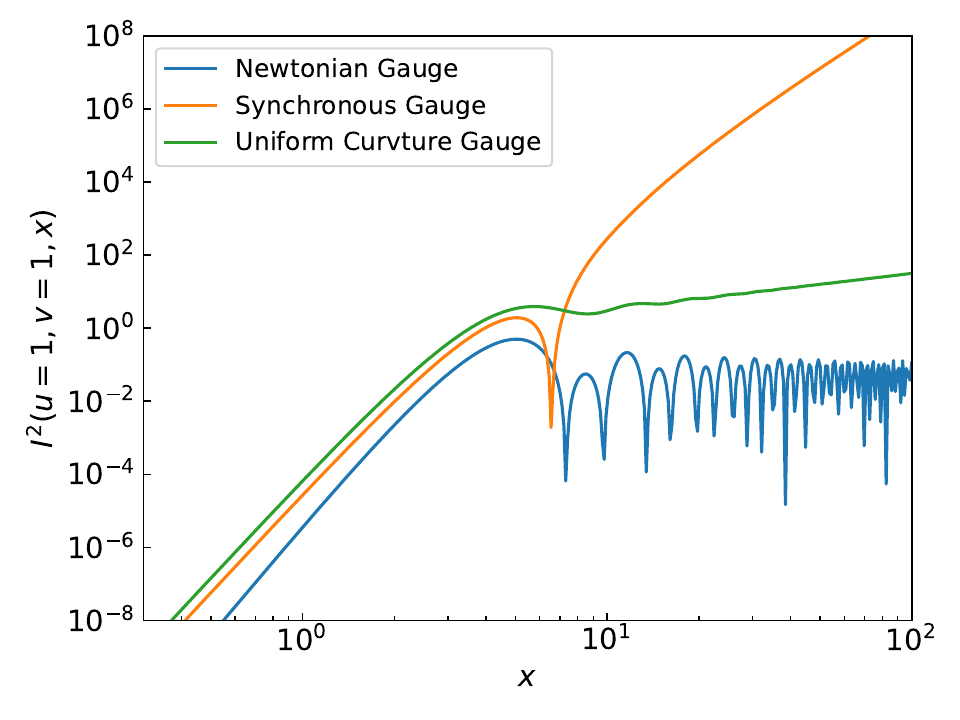}
    \caption{The same as the right panel of Fig.~\ref{fig:Fuv} but during kination stage where $w=1$.}
    \label{kination}
\end{figure}

We further present the result during the kination period where $w=1$. Firstly, we calculate the kernel functions of IGWs in Newtonian gauge, where the source term now becomes
\begin{equation}
        S_{ij} =2\phi \phi_{,ij} - \frac{2}{3}\left( \phi + {\phi'\over \mH} \right)_{,i}
        \left( \phi + {\phi'\over \mH} \right)_{,j},
\end{equation}
and we obtain the transfer function of $\phi$ by numerically solving the equation of motion:
\begin{equation}
    \begin{aligned}
        &{d^2T_S \over dx^2}+\left({1\over x}-{5\over 4\sqrt{2}\kappa} \right){dT_S \over dx}+{x \over 2\sqrt{2}\kappa }T_S-{x^2\over 3} T_\phi\simeq  0,\\
        &{d^2 T_\phi \over dx^2}+{6\over x}{d T_{\phi}\over dx}+\left(1+ {2\over x^2} \right){ T_\phi }
        +{1\over 2\sqrt{2}x \kappa}\left((1-x^2)T_{\phi} -3T_S \right)\simeq0,
    \end{aligned}
\end{equation}
The results of the kernel function are illustrated in Fig.~\ref{kination}. 
Similar to Fig.~\ref{fig:Fuv}, the kernel functions in both uniform curvature and synchronous gauges diverge in the sub-horizon limit, leading to persistent generation of tensor perturbations by isocurvature modes in these gauge choices.

\section{Summary and discussion}\label{summary}

In this paper, we have explored the gauge dependence of IGWs sourced by primordial isocurvature perturbations in the early Universe. We focused on three commonly used gauges: synchronous, Newtonian, and uniform curvature gauges, and derived the energy density spectra of IGWs in each case. Our results demonstrate that the choice of gauge significantly impacts the predicted IGW energy spectra, highlighting the importance of carefully considering gauge choice when studying IGWs from isocurvature perturbations.

We began by deriving the equations of motion for linear perturbations and the second-order source term in a general gauge during the radiation domination. We then applied these equations to the three specific gauges and determined the evolution of perturbations and the source term in each case. Our analysis revealed that the IGW energy density exhibits distinct behaviors in different gauges. In the synchronous and uniform curvature gauges, the IGW energy density grows with conformal time as $\eta^8$ and $\eta^4$, respectively. However, in the Newtonian gauge, the IGW energy density converges, yielding a finite result.
On the other hand, we derive a general gauge transformation from Newtonian gauge to an arbitrary gauge. {Since both curvature and isocurvature perturbations remain constant at leading order in the matter-dominated era, as shown in Ref. \cite{Domenech:2024wao}, the IGWs are highly suppressed. We then calculate the case $w=1$ as an example}. The IGWs for $w=1$ also exhibit divergence in the late-time limit in the uniform curvature gauge and the synchronous gauge while it converges in the Newtonian gauge.
These findings highlight the complexity of studying IGWs from isocurvature perturbations and the need for careful consideration of gauge choice in such analyses. The significant differences in IGW spectra across gauges suggest that the observable signatures of primordial isocurvature fluctuations through IGWs may be highly sensitive to the choice of gauge.


The gauge dependence of IGWs from isocurvature perturbations exhibits stronger divergences compared to the adiabatic case studied in Ref.~\cite{Lu:2020diy}. This enhanced gauge sensitivity stems from the distinct temporal evolution of scalar perturbations in the isocurvature scenario. Specifically, while the Bardeen potential $\psi$ in the adiabatic case decays as $\sim 1/x^2$ in synchronous gauge, the corresponding decay in the isocurvature scenario follows a slower $\sim 1/x$ behavior in the sub-horizon limit ($x \gg 1$). This slower decay of scalar perturbations leads to a more persistent source term for IGWs in the isocurvature case. Consequently, gauge choices that yield convergent results for adiabatic perturbations, such as the uniform curvature gauge, can produce unphysical divergences when applied to isocurvature perturbations. As emphasized in Ref.~\cite{Domenech:2020xin}, this persistent source term continuously generates new gravitational wave perturbations, rendering certain gauge choices unsuitable for physical calculations of IGWs from isocurvature modes.

In conclusion, our investigation of the gauge dependence of IGWs from primordial isocurvature perturbations has revealed significant differences in energy spectra across commonly used gauges. These results emphasize the importance of gauge choice in studying IGWs and offer new insights into the potential observational signatures of primordial fluctuations.
The analytical solutions we have derived for the perturbations contributing to the IGW spectra lay a solid foundation for future research into the intrinsic properties and behavior of isocurvature fluctuations in the early Universe.

\begin{acknowledgments}
Cosmological perturbations are derived using the \texttt{xPand}~\cite{Pitrou:2013hga} package.
C.Y. acknowledge the financial support provided under the European Union’s H2020 ERC Advanced Grant “Black holes: gravitational engines of discovery” grant agreement no. Gravitas–101052587. Views and opinions expressed are however those of the author only and do not necessarily reflect those of the European Union or the European Research Council. Neither the European Union nor the granting authority can be held responsible for them. We acknowledge support from the Villum Investigator program supported by the VILLUM Foundation (grant no. VIL37766) and the DNRF Chair program (grant no. DNRF162) by the Danish National Research Foundation.
ZCC is supported by the National Natural Science Foundation of China under Grant No.~12405056, the Natural Science Foundation of Hunan Province under Grant No.~2025JJ40006, and the Innovative Research Group of Hunan Province under Grant No.~2024JJ1006. 
LL is supported by the National Natural Science Foundation of China Grant under Grant No.~12433001. 
\end{acknowledgments}

\bibliography{ref}

\begin{thebibliography}{135}%
\makeatletter
\providecommand \@ifxundefined [1]{%
 \@ifx{#1\undefined}
}%
\providecommand \@ifnum [1]{%
 \ifnum #1\expandafter \@firstoftwo
 \else \expandafter \@secondoftwo
 \fi
}%
\providecommand \@ifx [1]{%
 \ifx #1\expandafter \@firstoftwo
 \else \expandafter \@secondoftwo
 \fi
}%
\providecommand \natexlab [1]{#1}%
\providecommand \enquote  [1]{``#1''}%
\providecommand \bibnamefont  [1]{#1}%
\providecommand \bibfnamefont [1]{#1}%
\providecommand \citenamefont [1]{#1}%
\providecommand \href@noop [0]{\@secondoftwo}%
\providecommand \href [0]{\begingroup \@sanitize@url \@href}%
\providecommand \@href[1]{\@@startlink{#1}\@@href}%
\providecommand \@@href[1]{\endgroup#1\@@endlink}%
\providecommand \@sanitize@url [0]{\catcode `\\12\catcode `\$12\catcode `\&12\catcode `\#12\catcode `\^12\catcode `\_12\catcode `\%12\relax}%
\providecommand \@@startlink[1]{}%
\providecommand \@@endlink[0]{}%
\providecommand \url  [0]{\begingroup\@sanitize@url \@url }%
\providecommand \@url [1]{\endgroup\@href {#1}{\urlprefix }}%
\providecommand \urlprefix  [0]{URL }%
\providecommand \Eprint [0]{\href }%
\providecommand \doibase [0]{http://dx.doi.org/}%
\providecommand \selectlanguage [0]{\@gobble}%
\providecommand \bibinfo  [0]{\@secondoftwo}%
\providecommand \bibfield  [0]{\@secondoftwo}%
\providecommand \translation [1]{[#1]}%
\providecommand \BibitemOpen [0]{}%
\providecommand \bibitemStop [0]{}%
\providecommand \bibitemNoStop [0]{.\EOS\space}%
\providecommand \EOS [0]{\spacefactor3000\relax}%
\providecommand \BibitemShut  [1]{\csname bibitem#1\endcsname}%
\let\auto@bib@innerbib\@empty
\bibitem [{\citenamefont {Kodama}\ and\ \citenamefont {Sasaki}(1984)}]{Kodama:1984ziu}%
  \BibitemOpen
  \bibfield  {author} {\bibinfo {author} {\bibfnamefont {Hideo}\ \bibnamefont {Kodama}}\ and\ \bibinfo {author} {\bibfnamefont {Misao}\ \bibnamefont {Sasaki}},\ }\bibfield  {title} {\enquote {\bibinfo {title} {{Cosmological Perturbation Theory}},}\ }\href {\doibase 10.1143/PTPS.78.1} {\bibfield  {journal} {\bibinfo  {journal} {Prog. Theor. Phys. Suppl.}\ }\textbf {\bibinfo {volume} {78}},\ \bibinfo {pages} {1--166} (\bibinfo {year} {1984})}\BibitemShut {NoStop}%
\bibitem [{\citenamefont {Bucher}\ \emph {et~al.}(2000)\citenamefont {Bucher}, \citenamefont {Moodley},\ and\ \citenamefont {Turok}}]{Bucher:1999re}%
  \BibitemOpen
  \bibfield  {author} {\bibinfo {author} {\bibfnamefont {Martin}\ \bibnamefont {Bucher}}, \bibinfo {author} {\bibfnamefont {Kavilan}\ \bibnamefont {Moodley}}, \ and\ \bibinfo {author} {\bibfnamefont {Neil}\ \bibnamefont {Turok}},\ }\bibfield  {title} {\enquote {\bibinfo {title} {{The General primordial cosmic perturbation}},}\ }\href {\doibase 10.1103/PhysRevD.62.083508} {\bibfield  {journal} {\bibinfo  {journal} {Phys. Rev. D}\ }\textbf {\bibinfo {volume} {62}},\ \bibinfo {pages} {083508} (\bibinfo {year} {2000})},\ \Eprint {http://arxiv.org/abs/astro-ph/9904231} {arXiv:astro-ph/9904231} \BibitemShut {NoStop}%
\bibitem [{\citenamefont {Akrami}\ \emph {et~al.}(2020{\natexlab{a}})\citenamefont {Akrami} \emph {et~al.}}]{Planck:2018jri}%
  \BibitemOpen
  \bibfield  {author} {\bibinfo {author} {\bibfnamefont {Y.}~\bibnamefont {Akrami}} \emph {et~al.} (\bibinfo {collaboration} {Planck}),\ }\bibfield  {title} {\enquote {\bibinfo {title} {{Planck 2018 results. X. Constraints on inflation}},}\ }\href {\doibase 10.1051/0004-6361/201833887} {\bibfield  {journal} {\bibinfo  {journal} {Astron. Astrophys.}\ }\textbf {\bibinfo {volume} {641}},\ \bibinfo {pages} {A10} (\bibinfo {year} {2020}{\natexlab{a}})},\ \Eprint {http://arxiv.org/abs/1807.06211} {arXiv:1807.06211 [astro-ph.CO]} \BibitemShut {NoStop}%
\bibitem [{\citenamefont {Linde}(1985)}]{Linde:1985yf}%
  \BibitemOpen
  \bibfield  {author} {\bibinfo {author} {\bibfnamefont {Andrei~D.}\ \bibnamefont {Linde}},\ }\bibfield  {title} {\enquote {\bibinfo {title} {{Generation of Isothermal Density Perturbations in the Inflationary Universe}},}\ }\href {\doibase 10.1016/0370-2693(85)90436-8} {\bibfield  {journal} {\bibinfo  {journal} {Phys. Lett. B}\ }\textbf {\bibinfo {volume} {158}},\ \bibinfo {pages} {375--380} (\bibinfo {year} {1985})}\BibitemShut {NoStop}%
\bibitem [{\citenamefont {Polarski}\ and\ \citenamefont {Starobinsky}(1994)}]{Polarski:1994rz}%
  \BibitemOpen
  \bibfield  {author} {\bibinfo {author} {\bibfnamefont {David}\ \bibnamefont {Polarski}}\ and\ \bibinfo {author} {\bibfnamefont {Alexei~A.}\ \bibnamefont {Starobinsky}},\ }\bibfield  {title} {\enquote {\bibinfo {title} {{Isocurvature perturbations in multiple inflationary models}},}\ }\href {\doibase 10.1103/PhysRevD.50.6123} {\bibfield  {journal} {\bibinfo  {journal} {Phys. Rev. D}\ }\textbf {\bibinfo {volume} {50}},\ \bibinfo {pages} {6123--6129} (\bibinfo {year} {1994})},\ \Eprint {http://arxiv.org/abs/astro-ph/9404061} {arXiv:astro-ph/9404061} \BibitemShut {NoStop}%
\bibitem [{\citenamefont {Akrami}\ \emph {et~al.}(2020{\natexlab{b}})\citenamefont {Akrami} \emph {et~al.}}]{Planck:2019kim}%
  \BibitemOpen
  \bibfield  {author} {\bibinfo {author} {\bibfnamefont {Y.}~\bibnamefont {Akrami}} \emph {et~al.} (\bibinfo {collaboration} {Planck}),\ }\bibfield  {title} {\enquote {\bibinfo {title} {{Planck 2018 results. IX. Constraints on primordial non-Gaussianity}},}\ }\href {\doibase 10.1051/0004-6361/201935891} {\bibfield  {journal} {\bibinfo  {journal} {Astron. Astrophys.}\ }\textbf {\bibinfo {volume} {641}},\ \bibinfo {pages} {A9} (\bibinfo {year} {2020}{\natexlab{b}})},\ \Eprint {http://arxiv.org/abs/1905.05697} {arXiv:1905.05697 [astro-ph.CO]} \BibitemShut {NoStop}%
\bibitem [{\citenamefont {Ananda}\ \emph {et~al.}(2007)\citenamefont {Ananda}, \citenamefont {Clarkson},\ and\ \citenamefont {Wands}}]{Ananda:2006af}%
  \BibitemOpen
  \bibfield  {author} {\bibinfo {author} {\bibfnamefont {Kishore~N.}\ \bibnamefont {Ananda}}, \bibinfo {author} {\bibfnamefont {Chris}\ \bibnamefont {Clarkson}}, \ and\ \bibinfo {author} {\bibfnamefont {David}\ \bibnamefont {Wands}},\ }\bibfield  {title} {\enquote {\bibinfo {title} {{The Cosmological gravitational wave background from primordial density perturbations}},}\ }\href {\doibase 10.1103/PhysRevD.75.123518} {\bibfield  {journal} {\bibinfo  {journal} {Phys. Rev. D}\ }\textbf {\bibinfo {volume} {75}},\ \bibinfo {pages} {123518} (\bibinfo {year} {2007})},\ \Eprint {http://arxiv.org/abs/gr-qc/0612013} {arXiv:gr-qc/0612013} \BibitemShut {NoStop}%
\bibitem [{\citenamefont {Baumann}\ \emph {et~al.}(2007)\citenamefont {Baumann}, \citenamefont {Steinhardt}, \citenamefont {Takahashi},\ and\ \citenamefont {Ichiki}}]{Baumann:2007zm}%
  \BibitemOpen
  \bibfield  {author} {\bibinfo {author} {\bibfnamefont {Daniel}\ \bibnamefont {Baumann}}, \bibinfo {author} {\bibfnamefont {Paul~J.}\ \bibnamefont {Steinhardt}}, \bibinfo {author} {\bibfnamefont {Keitaro}\ \bibnamefont {Takahashi}}, \ and\ \bibinfo {author} {\bibfnamefont {Kiyotomo}\ \bibnamefont {Ichiki}},\ }\bibfield  {title} {\enquote {\bibinfo {title} {{Gravitational Wave Spectrum Induced by Primordial Scalar Perturbations}},}\ }\href {\doibase 10.1103/PhysRevD.76.084019} {\bibfield  {journal} {\bibinfo  {journal} {Phys. Rev. D}\ }\textbf {\bibinfo {volume} {76}},\ \bibinfo {pages} {084019} (\bibinfo {year} {2007})},\ \Eprint {http://arxiv.org/abs/hep-th/0703290} {arXiv:hep-th/0703290} \BibitemShut {NoStop}%
\bibitem [{\citenamefont {Garcia-Bellido}\ \emph {et~al.}(2016)\citenamefont {Garcia-Bellido}, \citenamefont {Peloso},\ and\ \citenamefont {Unal}}]{Garcia-Bellido:2016dkw}%
  \BibitemOpen
  \bibfield  {author} {\bibinfo {author} {\bibfnamefont {Juan}\ \bibnamefont {Garcia-Bellido}}, \bibinfo {author} {\bibfnamefont {Marco}\ \bibnamefont {Peloso}}, \ and\ \bibinfo {author} {\bibfnamefont {Caner}\ \bibnamefont {Unal}},\ }\bibfield  {title} {\enquote {\bibinfo {title} {{Gravitational waves at interferometer scales and primordial black holes in axion inflation}},}\ }\href {\doibase 10.1088/1475-7516/2016/12/031} {\bibfield  {journal} {\bibinfo  {journal} {JCAP}\ }\textbf {\bibinfo {volume} {12}},\ \bibinfo {pages} {031} (\bibinfo {year} {2016})},\ \Eprint {http://arxiv.org/abs/1610.03763} {arXiv:1610.03763 [astro-ph.CO]} \BibitemShut {NoStop}%
\bibitem [{\citenamefont {Inomata}\ \emph {et~al.}(2017)\citenamefont {Inomata}, \citenamefont {Kawasaki}, \citenamefont {Mukaida}, \citenamefont {Tada},\ and\ \citenamefont {Yanagida}}]{Inomata:2016rbd}%
  \BibitemOpen
  \bibfield  {author} {\bibinfo {author} {\bibfnamefont {Keisuke}\ \bibnamefont {Inomata}}, \bibinfo {author} {\bibfnamefont {Masahiro}\ \bibnamefont {Kawasaki}}, \bibinfo {author} {\bibfnamefont {Kyohei}\ \bibnamefont {Mukaida}}, \bibinfo {author} {\bibfnamefont {Yuichiro}\ \bibnamefont {Tada}}, \ and\ \bibinfo {author} {\bibfnamefont {Tsutomu~T.}\ \bibnamefont {Yanagida}},\ }\bibfield  {title} {\enquote {\bibinfo {title} {{Inflationary primordial black holes for the LIGO gravitational wave events and pulsar timing array experiments}},}\ }\href {\doibase 10.1103/PhysRevD.95.123510} {\bibfield  {journal} {\bibinfo  {journal} {Phys. Rev. D}\ }\textbf {\bibinfo {volume} {95}},\ \bibinfo {pages} {123510} (\bibinfo {year} {2017})},\ \Eprint {http://arxiv.org/abs/1611.06130} {arXiv:1611.06130 [astro-ph.CO]} \BibitemShut {NoStop}%
\bibitem [{\citenamefont {Garcia-Bellido}\ \emph {et~al.}(2017)\citenamefont {Garcia-Bellido}, \citenamefont {Peloso},\ and\ \citenamefont {Unal}}]{Garcia-Bellido:2017aan}%
  \BibitemOpen
  \bibfield  {author} {\bibinfo {author} {\bibfnamefont {Juan}\ \bibnamefont {Garcia-Bellido}}, \bibinfo {author} {\bibfnamefont {Marco}\ \bibnamefont {Peloso}}, \ and\ \bibinfo {author} {\bibfnamefont {Caner}\ \bibnamefont {Unal}},\ }\bibfield  {title} {\enquote {\bibinfo {title} {{Gravitational Wave signatures of inflationary models from Primordial Black Hole Dark Matter}},}\ }\href {\doibase 10.1088/1475-7516/2017/09/013} {\bibfield  {journal} {\bibinfo  {journal} {JCAP}\ }\textbf {\bibinfo {volume} {09}},\ \bibinfo {pages} {013} (\bibinfo {year} {2017})},\ \Eprint {http://arxiv.org/abs/1707.02441} {arXiv:1707.02441 [astro-ph.CO]} \BibitemShut {NoStop}%
\bibitem [{\citenamefont {Kohri}\ and\ \citenamefont {Terada}(2018)}]{Kohri:2018awv}%
  \BibitemOpen
  \bibfield  {author} {\bibinfo {author} {\bibfnamefont {Kazunori}\ \bibnamefont {Kohri}}\ and\ \bibinfo {author} {\bibfnamefont {Takahiro}\ \bibnamefont {Terada}},\ }\bibfield  {title} {\enquote {\bibinfo {title} {{Semianalytic calculation of gravitational wave spectrum nonlinearly induced from primordial curvature perturbations}},}\ }\href {\doibase 10.1103/PhysRevD.97.123532} {\bibfield  {journal} {\bibinfo  {journal} {Phys. Rev. D}\ }\textbf {\bibinfo {volume} {97}},\ \bibinfo {pages} {123532} (\bibinfo {year} {2018})},\ \Eprint {http://arxiv.org/abs/1804.08577} {arXiv:1804.08577 [gr-qc]} \BibitemShut {NoStop}%
\bibitem [{\citenamefont {Cai}\ \emph {et~al.}(2019{\natexlab{a}})\citenamefont {Cai}, \citenamefont {Pi},\ and\ \citenamefont {Sasaki}}]{Cai:2018dig}%
  \BibitemOpen
  \bibfield  {author} {\bibinfo {author} {\bibfnamefont {Rong-gen}\ \bibnamefont {Cai}}, \bibinfo {author} {\bibfnamefont {Shi}\ \bibnamefont {Pi}}, \ and\ \bibinfo {author} {\bibfnamefont {Misao}\ \bibnamefont {Sasaki}},\ }\bibfield  {title} {\enquote {\bibinfo {title} {{Gravitational Waves Induced by non-Gaussian Scalar Perturbations}},}\ }\href {\doibase 10.1103/PhysRevLett.122.201101} {\bibfield  {journal} {\bibinfo  {journal} {Phys. Rev. Lett.}\ }\textbf {\bibinfo {volume} {122}},\ \bibinfo {pages} {201101} (\bibinfo {year} {2019}{\natexlab{a}})},\ \Eprint {http://arxiv.org/abs/1810.11000} {arXiv:1810.11000 [astro-ph.CO]} \BibitemShut {NoStop}%
\bibitem [{\citenamefont {Zel'dovich}\ and\ \citenamefont {Novikov}(1967)}]{Zeldovich:1967lct}%
  \BibitemOpen
  \bibfield  {author} {\bibinfo {author} {\bibfnamefont {Ya.~B.}\ \bibnamefont {Zel'dovich}}\ and\ \bibinfo {author} {\bibfnamefont {I.~D.}\ \bibnamefont {Novikov}},\ }\bibfield  {title} {\enquote {\bibinfo {title} {{The Hypothesis of Cores Retarded during Expansion and the Hot Cosmological Model}},}\ }\href@noop {} {\bibfield  {journal} {\bibinfo  {journal} {Sov. Astron.}\ }\textbf {\bibinfo {volume} {10}},\ \bibinfo {pages} {602} (\bibinfo {year} {1967})}\BibitemShut {NoStop}%
\bibitem [{\citenamefont {Hawking}(1971)}]{Hawking:1971ei}%
  \BibitemOpen
  \bibfield  {author} {\bibinfo {author} {\bibfnamefont {Stephen}\ \bibnamefont {Hawking}},\ }\bibfield  {title} {\enquote {\bibinfo {title} {{Gravitationally collapsed objects of very low mass}},}\ }\href {\doibase 10.1093/mnras/152.1.75} {\bibfield  {journal} {\bibinfo  {journal} {Mon. Not. Roy. Astron. Soc.}\ }\textbf {\bibinfo {volume} {152}},\ \bibinfo {pages} {75} (\bibinfo {year} {1971})}\BibitemShut {NoStop}%
\bibitem [{\citenamefont {Carr}\ and\ \citenamefont {Hawking}(1974)}]{Carr:1974nx}%
  \BibitemOpen
  \bibfield  {author} {\bibinfo {author} {\bibfnamefont {Bernard~J.}\ \bibnamefont {Carr}}\ and\ \bibinfo {author} {\bibfnamefont {S.~W.}\ \bibnamefont {Hawking}},\ }\bibfield  {title} {\enquote {\bibinfo {title} {{Black holes in the early Universe}},}\ }\href {\doibase 10.1093/mnras/168.2.399} {\bibfield  {journal} {\bibinfo  {journal} {Mon. Not. Roy. Astron. Soc.}\ }\textbf {\bibinfo {volume} {168}},\ \bibinfo {pages} {399--415} (\bibinfo {year} {1974})}\BibitemShut {NoStop}%
\bibitem [{\citenamefont {Saito}\ and\ \citenamefont {Yokoyama}(2009)}]{Saito:2008jc}%
  \BibitemOpen
  \bibfield  {author} {\bibinfo {author} {\bibfnamefont {Ryo}\ \bibnamefont {Saito}}\ and\ \bibinfo {author} {\bibfnamefont {Jun'ichi}\ \bibnamefont {Yokoyama}},\ }\bibfield  {title} {\enquote {\bibinfo {title} {{Gravitational wave background as a probe of the primordial black hole abundance}},}\ }\href {\doibase 10.1103/PhysRevLett.102.161101} {\bibfield  {journal} {\bibinfo  {journal} {Phys. Rev. Lett.}\ }\textbf {\bibinfo {volume} {102}},\ \bibinfo {pages} {161101} (\bibinfo {year} {2009})},\ \bibinfo {note} {[Erratum: Phys.Rev.Lett. 107, 069901 (2011)]},\ \Eprint {http://arxiv.org/abs/0812.4339} {arXiv:0812.4339 [astro-ph]} \BibitemShut {NoStop}%
\bibitem [{\citenamefont {Belotsky}\ \emph {et~al.}(2014)\citenamefont {Belotsky}, \citenamefont {Dmitriev}, \citenamefont {Esipova}, \citenamefont {Gani}, \citenamefont {Grobov}, \citenamefont {Khlopov}, \citenamefont {Kirillov}, \citenamefont {Rubin},\ and\ \citenamefont {Svadkovsky}}]{Belotsky:2014kca}%
  \BibitemOpen
  \bibfield  {author} {\bibinfo {author} {\bibfnamefont {K.~M.}\ \bibnamefont {Belotsky}}, \bibinfo {author} {\bibfnamefont {A.~D.}\ \bibnamefont {Dmitriev}}, \bibinfo {author} {\bibfnamefont {E.~A.}\ \bibnamefont {Esipova}}, \bibinfo {author} {\bibfnamefont {V.~A.}\ \bibnamefont {Gani}}, \bibinfo {author} {\bibfnamefont {A.~V.}\ \bibnamefont {Grobov}}, \bibinfo {author} {\bibfnamefont {M.~Yu.}\ \bibnamefont {Khlopov}}, \bibinfo {author} {\bibfnamefont {A.~A.}\ \bibnamefont {Kirillov}}, \bibinfo {author} {\bibfnamefont {S.~G.}\ \bibnamefont {Rubin}}, \ and\ \bibinfo {author} {\bibfnamefont {I.~V.}\ \bibnamefont {Svadkovsky}},\ }\bibfield  {title} {\enquote {\bibinfo {title} {{Signatures of primordial black hole dark matter}},}\ }\href {\doibase 10.1142/S0217732314400057} {\bibfield  {journal} {\bibinfo  {journal} {Mod. Phys. Lett. A}\ }\textbf {\bibinfo {volume} {29}},\ \bibinfo {pages} {1440005} (\bibinfo {year} {2014})},\ \Eprint {http://arxiv.org/abs/1410.0203} {arXiv:1410.0203 [astro-ph.CO]} \BibitemShut
  {NoStop}%
\bibitem [{\citenamefont {Carr}\ \emph {et~al.}(2016)\citenamefont {Carr}, \citenamefont {Kuhnel},\ and\ \citenamefont {Sandstad}}]{Carr:2016drx}%
  \BibitemOpen
  \bibfield  {author} {\bibinfo {author} {\bibfnamefont {Bernard}\ \bibnamefont {Carr}}, \bibinfo {author} {\bibfnamefont {Florian}\ \bibnamefont {Kuhnel}}, \ and\ \bibinfo {author} {\bibfnamefont {Marit}\ \bibnamefont {Sandstad}},\ }\bibfield  {title} {\enquote {\bibinfo {title} {{Primordial Black Holes as Dark Matter}},}\ }\href {\doibase 10.1103/PhysRevD.94.083504} {\bibfield  {journal} {\bibinfo  {journal} {Phys. Rev. D}\ }\textbf {\bibinfo {volume} {94}},\ \bibinfo {pages} {083504} (\bibinfo {year} {2016})},\ \Eprint {http://arxiv.org/abs/1607.06077} {arXiv:1607.06077 [astro-ph.CO]} \BibitemShut {NoStop}%
\bibitem [{\citenamefont {Garcia-Bellido}\ and\ \citenamefont {Ruiz~Morales}(2017)}]{Garcia-Bellido:2017mdw}%
  \BibitemOpen
  \bibfield  {author} {\bibinfo {author} {\bibfnamefont {Juan}\ \bibnamefont {Garcia-Bellido}}\ and\ \bibinfo {author} {\bibfnamefont {Ester}\ \bibnamefont {Ruiz~Morales}},\ }\bibfield  {title} {\enquote {\bibinfo {title} {{Primordial black holes from single field models of inflation}},}\ }\href {\doibase 10.1016/j.dark.2017.09.007} {\bibfield  {journal} {\bibinfo  {journal} {Phys. Dark Univ.}\ }\textbf {\bibinfo {volume} {18}},\ \bibinfo {pages} {47--54} (\bibinfo {year} {2017})},\ \Eprint {http://arxiv.org/abs/1702.03901} {arXiv:1702.03901 [astro-ph.CO]} \BibitemShut {NoStop}%
\bibitem [{\citenamefont {Carr}\ \emph {et~al.}(2017)\citenamefont {Carr}, \citenamefont {Raidal}, \citenamefont {Tenkanen}, \citenamefont {Vaskonen},\ and\ \citenamefont {Veerm\"ae}}]{Carr:2017jsz}%
  \BibitemOpen
  \bibfield  {author} {\bibinfo {author} {\bibfnamefont {Bernard}\ \bibnamefont {Carr}}, \bibinfo {author} {\bibfnamefont {Martti}\ \bibnamefont {Raidal}}, \bibinfo {author} {\bibfnamefont {Tommi}\ \bibnamefont {Tenkanen}}, \bibinfo {author} {\bibfnamefont {Ville}\ \bibnamefont {Vaskonen}}, \ and\ \bibinfo {author} {\bibfnamefont {Hardi}\ \bibnamefont {Veerm\"ae}},\ }\bibfield  {title} {\enquote {\bibinfo {title} {{Primordial black hole constraints for extended mass functions}},}\ }\href {\doibase 10.1103/PhysRevD.96.023514} {\bibfield  {journal} {\bibinfo  {journal} {Phys. Rev. D}\ }\textbf {\bibinfo {volume} {96}},\ \bibinfo {pages} {023514} (\bibinfo {year} {2017})},\ \Eprint {http://arxiv.org/abs/1705.05567} {arXiv:1705.05567 [astro-ph.CO]} \BibitemShut {NoStop}%
\bibitem [{\citenamefont {Germani}\ and\ \citenamefont {Prokopec}(2017)}]{Germani:2017bcs}%
  \BibitemOpen
  \bibfield  {author} {\bibinfo {author} {\bibfnamefont {Cristiano}\ \bibnamefont {Germani}}\ and\ \bibinfo {author} {\bibfnamefont {Tomislav}\ \bibnamefont {Prokopec}},\ }\bibfield  {title} {\enquote {\bibinfo {title} {{On primordial black holes from an inflection point}},}\ }\href {\doibase 10.1016/j.dark.2017.09.001} {\bibfield  {journal} {\bibinfo  {journal} {Phys. Dark Univ.}\ }\textbf {\bibinfo {volume} {18}},\ \bibinfo {pages} {6--10} (\bibinfo {year} {2017})},\ \Eprint {http://arxiv.org/abs/1706.04226} {arXiv:1706.04226 [astro-ph.CO]} \BibitemShut {NoStop}%
\bibitem [{\citenamefont {Chen}\ and\ \citenamefont {Huang}(2018)}]{Chen:2018czv}%
  \BibitemOpen
  \bibfield  {author} {\bibinfo {author} {\bibfnamefont {Zu-Cheng}\ \bibnamefont {Chen}}\ and\ \bibinfo {author} {\bibfnamefont {Qing-Guo}\ \bibnamefont {Huang}},\ }\bibfield  {title} {\enquote {\bibinfo {title} {{Merger Rate Distribution of Primordial-Black-Hole Binaries}},}\ }\href {\doibase 10.3847/1538-4357/aad6e2} {\bibfield  {journal} {\bibinfo  {journal} {Astrophys. J.}\ }\textbf {\bibinfo {volume} {864}},\ \bibinfo {pages} {61} (\bibinfo {year} {2018})},\ \Eprint {http://arxiv.org/abs/1801.10327} {arXiv:1801.10327 [astro-ph.CO]} \BibitemShut {NoStop}%
\bibitem [{\citenamefont {Liu}\ \emph {et~al.}(2019{\natexlab{a}})\citenamefont {Liu}, \citenamefont {Guo},\ and\ \citenamefont {Cai}}]{Liu:2018ess}%
  \BibitemOpen
  \bibfield  {author} {\bibinfo {author} {\bibfnamefont {Lang}\ \bibnamefont {Liu}}, \bibinfo {author} {\bibfnamefont {Zong-Kuan}\ \bibnamefont {Guo}}, \ and\ \bibinfo {author} {\bibfnamefont {Rong-Gen}\ \bibnamefont {Cai}},\ }\bibfield  {title} {\enquote {\bibinfo {title} {{Effects of the surrounding primordial black holes on the merger rate of primordial black hole binaries}},}\ }\href {\doibase 10.1103/PhysRevD.99.063523} {\bibfield  {journal} {\bibinfo  {journal} {Phys. Rev. D}\ }\textbf {\bibinfo {volume} {99}},\ \bibinfo {pages} {063523} (\bibinfo {year} {2019}{\natexlab{a}})},\ \Eprint {http://arxiv.org/abs/1812.05376} {arXiv:1812.05376 [astro-ph.CO]} \BibitemShut {NoStop}%
\bibitem [{\citenamefont {Liu}\ \emph {et~al.}(2019{\natexlab{b}})\citenamefont {Liu}, \citenamefont {Guo},\ and\ \citenamefont {Cai}}]{Liu:2019rnx}%
  \BibitemOpen
  \bibfield  {author} {\bibinfo {author} {\bibfnamefont {Lang}\ \bibnamefont {Liu}}, \bibinfo {author} {\bibfnamefont {Zong-Kuan}\ \bibnamefont {Guo}}, \ and\ \bibinfo {author} {\bibfnamefont {Rong-Gen}\ \bibnamefont {Cai}},\ }\bibfield  {title} {\enquote {\bibinfo {title} {{Effects of the merger history on the merger rate density of primordial black hole binaries}},}\ }\href {\doibase 10.1140/epjc/s10052-019-7227-0} {\bibfield  {journal} {\bibinfo  {journal} {Eur. Phys. J. C}\ }\textbf {\bibinfo {volume} {79}},\ \bibinfo {pages} {717} (\bibinfo {year} {2019}{\natexlab{b}})},\ \Eprint {http://arxiv.org/abs/1901.07672} {arXiv:1901.07672 [astro-ph.CO]} \BibitemShut {NoStop}%
\bibitem [{\citenamefont {Chen}\ \emph {et~al.}(2019)\citenamefont {Chen}, \citenamefont {Huang},\ and\ \citenamefont {Huang}}]{Chen:2018rzo}%
  \BibitemOpen
  \bibfield  {author} {\bibinfo {author} {\bibfnamefont {Zu-Cheng}\ \bibnamefont {Chen}}, \bibinfo {author} {\bibfnamefont {Fan}\ \bibnamefont {Huang}}, \ and\ \bibinfo {author} {\bibfnamefont {Qing-Guo}\ \bibnamefont {Huang}},\ }\bibfield  {title} {\enquote {\bibinfo {title} {{Stochastic Gravitational-wave Background from Binary Black Holes and Binary Neutron Stars and Implications for LISA}},}\ }\href {\doibase 10.3847/1538-4357/aaf581} {\bibfield  {journal} {\bibinfo  {journal} {Astrophys. J.}\ }\textbf {\bibinfo {volume} {871}},\ \bibinfo {pages} {97} (\bibinfo {year} {2019})},\ \Eprint {http://arxiv.org/abs/1809.10360} {arXiv:1809.10360 [gr-qc]} \BibitemShut {NoStop}%
\bibitem [{\citenamefont {Chen}\ and\ \citenamefont {Huang}(2020)}]{Chen:2019irf}%
  \BibitemOpen
  \bibfield  {author} {\bibinfo {author} {\bibfnamefont {Zu-Cheng}\ \bibnamefont {Chen}}\ and\ \bibinfo {author} {\bibfnamefont {Qing-Guo}\ \bibnamefont {Huang}},\ }\bibfield  {title} {\enquote {\bibinfo {title} {{Distinguishing Primordial Black Holes from Astrophysical Black Holes by Einstein Telescope and Cosmic Explorer}},}\ }\href {\doibase 10.1088/1475-7516/2020/08/039} {\bibfield  {journal} {\bibinfo  {journal} {JCAP}\ }\textbf {\bibinfo {volume} {08}},\ \bibinfo {pages} {039} (\bibinfo {year} {2020})},\ \Eprint {http://arxiv.org/abs/1904.02396} {arXiv:1904.02396 [astro-ph.CO]} \BibitemShut {NoStop}%
\bibitem [{\citenamefont {Wang}\ \emph {et~al.}(2019)\citenamefont {Wang}, \citenamefont {Terada},\ and\ \citenamefont {Kohri}}]{Wang:2019kaf}%
  \BibitemOpen
  \bibfield  {author} {\bibinfo {author} {\bibfnamefont {Sai}\ \bibnamefont {Wang}}, \bibinfo {author} {\bibfnamefont {Takahiro}\ \bibnamefont {Terada}}, \ and\ \bibinfo {author} {\bibfnamefont {Kazunori}\ \bibnamefont {Kohri}},\ }\bibfield  {title} {\enquote {\bibinfo {title} {{Prospective constraints on the primordial black hole abundance from the stochastic gravitational-wave backgrounds produced by coalescing events and curvature perturbations}},}\ }\href {\doibase 10.1103/PhysRevD.99.103531} {\bibfield  {journal} {\bibinfo  {journal} {Phys. Rev. D}\ }\textbf {\bibinfo {volume} {99}},\ \bibinfo {pages} {103531} (\bibinfo {year} {2019})},\ \bibinfo {note} {[Erratum: Phys.Rev.D 101, 069901 (2020)]},\ \Eprint {http://arxiv.org/abs/1903.05924} {arXiv:1903.05924 [astro-ph.CO]} \BibitemShut {NoStop}%
\bibitem [{\citenamefont {Cai}\ \emph {et~al.}(2020)\citenamefont {Cai}, \citenamefont {Guo}, \citenamefont {Liu}, \citenamefont {Liu},\ and\ \citenamefont {Yang}}]{Cai:2019bmk}%
  \BibitemOpen
  \bibfield  {author} {\bibinfo {author} {\bibfnamefont {Rong-Gen}\ \bibnamefont {Cai}}, \bibinfo {author} {\bibfnamefont {Zong-Kuan}\ \bibnamefont {Guo}}, \bibinfo {author} {\bibfnamefont {Jing}\ \bibnamefont {Liu}}, \bibinfo {author} {\bibfnamefont {Lang}\ \bibnamefont {Liu}}, \ and\ \bibinfo {author} {\bibfnamefont {Xing-Yu}\ \bibnamefont {Yang}},\ }\bibfield  {title} {\enquote {\bibinfo {title} {{Primordial black holes and gravitational waves from parametric amplification of curvature perturbations}},}\ }\href {\doibase 10.1088/1475-7516/2020/06/013} {\bibfield  {journal} {\bibinfo  {journal} {JCAP}\ }\textbf {\bibinfo {volume} {06}},\ \bibinfo {pages} {013} (\bibinfo {year} {2020})},\ \Eprint {http://arxiv.org/abs/1912.10437} {arXiv:1912.10437 [astro-ph.CO]} \BibitemShut {NoStop}%
\bibitem [{\citenamefont {Liu}\ \emph {et~al.}(2020{\natexlab{a}})\citenamefont {Liu}, \citenamefont {Guo}, \citenamefont {Cai},\ and\ \citenamefont {Kim}}]{Liu:2020cds}%
  \BibitemOpen
  \bibfield  {author} {\bibinfo {author} {\bibfnamefont {Lang}\ \bibnamefont {Liu}}, \bibinfo {author} {\bibfnamefont {Zong-Kuan}\ \bibnamefont {Guo}}, \bibinfo {author} {\bibfnamefont {Rong-Gen}\ \bibnamefont {Cai}}, \ and\ \bibinfo {author} {\bibfnamefont {Sang~Pyo}\ \bibnamefont {Kim}},\ }\bibfield  {title} {\enquote {\bibinfo {title} {{Merger rate distribution of primordial black hole binaries with electric charges}},}\ }\href {\doibase 10.1103/PhysRevD.102.043508} {\bibfield  {journal} {\bibinfo  {journal} {Phys. Rev. D}\ }\textbf {\bibinfo {volume} {102}},\ \bibinfo {pages} {043508} (\bibinfo {year} {2020}{\natexlab{a}})},\ \Eprint {http://arxiv.org/abs/2001.02984} {arXiv:2001.02984 [astro-ph.CO]} \BibitemShut {NoStop}%
\bibitem [{\citenamefont {Wu}(2020)}]{Wu:2020drm}%
  \BibitemOpen
  \bibfield  {author} {\bibinfo {author} {\bibfnamefont {You}\ \bibnamefont {Wu}},\ }\bibfield  {title} {\enquote {\bibinfo {title} {{Merger history of primordial black-hole binaries}},}\ }\href {\doibase 10.1103/PhysRevD.101.083008} {\bibfield  {journal} {\bibinfo  {journal} {Phys. Rev. D}\ }\textbf {\bibinfo {volume} {101}},\ \bibinfo {pages} {083008} (\bibinfo {year} {2020})},\ \Eprint {http://arxiv.org/abs/2001.03833} {arXiv:2001.03833 [astro-ph.CO]} \BibitemShut {NoStop}%
\bibitem [{\citenamefont {De~Luca}\ \emph {et~al.}(2021{\natexlab{a}})\citenamefont {De~Luca}, \citenamefont {Desjacques}, \citenamefont {Franciolini}, \citenamefont {Pani},\ and\ \citenamefont {Riotto}}]{DeLuca:2020sae}%
  \BibitemOpen
  \bibfield  {author} {\bibinfo {author} {\bibfnamefont {V.}~\bibnamefont {De~Luca}}, \bibinfo {author} {\bibfnamefont {V.}~\bibnamefont {Desjacques}}, \bibinfo {author} {\bibfnamefont {G.}~\bibnamefont {Franciolini}}, \bibinfo {author} {\bibfnamefont {P.}~\bibnamefont {Pani}}, \ and\ \bibinfo {author} {\bibfnamefont {A.}~\bibnamefont {Riotto}},\ }\bibfield  {title} {\enquote {\bibinfo {title} {{GW190521 Mass Gap Event and the Primordial Black Hole Scenario}},}\ }\href {\doibase 10.1103/PhysRevLett.126.051101} {\bibfield  {journal} {\bibinfo  {journal} {Phys. Rev. Lett.}\ }\textbf {\bibinfo {volume} {126}},\ \bibinfo {pages} {051101} (\bibinfo {year} {2021}{\natexlab{a}})},\ \Eprint {http://arxiv.org/abs/2009.01728} {arXiv:2009.01728 [astro-ph.CO]} \BibitemShut {NoStop}%
\bibitem [{\citenamefont {Vaskonen}\ and\ \citenamefont {Veerm\"ae}(2021)}]{Vaskonen:2020lbd}%
  \BibitemOpen
  \bibfield  {author} {\bibinfo {author} {\bibfnamefont {Ville}\ \bibnamefont {Vaskonen}}\ and\ \bibinfo {author} {\bibfnamefont {Hardi}\ \bibnamefont {Veerm\"ae}},\ }\bibfield  {title} {\enquote {\bibinfo {title} {{Did NANOGrav see a signal from primordial black hole formation?}}}\ }\href {\doibase 10.1103/PhysRevLett.126.051303} {\bibfield  {journal} {\bibinfo  {journal} {Phys. Rev. Lett.}\ }\textbf {\bibinfo {volume} {126}},\ \bibinfo {pages} {051303} (\bibinfo {year} {2021})},\ \Eprint {http://arxiv.org/abs/2009.07832} {arXiv:2009.07832 [astro-ph.CO]} \BibitemShut {NoStop}%
\bibitem [{\citenamefont {Chen}\ \emph {et~al.}(2022)\citenamefont {Chen}, \citenamefont {Yuan},\ and\ \citenamefont {Huang}}]{Chen:2021nxo}%
  \BibitemOpen
  \bibfield  {author} {\bibinfo {author} {\bibfnamefont {Zu-Cheng}\ \bibnamefont {Chen}}, \bibinfo {author} {\bibfnamefont {Chen}\ \bibnamefont {Yuan}}, \ and\ \bibinfo {author} {\bibfnamefont {Qing-Guo}\ \bibnamefont {Huang}},\ }\bibfield  {title} {\enquote {\bibinfo {title} {{Confronting the primordial black hole scenario with the gravitational-wave events detected by LIGO-Virgo}},}\ }\href {\doibase 10.1016/j.physletb.2022.137040} {\bibfield  {journal} {\bibinfo  {journal} {Phys. Lett. B}\ }\textbf {\bibinfo {volume} {829}},\ \bibinfo {pages} {137040} (\bibinfo {year} {2022})},\ \Eprint {http://arxiv.org/abs/2108.11740} {arXiv:2108.11740 [astro-ph.CO]} \BibitemShut {NoStop}%
\bibitem [{\citenamefont {De~Luca}\ \emph {et~al.}(2021{\natexlab{b}})\citenamefont {De~Luca}, \citenamefont {Franciolini},\ and\ \citenamefont {Riotto}}]{DeLuca:2020agl}%
  \BibitemOpen
  \bibfield  {author} {\bibinfo {author} {\bibfnamefont {V.}~\bibnamefont {De~Luca}}, \bibinfo {author} {\bibfnamefont {G.}~\bibnamefont {Franciolini}}, \ and\ \bibinfo {author} {\bibfnamefont {A.}~\bibnamefont {Riotto}},\ }\bibfield  {title} {\enquote {\bibinfo {title} {{NANOGrav Data Hints at Primordial Black Holes as Dark Matter}},}\ }\href {\doibase 10.1103/PhysRevLett.126.041303} {\bibfield  {journal} {\bibinfo  {journal} {Phys. Rev. Lett.}\ }\textbf {\bibinfo {volume} {126}},\ \bibinfo {pages} {041303} (\bibinfo {year} {2021}{\natexlab{b}})},\ \Eprint {http://arxiv.org/abs/2009.08268} {arXiv:2009.08268 [astro-ph.CO]} \BibitemShut {NoStop}%
\bibitem [{\citenamefont {Dom\`enech}\ and\ \citenamefont {Pi}(2022)}]{Domenech:2020ers}%
  \BibitemOpen
  \bibfield  {author} {\bibinfo {author} {\bibfnamefont {Guillem}\ \bibnamefont {Dom\`enech}}\ and\ \bibinfo {author} {\bibfnamefont {Shi}\ \bibnamefont {Pi}},\ }\bibfield  {title} {\enquote {\bibinfo {title} {{NANOGrav hints on planet-mass primordial black holes}},}\ }\href {\doibase 10.1007/s11433-021-1839-6} {\bibfield  {journal} {\bibinfo  {journal} {Sci. China Phys. Mech. Astron.}\ }\textbf {\bibinfo {volume} {65}},\ \bibinfo {pages} {230411} (\bibinfo {year} {2022})},\ \Eprint {http://arxiv.org/abs/2010.03976} {arXiv:2010.03976 [astro-ph.CO]} \BibitemShut {NoStop}%
\bibitem [{\citenamefont {Liu}\ \emph {et~al.}(2020{\natexlab{b}})\citenamefont {Liu}, \citenamefont {Christiansen}, \citenamefont {Guo}, \citenamefont {Cai},\ and\ \citenamefont {Kim}}]{Liu:2020vsy}%
  \BibitemOpen
  \bibfield  {author} {\bibinfo {author} {\bibfnamefont {Lang}\ \bibnamefont {Liu}}, \bibinfo {author} {\bibfnamefont {\O{}yvind}\ \bibnamefont {Christiansen}}, \bibinfo {author} {\bibfnamefont {Zong-Kuan}\ \bibnamefont {Guo}}, \bibinfo {author} {\bibfnamefont {Rong-Gen}\ \bibnamefont {Cai}}, \ and\ \bibinfo {author} {\bibfnamefont {Sang~Pyo}\ \bibnamefont {Kim}},\ }\bibfield  {title} {\enquote {\bibinfo {title} {{Gravitational and electromagnetic radiation from binary black holes with electric and magnetic charges: Circular orbits on a cone}},}\ }\href {\doibase 10.1103/PhysRevD.102.103520} {\bibfield  {journal} {\bibinfo  {journal} {Phys. Rev. D}\ }\textbf {\bibinfo {volume} {102}},\ \bibinfo {pages} {103520} (\bibinfo {year} {2020}{\natexlab{b}})},\ \Eprint {http://arxiv.org/abs/2008.02326} {arXiv:2008.02326 [gr-qc]} \BibitemShut {NoStop}%
\bibitem [{\citenamefont {Liu}\ \emph {et~al.}(2021)\citenamefont {Liu}, \citenamefont {Christiansen}, \citenamefont {Ruan}, \citenamefont {Guo}, \citenamefont {Cai},\ and\ \citenamefont {Kim}}]{Liu:2020bag}%
  \BibitemOpen
  \bibfield  {author} {\bibinfo {author} {\bibfnamefont {Lang}\ \bibnamefont {Liu}}, \bibinfo {author} {\bibfnamefont {\O{}yvind}\ \bibnamefont {Christiansen}}, \bibinfo {author} {\bibfnamefont {Wen-Hong}\ \bibnamefont {Ruan}}, \bibinfo {author} {\bibfnamefont {Zong-Kuan}\ \bibnamefont {Guo}}, \bibinfo {author} {\bibfnamefont {Rong-Gen}\ \bibnamefont {Cai}}, \ and\ \bibinfo {author} {\bibfnamefont {Sang~Pyo}\ \bibnamefont {Kim}},\ }\bibfield  {title} {\enquote {\bibinfo {title} {{Gravitational and electromagnetic radiation from binary black holes with electric and magnetic charges: elliptical orbits on a cone}},}\ }\href {\doibase 10.1140/epjc/s10052-021-09849-4} {\bibfield  {journal} {\bibinfo  {journal} {Eur. Phys. J. C}\ }\textbf {\bibinfo {volume} {81}},\ \bibinfo {pages} {1048} (\bibinfo {year} {2021})},\ \Eprint {http://arxiv.org/abs/2011.13586} {arXiv:2011.13586 [gr-qc]} \BibitemShut {NoStop}%
\bibitem [{\citenamefont {Cai}\ \emph {et~al.}(2021{\natexlab{a}})\citenamefont {Cai}, \citenamefont {Chen},\ and\ \citenamefont {Fu}}]{Cai:2021wzd}%
  \BibitemOpen
  \bibfield  {author} {\bibinfo {author} {\bibfnamefont {Rong-Gen}\ \bibnamefont {Cai}}, \bibinfo {author} {\bibfnamefont {Chao}\ \bibnamefont {Chen}}, \ and\ \bibinfo {author} {\bibfnamefont {Chengjie}\ \bibnamefont {Fu}},\ }\bibfield  {title} {\enquote {\bibinfo {title} {{Primordial black holes and stochastic gravitational wave background from inflation with a noncanonical spectator field}},}\ }\href {\doibase 10.1103/PhysRevD.104.083537} {\bibfield  {journal} {\bibinfo  {journal} {Phys. Rev. D}\ }\textbf {\bibinfo {volume} {104}},\ \bibinfo {pages} {083537} (\bibinfo {year} {2021}{\natexlab{a}})},\ \Eprint {http://arxiv.org/abs/2108.03422} {arXiv:2108.03422 [astro-ph.CO]} \BibitemShut {NoStop}%
\bibitem [{\citenamefont {Yuan}\ and\ \citenamefont {Huang}(2021)}]{Yuan:2021qgz}%
  \BibitemOpen
  \bibfield  {author} {\bibinfo {author} {\bibfnamefont {Chen}\ \bibnamefont {Yuan}}\ and\ \bibinfo {author} {\bibfnamefont {Qing-Guo}\ \bibnamefont {Huang}},\ }\bibfield  {title} {\enquote {\bibinfo {title} {{A topic review on probing primordial black hole dark matter with scalar induced gravitational waves}},}\ }\href {\doibase 10.1016/j.isci.2021.102860} {\bibfield  {journal} {\bibinfo  {journal} {iScience}\ }\textbf {\bibinfo {volume} {24}},\ \bibinfo {pages} {102860} (\bibinfo {year} {2021})},\ \Eprint {http://arxiv.org/abs/2103.04739} {arXiv:2103.04739 [astro-ph.GA]} \BibitemShut {NoStop}%
\bibitem [{\citenamefont {Liu}\ \emph {et~al.}(2023{\natexlab{a}})\citenamefont {Liu}, \citenamefont {Yang}, \citenamefont {Guo},\ and\ \citenamefont {Cai}}]{Liu:2021jnw}%
  \BibitemOpen
  \bibfield  {author} {\bibinfo {author} {\bibfnamefont {Lang}\ \bibnamefont {Liu}}, \bibinfo {author} {\bibfnamefont {Xing-Yu}\ \bibnamefont {Yang}}, \bibinfo {author} {\bibfnamefont {Zong-Kuan}\ \bibnamefont {Guo}}, \ and\ \bibinfo {author} {\bibfnamefont {Rong-Gen}\ \bibnamefont {Cai}},\ }\bibfield  {title} {\enquote {\bibinfo {title} {{Testing primordial black hole and measuring the Hubble constant with multiband gravitational-wave observations}},}\ }\href {\doibase 10.1088/1475-7516/2023/01/006} {\bibfield  {journal} {\bibinfo  {journal} {JCAP}\ }\textbf {\bibinfo {volume} {01}},\ \bibinfo {pages} {006} (\bibinfo {year} {2023}{\natexlab{a}})},\ \Eprint {http://arxiv.org/abs/2112.05473} {arXiv:2112.05473 [astro-ph.CO]} \BibitemShut {NoStop}%
\bibitem [{\citenamefont {Liu}\ and\ \citenamefont {Kim}(2022)}]{Liu:2022wtq}%
  \BibitemOpen
  \bibfield  {author} {\bibinfo {author} {\bibfnamefont {Lang}\ \bibnamefont {Liu}}\ and\ \bibinfo {author} {\bibfnamefont {Sang~Pyo}\ \bibnamefont {Kim}},\ }\bibfield  {title} {\enquote {\bibinfo {title} {{Merger rate of charged black holes from the two-body dynamical capture}},}\ }\href {\doibase 10.1088/1475-7516/2022/03/059} {\bibfield  {journal} {\bibinfo  {journal} {JCAP}\ }\textbf {\bibinfo {volume} {03}},\ \bibinfo {pages} {059} (\bibinfo {year} {2022})},\ \Eprint {http://arxiv.org/abs/2201.02581} {arXiv:2201.02581 [gr-qc]} \BibitemShut {NoStop}%
\bibitem [{\citenamefont {Inomata}\ \emph {et~al.}(2023)\citenamefont {Inomata}, \citenamefont {Braglia}, \citenamefont {Chen},\ and\ \citenamefont {Renaux-Petel}}]{Inomata:2022yte}%
  \BibitemOpen
  \bibfield  {author} {\bibinfo {author} {\bibfnamefont {Keisuke}\ \bibnamefont {Inomata}}, \bibinfo {author} {\bibfnamefont {Matteo}\ \bibnamefont {Braglia}}, \bibinfo {author} {\bibfnamefont {Xingang}\ \bibnamefont {Chen}}, \ and\ \bibinfo {author} {\bibfnamefont {S\'ebastien}\ \bibnamefont {Renaux-Petel}},\ }\bibfield  {title} {\enquote {\bibinfo {title} {{Questions on calculation of primordial power spectrum with large spikes: the resonance model case}},}\ }\href {\doibase 10.1088/1475-7516/2023/04/011} {\bibfield  {journal} {\bibinfo  {journal} {JCAP}\ }\textbf {\bibinfo {volume} {04}},\ \bibinfo {pages} {011} (\bibinfo {year} {2023})},\ \bibinfo {note} {[Erratum: JCAP 09, E01 (2023)]},\ \Eprint {http://arxiv.org/abs/2211.02586} {arXiv:2211.02586 [astro-ph.CO]} \BibitemShut {NoStop}%
\bibitem [{\citenamefont {Meng}\ \emph {et~al.}(2023)\citenamefont {Meng}, \citenamefont {Yuan},\ and\ \citenamefont {Huang}}]{Meng:2022low}%
  \BibitemOpen
  \bibfield  {author} {\bibinfo {author} {\bibfnamefont {De-Shuang}\ \bibnamefont {Meng}}, \bibinfo {author} {\bibfnamefont {Chen}\ \bibnamefont {Yuan}}, \ and\ \bibinfo {author} {\bibfnamefont {Qing-Guo}\ \bibnamefont {Huang}},\ }\bibfield  {title} {\enquote {\bibinfo {title} {{Primordial black holes generated by the non-minimal spectator field}},}\ }\href {\doibase 10.1007/s11433-022-2095-5} {\bibfield  {journal} {\bibinfo  {journal} {Sci. China Phys. Mech. Astron.}\ }\textbf {\bibinfo {volume} {66}},\ \bibinfo {pages} {280411} (\bibinfo {year} {2023})},\ \Eprint {http://arxiv.org/abs/2212.03577} {arXiv:2212.03577 [astro-ph.CO]} \BibitemShut {NoStop}%
\bibitem [{\citenamefont {Chen}\ \emph {et~al.}(2023{\natexlab{a}})\citenamefont {Chen}, \citenamefont {Kim},\ and\ \citenamefont {Liu}}]{Chen:2022qvg}%
  \BibitemOpen
  \bibfield  {author} {\bibinfo {author} {\bibfnamefont {Zu-Cheng}\ \bibnamefont {Chen}}, \bibinfo {author} {\bibfnamefont {Sang~Pyo}\ \bibnamefont {Kim}}, \ and\ \bibinfo {author} {\bibfnamefont {Lang}\ \bibnamefont {Liu}},\ }\bibfield  {title} {\enquote {\bibinfo {title} {{Gravitational and electromagnetic radiation from binary black holes with electric and magnetic charges: hyperbolic orbits on a cone}},}\ }\href {\doibase 10.1088/1572-9494/acce98} {\bibfield  {journal} {\bibinfo  {journal} {Commun. Theor. Phys.}\ }\textbf {\bibinfo {volume} {75}},\ \bibinfo {pages} {065401} (\bibinfo {year} {2023}{\natexlab{a}})},\ \Eprint {http://arxiv.org/abs/2210.15564} {arXiv:2210.15564 [gr-qc]} \BibitemShut {NoStop}%
\bibitem [{\citenamefont {Liu}\ \emph {et~al.}(2023{\natexlab{b}})\citenamefont {Liu}, \citenamefont {You}, \citenamefont {Wu},\ and\ \citenamefont {Chen}}]{Liu:2022iuf}%
  \BibitemOpen
  \bibfield  {author} {\bibinfo {author} {\bibfnamefont {Lang}\ \bibnamefont {Liu}}, \bibinfo {author} {\bibfnamefont {Zhi-Qiang}\ \bibnamefont {You}}, \bibinfo {author} {\bibfnamefont {You}\ \bibnamefont {Wu}}, \ and\ \bibinfo {author} {\bibfnamefont {Zu-Cheng}\ \bibnamefont {Chen}},\ }\bibfield  {title} {\enquote {\bibinfo {title} {{Constraining the merger history of primordial-black-hole binaries from GWTC-3}},}\ }\href {\doibase 10.1103/PhysRevD.107.063035} {\bibfield  {journal} {\bibinfo  {journal} {Phys. Rev. D}\ }\textbf {\bibinfo {volume} {107}},\ \bibinfo {pages} {063035} (\bibinfo {year} {2023}{\natexlab{b}})},\ \Eprint {http://arxiv.org/abs/2210.16094} {arXiv:2210.16094 [astro-ph.CO]} \BibitemShut {NoStop}%
\bibitem [{\citenamefont {Zheng}\ \emph {et~al.}(2023)\citenamefont {Zheng}, \citenamefont {Li}, \citenamefont {Chen}, \citenamefont {Zhou},\ and\ \citenamefont {Zhu}}]{Zheng:2022wqo}%
  \BibitemOpen
  \bibfield  {author} {\bibinfo {author} {\bibfnamefont {Li-Ming}\ \bibnamefont {Zheng}}, \bibinfo {author} {\bibfnamefont {Zhengxiang}\ \bibnamefont {Li}}, \bibinfo {author} {\bibfnamefont {Zu-Cheng}\ \bibnamefont {Chen}}, \bibinfo {author} {\bibfnamefont {Huan}\ \bibnamefont {Zhou}}, \ and\ \bibinfo {author} {\bibfnamefont {Zong-Hong}\ \bibnamefont {Zhu}},\ }\bibfield  {title} {\enquote {\bibinfo {title} {{Towards a reliable reconstruction of the power spectrum of primordial curvature perturbation on small scales from GWTC-3}},}\ }\href {\doibase 10.1016/j.physletb.2023.137720} {\bibfield  {journal} {\bibinfo  {journal} {Phys. Lett. B}\ }\textbf {\bibinfo {volume} {838}},\ \bibinfo {pages} {137720} (\bibinfo {year} {2023})},\ \Eprint {http://arxiv.org/abs/2212.05516} {arXiv:2212.05516 [astro-ph.CO]} \BibitemShut {NoStop}%
\bibitem [{\citenamefont {Choudhury}\ \emph {et~al.}(2023{\natexlab{a}})\citenamefont {Choudhury}, \citenamefont {Panda},\ and\ \citenamefont {Sami}}]{Choudhury:2023jlt}%
  \BibitemOpen
  \bibfield  {author} {\bibinfo {author} {\bibfnamefont {Sayantan}\ \bibnamefont {Choudhury}}, \bibinfo {author} {\bibfnamefont {Sudhakar}\ \bibnamefont {Panda}}, \ and\ \bibinfo {author} {\bibfnamefont {M.}~\bibnamefont {Sami}},\ }\bibfield  {title} {\enquote {\bibinfo {title} {{PBH formation in EFT of single field inflation with sharp transition}},}\ }\href {\doibase 10.1016/j.physletb.2023.138123} {\bibfield  {journal} {\bibinfo  {journal} {Phys. Lett. B}\ }\textbf {\bibinfo {volume} {845}},\ \bibinfo {pages} {138123} (\bibinfo {year} {2023}{\natexlab{a}})},\ \Eprint {http://arxiv.org/abs/2302.05655} {arXiv:2302.05655 [astro-ph.CO]} \BibitemShut {NoStop}%
\bibitem [{\citenamefont {Chen}\ \emph {et~al.}(2023{\natexlab{b}})\citenamefont {Chen}, \citenamefont {Du}, \citenamefont {Huang},\ and\ \citenamefont {You}}]{Chen:2022fda}%
  \BibitemOpen
  \bibfield  {author} {\bibinfo {author} {\bibfnamefont {Zu-Cheng}\ \bibnamefont {Chen}}, \bibinfo {author} {\bibfnamefont {Shen-Shi}\ \bibnamefont {Du}}, \bibinfo {author} {\bibfnamefont {Qing-Guo}\ \bibnamefont {Huang}}, \ and\ \bibinfo {author} {\bibfnamefont {Zhi-Qiang}\ \bibnamefont {You}},\ }\bibfield  {title} {\enquote {\bibinfo {title} {{Constraints on primordial-black-hole population and cosmic expansion history from GWTC-3}},}\ }\href {\doibase 10.1088/1475-7516/2023/03/024} {\bibfield  {journal} {\bibinfo  {journal} {JCAP}\ }\textbf {\bibinfo {volume} {03}},\ \bibinfo {pages} {024} (\bibinfo {year} {2023}{\natexlab{b}})},\ \Eprint {http://arxiv.org/abs/2205.11278} {arXiv:2205.11278 [astro-ph.CO]} \BibitemShut {NoStop}%
\bibitem [{\citenamefont {Choudhury}\ \emph {et~al.}(2024{\natexlab{a}})\citenamefont {Choudhury}, \citenamefont {Gangopadhyay},\ and\ \citenamefont {Sami}}]{Choudhury:2023vuj}%
  \BibitemOpen
  \bibfield  {author} {\bibinfo {author} {\bibfnamefont {Sayantan}\ \bibnamefont {Choudhury}}, \bibinfo {author} {\bibfnamefont {Mayukh~R.}\ \bibnamefont {Gangopadhyay}}, \ and\ \bibinfo {author} {\bibfnamefont {M.}~\bibnamefont {Sami}},\ }\bibfield  {title} {\enquote {\bibinfo {title} {{No-go for the formation of heavy mass Primordial Black Holes in Single Field Inflation}},}\ }\href {\doibase 10.1140/epjc/s10052-024-13218-2} {\bibfield  {journal} {\bibinfo  {journal} {Eur. Phys. J. C}\ }\textbf {\bibinfo {volume} {84}},\ \bibinfo {pages} {884} (\bibinfo {year} {2024}{\natexlab{a}})},\ \Eprint {http://arxiv.org/abs/2301.10000} {arXiv:2301.10000 [astro-ph.CO]} \BibitemShut {NoStop}%
\bibitem [{\citenamefont {Huang}\ \emph {et~al.}(2024{\natexlab{a}})\citenamefont {Huang}, \citenamefont {Jiang},\ and\ \citenamefont {Piao}}]{Hai-LongHuang:2023atg}%
  \BibitemOpen
  \bibfield  {author} {\bibinfo {author} {\bibfnamefont {Hai-Long}\ \bibnamefont {Huang}}, \bibinfo {author} {\bibfnamefont {Jun-Qian}\ \bibnamefont {Jiang}}, \ and\ \bibinfo {author} {\bibfnamefont {Yun-Song}\ \bibnamefont {Piao}},\ }\bibfield  {title} {\enquote {\bibinfo {title} {{Merger rate of supermassive primordial black hole binaries}},}\ }\href {\doibase 10.1103/PhysRevD.109.063515} {\bibfield  {journal} {\bibinfo  {journal} {Phys. Rev. D}\ }\textbf {\bibinfo {volume} {109}},\ \bibinfo {pages} {063515} (\bibinfo {year} {2024}{\natexlab{a}})},\ \Eprint {http://arxiv.org/abs/2312.00338} {arXiv:2312.00338 [astro-ph.CO]} \BibitemShut {NoStop}%
\bibitem [{\citenamefont {Choudhury}\ \emph {et~al.}(2024{\natexlab{b}})\citenamefont {Choudhury}, \citenamefont {Karde}, \citenamefont {Panda},\ and\ \citenamefont {Sami}}]{Choudhury:2023kdb}%
  \BibitemOpen
  \bibfield  {author} {\bibinfo {author} {\bibfnamefont {Sayantan}\ \bibnamefont {Choudhury}}, \bibinfo {author} {\bibfnamefont {Ahaskar}\ \bibnamefont {Karde}}, \bibinfo {author} {\bibfnamefont {Sudhakar}\ \bibnamefont {Panda}}, \ and\ \bibinfo {author} {\bibfnamefont {M.}~\bibnamefont {Sami}},\ }\bibfield  {title} {\enquote {\bibinfo {title} {{Primordial non-Gaussianity from ultra slow-roll Galileon inflation}},}\ }\href {\doibase 10.1088/1475-7516/2024/01/012} {\bibfield  {journal} {\bibinfo  {journal} {JCAP}\ }\textbf {\bibinfo {volume} {01}},\ \bibinfo {pages} {012} (\bibinfo {year} {2024}{\natexlab{b}})},\ \Eprint {http://arxiv.org/abs/2306.12334} {arXiv:2306.12334 [astro-ph.CO]} \BibitemShut {NoStop}%
\bibitem [{\citenamefont {Choudhury}\ \emph {et~al.}(2023{\natexlab{b}})\citenamefont {Choudhury}, \citenamefont {Panda},\ and\ \citenamefont {Sami}}]{Choudhury:2023hvf}%
  \BibitemOpen
  \bibfield  {author} {\bibinfo {author} {\bibfnamefont {Sayantan}\ \bibnamefont {Choudhury}}, \bibinfo {author} {\bibfnamefont {Sudhakar}\ \bibnamefont {Panda}}, \ and\ \bibinfo {author} {\bibfnamefont {M.}~\bibnamefont {Sami}},\ }\bibfield  {title} {\enquote {\bibinfo {title} {{Galileon inflation evades the no-go for PBH formation in the single-field framework}},}\ }\href {\doibase 10.1088/1475-7516/2023/08/078} {\bibfield  {journal} {\bibinfo  {journal} {JCAP}\ }\textbf {\bibinfo {volume} {08}},\ \bibinfo {pages} {078} (\bibinfo {year} {2023}{\natexlab{b}})},\ \Eprint {http://arxiv.org/abs/2304.04065} {arXiv:2304.04065 [astro-ph.CO]} \BibitemShut {NoStop}%
\bibitem [{\citenamefont {Choudhury}\ \emph {et~al.}(2023{\natexlab{c}})\citenamefont {Choudhury}, \citenamefont {Panda},\ and\ \citenamefont {Sami}}]{Choudhury:2023rks}%
  \BibitemOpen
  \bibfield  {author} {\bibinfo {author} {\bibfnamefont {Sayantan}\ \bibnamefont {Choudhury}}, \bibinfo {author} {\bibfnamefont {Sudhakar}\ \bibnamefont {Panda}}, \ and\ \bibinfo {author} {\bibfnamefont {M.}~\bibnamefont {Sami}},\ }\bibfield  {title} {\enquote {\bibinfo {title} {{Quantum loop effects on the power spectrum and constraints on primordial black holes}},}\ }\href {\doibase 10.1088/1475-7516/2023/11/066} {\bibfield  {journal} {\bibinfo  {journal} {JCAP}\ }\textbf {\bibinfo {volume} {11}},\ \bibinfo {pages} {066} (\bibinfo {year} {2023}{\natexlab{c}})},\ \Eprint {http://arxiv.org/abs/2303.06066} {arXiv:2303.06066 [astro-ph.CO]} \BibitemShut {NoStop}%
\bibitem [{\citenamefont {Wang}\ \emph {et~al.}(2024)\citenamefont {Wang}, \citenamefont {Zhang},\ and\ \citenamefont {Sasaki}}]{Wang:2024vfv}%
  \BibitemOpen
  \bibfield  {author} {\bibinfo {author} {\bibfnamefont {Xinpeng}\ \bibnamefont {Wang}}, \bibinfo {author} {\bibfnamefont {Ying-li}\ \bibnamefont {Zhang}}, \ and\ \bibinfo {author} {\bibfnamefont {Misao}\ \bibnamefont {Sasaki}},\ }\bibfield  {title} {\enquote {\bibinfo {title} {{Enhanced curvature perturbation and primordial black hole formation in two-stage inflation with a break}},}\ }\href {\doibase 10.1088/1475-7516/2024/07/076} {\bibfield  {journal} {\bibinfo  {journal} {JCAP}\ }\textbf {\bibinfo {volume} {07}},\ \bibinfo {pages} {076} (\bibinfo {year} {2024})},\ \Eprint {http://arxiv.org/abs/2404.02492} {arXiv:2404.02492 [astro-ph.CO]} \BibitemShut {NoStop}%
\bibitem [{\citenamefont {Choudhury}\ \emph {et~al.}(2024{\natexlab{c}})\citenamefont {Choudhury}, \citenamefont {Karde}, \citenamefont {Padiyar},\ and\ \citenamefont {Sami}}]{Choudhury:2024jlz}%
  \BibitemOpen
  \bibfield  {author} {\bibinfo {author} {\bibfnamefont {Sayantan}\ \bibnamefont {Choudhury}}, \bibinfo {author} {\bibfnamefont {Ahaskar}\ \bibnamefont {Karde}}, \bibinfo {author} {\bibfnamefont {Pankaj}\ \bibnamefont {Padiyar}}, \ and\ \bibinfo {author} {\bibfnamefont {M.}~\bibnamefont {Sami}},\ }\bibfield  {title} {\enquote {\bibinfo {title} {{Primordial Black Holes from Effective Field Theory of Stochastic Single Field Inflation at NNNLO}},}\ }\href@noop {} {\  (\bibinfo {year} {2024}{\natexlab{c}})},\ \Eprint {http://arxiv.org/abs/2403.13484} {arXiv:2403.13484 [astro-ph.CO]} \BibitemShut {NoStop}%
\bibitem [{\citenamefont {Yuan}\ and\ \citenamefont {Huang}(2024)}]{Yuan:2024yyo}%
  \BibitemOpen
  \bibfield  {author} {\bibinfo {author} {\bibfnamefont {Chen}\ \bibnamefont {Yuan}}\ and\ \bibinfo {author} {\bibfnamefont {Qing-Guo}\ \bibnamefont {Huang}},\ }\bibfield  {title} {\enquote {\bibinfo {title} {{Primordial black hole interpretation in subsolar mass gravitational wave candidate SSM200308}},}\ }\href {\doibase 10.1088/1475-7516/2024/09/051} {\bibfield  {journal} {\bibinfo  {journal} {JCAP}\ }\textbf {\bibinfo {volume} {09}},\ \bibinfo {pages} {051} (\bibinfo {year} {2024})},\ \Eprint {http://arxiv.org/abs/2404.03328} {arXiv:2404.03328 [astro-ph.CO]} \BibitemShut {NoStop}%
\bibitem [{\citenamefont {Chen}\ and\ \citenamefont {Hall}(2024)}]{Chen:2024dxh}%
  \BibitemOpen
  \bibfield  {author} {\bibinfo {author} {\bibfnamefont {Zu-Cheng}\ \bibnamefont {Chen}}\ and\ \bibinfo {author} {\bibfnamefont {Alex}\ \bibnamefont {Hall}},\ }\bibfield  {title} {\enquote {\bibinfo {title} {{Confronting primordial black holes with LIGO-Virgo-KAGRA and the Einstein Telescope}},}\ }\href@noop {} {\  (\bibinfo {year} {2024})},\ \Eprint {http://arxiv.org/abs/2402.03934} {arXiv:2402.03934 [astro-ph.CO]} \BibitemShut {NoStop}%
\bibitem [{\citenamefont {Chen}\ and\ \citenamefont {Liu}(2024{\natexlab{a}})}]{Chen:2024joj}%
  \BibitemOpen
  \bibfield  {author} {\bibinfo {author} {\bibfnamefont {Zu-Cheng}\ \bibnamefont {Chen}}\ and\ \bibinfo {author} {\bibfnamefont {Lang}\ \bibnamefont {Liu}},\ }\bibfield  {title} {\enquote {\bibinfo {title} {{Is PSR J0514$-$4002E in a PBH-NS binary?}}}\ }\href@noop {} {\  (\bibinfo {year} {2024}{\natexlab{a}})},\ \Eprint {http://arxiv.org/abs/2401.12889} {arXiv:2401.12889 [astro-ph.HE]} \BibitemShut {NoStop}%
\bibitem [{\citenamefont {Choudhury}\ \emph {et~al.}(2024{\natexlab{d}})\citenamefont {Choudhury}, \citenamefont {Dey}, \citenamefont {Ganguly}, \citenamefont {Karde}, \citenamefont {Singh},\ and\ \citenamefont {Tiwari}}]{Choudhury:2024kjj}%
  \BibitemOpen
  \bibfield  {author} {\bibinfo {author} {\bibfnamefont {Sayantan}\ \bibnamefont {Choudhury}}, \bibinfo {author} {\bibfnamefont {Kritartha}\ \bibnamefont {Dey}}, \bibinfo {author} {\bibfnamefont {Siddhant}\ \bibnamefont {Ganguly}}, \bibinfo {author} {\bibfnamefont {Ahaskar}\ \bibnamefont {Karde}}, \bibinfo {author} {\bibfnamefont {Swapnil~Kumar}\ \bibnamefont {Singh}}, \ and\ \bibinfo {author} {\bibfnamefont {Pranjal}\ \bibnamefont {Tiwari}},\ }\bibfield  {title} {\enquote {\bibinfo {title} {{Negative non-Gaussianity as a salvager for PBHs with PTAs in bounce}},}\ }\href@noop {} {\  (\bibinfo {year} {2024}{\natexlab{d}})},\ \Eprint {http://arxiv.org/abs/2409.18983} {arXiv:2409.18983 [astro-ph.CO]} \BibitemShut {NoStop}%
\bibitem [{\citenamefont {Huang}\ \emph {et~al.}(2024{\natexlab{b}})\citenamefont {Huang}, \citenamefont {Yuan}, \citenamefont {Chen},\ and\ \citenamefont {Liu}}]{Huang:2024wse}%
  \BibitemOpen
  \bibfield  {author} {\bibinfo {author} {\bibfnamefont {Qing-Guo}\ \bibnamefont {Huang}}, \bibinfo {author} {\bibfnamefont {Chen}\ \bibnamefont {Yuan}}, \bibinfo {author} {\bibfnamefont {Zu-Cheng}\ \bibnamefont {Chen}}, \ and\ \bibinfo {author} {\bibfnamefont {Lang}\ \bibnamefont {Liu}},\ }\bibfield  {title} {\enquote {\bibinfo {title} {{GW230529\_181500: a potential primordial binary black hole merger in the mass gap}},}\ }\href {\doibase 10.1088/1475-7516/2024/08/030} {\bibfield  {journal} {\bibinfo  {journal} {JCAP}\ }\textbf {\bibinfo {volume} {08}},\ \bibinfo {pages} {030} (\bibinfo {year} {2024}{\natexlab{b}})},\ \Eprint {http://arxiv.org/abs/2404.05691} {arXiv:2404.05691 [gr-qc]} \BibitemShut {NoStop}%
\bibitem [{\citenamefont {Ding}\ \emph {et~al.}(2024)\citenamefont {Ding}, \citenamefont {He},\ and\ \citenamefont {Takhistov}}]{Ding:2024mro}%
  \BibitemOpen
  \bibfield  {author} {\bibinfo {author} {\bibfnamefont {Qianhang}\ \bibnamefont {Ding}}, \bibinfo {author} {\bibfnamefont {Minxi}\ \bibnamefont {He}}, \ and\ \bibinfo {author} {\bibfnamefont {Volodymyr}\ \bibnamefont {Takhistov}},\ }\bibfield  {title} {\enquote {\bibinfo {title} {{Primordial Black Hole Mergers as Probes of Dark Matter in Galactic Center}},}\ }\href@noop {} {\  (\bibinfo {year} {2024})},\ \Eprint {http://arxiv.org/abs/2410.02591} {arXiv:2410.02591 [astro-ph.CO]} \BibitemShut {NoStop}%
\bibitem [{\citenamefont {Choudhury}\ \emph {et~al.}(2024{\natexlab{e}})\citenamefont {Choudhury}, \citenamefont {Karde}, \citenamefont {Panda},\ and\ \citenamefont {SenGupta}}]{Choudhury:2024dei}%
  \BibitemOpen
  \bibfield  {author} {\bibinfo {author} {\bibfnamefont {Sayantan}\ \bibnamefont {Choudhury}}, \bibinfo {author} {\bibfnamefont {Ahaskar}\ \bibnamefont {Karde}}, \bibinfo {author} {\bibfnamefont {Sudhakar}\ \bibnamefont {Panda}}, \ and\ \bibinfo {author} {\bibfnamefont {Soumitra}\ \bibnamefont {SenGupta}},\ }\bibfield  {title} {\enquote {\bibinfo {title} {{Regularized-Renormalized-Resummed loop corrected power spectrum of non-singular bounce with Primordial Black Hole formation}},}\ }\href@noop {} {\  (\bibinfo {year} {2024}{\natexlab{e}})},\ \Eprint {http://arxiv.org/abs/2405.06882} {arXiv:2405.06882 [astro-ph.CO]} \BibitemShut {NoStop}%
\bibitem [{\citenamefont {Calz\`a}\ \emph {et~al.}(2024{\natexlab{a}})\citenamefont {Calz\`a}, \citenamefont {Pedrotti},\ and\ \citenamefont {Vagnozzi}}]{Calza:2024fzo}%
  \BibitemOpen
  \bibfield  {author} {\bibinfo {author} {\bibfnamefont {Marco}\ \bibnamefont {Calz\`a}}, \bibinfo {author} {\bibfnamefont {Davide}\ \bibnamefont {Pedrotti}}, \ and\ \bibinfo {author} {\bibfnamefont {Sunny}\ \bibnamefont {Vagnozzi}},\ }\bibfield  {title} {\enquote {\bibinfo {title} {{Primordial regular black holes as all the dark matter (I): tr-symmetric metrics}},}\ }\href@noop {} {\  (\bibinfo {year} {2024}{\natexlab{a}})},\ \Eprint {http://arxiv.org/abs/2409.02804} {arXiv:2409.02804 [gr-qc]} \BibitemShut {NoStop}%
\bibitem [{\citenamefont {Calz\`a}\ \emph {et~al.}(2024{\natexlab{b}})\citenamefont {Calz\`a}, \citenamefont {Pedrotti},\ and\ \citenamefont {Vagnozzi}}]{Calza:2024xdh}%
  \BibitemOpen
  \bibfield  {author} {\bibinfo {author} {\bibfnamefont {Marco}\ \bibnamefont {Calz\`a}}, \bibinfo {author} {\bibfnamefont {Davide}\ \bibnamefont {Pedrotti}}, \ and\ \bibinfo {author} {\bibfnamefont {Sunny}\ \bibnamefont {Vagnozzi}},\ }\bibfield  {title} {\enquote {\bibinfo {title} {{Primordial regular black holes as all the dark matter (II): non-tr-symmetric and loop quantum gravity-inspired metrics}},}\ }\href@noop {} {\  (\bibinfo {year} {2024}{\natexlab{b}})},\ \Eprint {http://arxiv.org/abs/2409.02807} {arXiv:2409.02807 [gr-qc]} \BibitemShut {NoStop}%
\bibitem [{\citenamefont {Sasaki}\ \emph {et~al.}(2018)\citenamefont {Sasaki}, \citenamefont {Suyama}, \citenamefont {Tanaka},\ and\ \citenamefont {Yokoyama}}]{Sasaki:2018dmp}%
  \BibitemOpen
  \bibfield  {author} {\bibinfo {author} {\bibfnamefont {Misao}\ \bibnamefont {Sasaki}}, \bibinfo {author} {\bibfnamefont {Teruaki}\ \bibnamefont {Suyama}}, \bibinfo {author} {\bibfnamefont {Takahiro}\ \bibnamefont {Tanaka}}, \ and\ \bibinfo {author} {\bibfnamefont {Shuichiro}\ \bibnamefont {Yokoyama}},\ }\bibfield  {title} {\enquote {\bibinfo {title} {{Primordial black holes\textemdash{}perspectives in gravitational wave astronomy}},}\ }\href {\doibase 10.1088/1361-6382/aaa7b4} {\bibfield  {journal} {\bibinfo  {journal} {Class. Quant. Grav.}\ }\textbf {\bibinfo {volume} {35}},\ \bibinfo {pages} {063001} (\bibinfo {year} {2018})},\ \Eprint {http://arxiv.org/abs/1801.05235} {arXiv:1801.05235 [astro-ph.CO]} \BibitemShut {NoStop}%
\bibitem [{\citenamefont {Carr}\ \emph {et~al.}(2021)\citenamefont {Carr}, \citenamefont {Kohri}, \citenamefont {Sendouda},\ and\ \citenamefont {Yokoyama}}]{Carr:2020gox}%
  \BibitemOpen
  \bibfield  {author} {\bibinfo {author} {\bibfnamefont {Bernard}\ \bibnamefont {Carr}}, \bibinfo {author} {\bibfnamefont {Kazunori}\ \bibnamefont {Kohri}}, \bibinfo {author} {\bibfnamefont {Yuuiti}\ \bibnamefont {Sendouda}}, \ and\ \bibinfo {author} {\bibfnamefont {Jun'ichi}\ \bibnamefont {Yokoyama}},\ }\bibfield  {title} {\enquote {\bibinfo {title} {{Constraints on primordial black holes}},}\ }\href {\doibase 10.1088/1361-6633/ac1e31} {\bibfield  {journal} {\bibinfo  {journal} {Rept. Prog. Phys.}\ }\textbf {\bibinfo {volume} {84}},\ \bibinfo {pages} {116902} (\bibinfo {year} {2021})},\ \Eprint {http://arxiv.org/abs/2002.12778} {arXiv:2002.12778 [astro-ph.CO]} \BibitemShut {NoStop}%
\bibitem [{\citenamefont {Carr}\ and\ \citenamefont {Kuhnel}(2020)}]{Carr:2020xqk}%
  \BibitemOpen
  \bibfield  {author} {\bibinfo {author} {\bibfnamefont {Bernard}\ \bibnamefont {Carr}}\ and\ \bibinfo {author} {\bibfnamefont {Florian}\ \bibnamefont {Kuhnel}},\ }\bibfield  {title} {\enquote {\bibinfo {title} {{Primordial Black Holes as Dark Matter: Recent Developments}},}\ }\href {\doibase 10.1146/annurev-nucl-050520-125911} {\bibfield  {journal} {\bibinfo  {journal} {Ann. Rev. Nucl. Part. Sci.}\ }\textbf {\bibinfo {volume} {70}},\ \bibinfo {pages} {355--394} (\bibinfo {year} {2020})},\ \Eprint {http://arxiv.org/abs/2006.02838} {arXiv:2006.02838 [astro-ph.CO]} \BibitemShut {NoStop}%
\bibitem [{\citenamefont {Choudhury}\ and\ \citenamefont {Sami}(2024)}]{Choudhury:2024aji}%
  \BibitemOpen
  \bibfield  {author} {\bibinfo {author} {\bibfnamefont {Sayantan}\ \bibnamefont {Choudhury}}\ and\ \bibinfo {author} {\bibfnamefont {M.}~\bibnamefont {Sami}},\ }\bibfield  {title} {\enquote {\bibinfo {title} {{Large fluctuations and Primordial Black Holes}},}\ }\href@noop {} {\  (\bibinfo {year} {2024})},\ \Eprint {http://arxiv.org/abs/2407.17006} {arXiv:2407.17006 [gr-qc]} \BibitemShut {NoStop}%
\bibitem [{\citenamefont {Bird}\ \emph {et~al.}(2016)\citenamefont {Bird}, \citenamefont {Cholis}, \citenamefont {Mu\~noz}, \citenamefont {Ali-Ha\"\i{}moud}, \citenamefont {Kamionkowski}, \citenamefont {Kovetz}, \citenamefont {Raccanelli},\ and\ \citenamefont {Riess}}]{Bird:2016dcv}%
  \BibitemOpen
  \bibfield  {author} {\bibinfo {author} {\bibfnamefont {Simeon}\ \bibnamefont {Bird}}, \bibinfo {author} {\bibfnamefont {Ilias}\ \bibnamefont {Cholis}}, \bibinfo {author} {\bibfnamefont {Julian~B.}\ \bibnamefont {Mu\~noz}}, \bibinfo {author} {\bibfnamefont {Yacine}\ \bibnamefont {Ali-Ha\"\i{}moud}}, \bibinfo {author} {\bibfnamefont {Marc}\ \bibnamefont {Kamionkowski}}, \bibinfo {author} {\bibfnamefont {Ely~D.}\ \bibnamefont {Kovetz}}, \bibinfo {author} {\bibfnamefont {Alvise}\ \bibnamefont {Raccanelli}}, \ and\ \bibinfo {author} {\bibfnamefont {Adam~G.}\ \bibnamefont {Riess}},\ }\bibfield  {title} {\enquote {\bibinfo {title} {{Did LIGO detect dark matter?}}}\ }\href {\doibase 10.1103/PhysRevLett.116.201301} {\bibfield  {journal} {\bibinfo  {journal} {Phys. Rev. Lett.}\ }\textbf {\bibinfo {volume} {116}},\ \bibinfo {pages} {201301} (\bibinfo {year} {2016})},\ \Eprint {http://arxiv.org/abs/1603.00464} {arXiv:1603.00464 [astro-ph.CO]} \BibitemShut {NoStop}%
\bibitem [{\citenamefont {Sasaki}\ \emph {et~al.}(2016)\citenamefont {Sasaki}, \citenamefont {Suyama}, \citenamefont {Tanaka},\ and\ \citenamefont {Yokoyama}}]{Sasaki:2016jop}%
  \BibitemOpen
  \bibfield  {author} {\bibinfo {author} {\bibfnamefont {Misao}\ \bibnamefont {Sasaki}}, \bibinfo {author} {\bibfnamefont {Teruaki}\ \bibnamefont {Suyama}}, \bibinfo {author} {\bibfnamefont {Takahiro}\ \bibnamefont {Tanaka}}, \ and\ \bibinfo {author} {\bibfnamefont {Shuichiro}\ \bibnamefont {Yokoyama}},\ }\bibfield  {title} {\enquote {\bibinfo {title} {{Primordial Black Hole Scenario for the Gravitational-Wave Event GW150914}},}\ }\href {\doibase 10.1103/PhysRevLett.117.061101} {\bibfield  {journal} {\bibinfo  {journal} {Phys. Rev. Lett.}\ }\textbf {\bibinfo {volume} {117}},\ \bibinfo {pages} {061101} (\bibinfo {year} {2016})},\ \bibinfo {note} {[Erratum: Phys.Rev.Lett. 121, 059901 (2018)]},\ \Eprint {http://arxiv.org/abs/1603.08338} {arXiv:1603.08338 [astro-ph.CO]} \BibitemShut {NoStop}%
\bibitem [{\citenamefont {Alabidi}\ \emph {et~al.}(2012)\citenamefont {Alabidi}, \citenamefont {Kohri}, \citenamefont {Sasaki},\ and\ \citenamefont {Sendouda}}]{Alabidi:2012ex}%
  \BibitemOpen
  \bibfield  {author} {\bibinfo {author} {\bibfnamefont {Laila}\ \bibnamefont {Alabidi}}, \bibinfo {author} {\bibfnamefont {Kazunori}\ \bibnamefont {Kohri}}, \bibinfo {author} {\bibfnamefont {Misao}\ \bibnamefont {Sasaki}}, \ and\ \bibinfo {author} {\bibfnamefont {Yuuiti}\ \bibnamefont {Sendouda}},\ }\bibfield  {title} {\enquote {\bibinfo {title} {{Observable Spectra of Induced Gravitational Waves from Inflation}},}\ }\href {\doibase 10.1088/1475-7516/2012/09/017} {\bibfield  {journal} {\bibinfo  {journal} {JCAP}\ }\textbf {\bibinfo {volume} {09}},\ \bibinfo {pages} {017} (\bibinfo {year} {2012})},\ \Eprint {http://arxiv.org/abs/1203.4663} {arXiv:1203.4663 [astro-ph.CO]} \BibitemShut {NoStop}%
\bibitem [{\citenamefont {Alabidi}\ \emph {et~al.}(2013)\citenamefont {Alabidi}, \citenamefont {Kohri}, \citenamefont {Sasaki},\ and\ \citenamefont {Sendouda}}]{Alabidi:2013lya}%
  \BibitemOpen
  \bibfield  {author} {\bibinfo {author} {\bibfnamefont {Laila}\ \bibnamefont {Alabidi}}, \bibinfo {author} {\bibfnamefont {Kazunori}\ \bibnamefont {Kohri}}, \bibinfo {author} {\bibfnamefont {Misao}\ \bibnamefont {Sasaki}}, \ and\ \bibinfo {author} {\bibfnamefont {Yuuiti}\ \bibnamefont {Sendouda}},\ }\bibfield  {title} {\enquote {\bibinfo {title} {{Observable induced gravitational waves from an early matter phase}},}\ }\href {\doibase 10.1088/1475-7516/2013/05/033} {\bibfield  {journal} {\bibinfo  {journal} {JCAP}\ }\textbf {\bibinfo {volume} {05}},\ \bibinfo {pages} {033} (\bibinfo {year} {2013})},\ \Eprint {http://arxiv.org/abs/1303.4519} {arXiv:1303.4519 [astro-ph.CO]} \BibitemShut {NoStop}%
\bibitem [{\citenamefont {Choudhury}\ and\ \citenamefont {Mazumdar}(2014)}]{Choudhury:2013woa}%
  \BibitemOpen
  \bibfield  {author} {\bibinfo {author} {\bibfnamefont {Sayantan}\ \bibnamefont {Choudhury}}\ and\ \bibinfo {author} {\bibfnamefont {Anupam}\ \bibnamefont {Mazumdar}},\ }\bibfield  {title} {\enquote {\bibinfo {title} {{Primordial blackholes and gravitational waves for an inflection-point model of inflation}},}\ }\href {\doibase 10.1016/j.physletb.2014.04.050} {\bibfield  {journal} {\bibinfo  {journal} {Phys. Lett. B}\ }\textbf {\bibinfo {volume} {733}},\ \bibinfo {pages} {270--275} (\bibinfo {year} {2014})},\ \Eprint {http://arxiv.org/abs/1307.5119} {arXiv:1307.5119 [astro-ph.CO]} \BibitemShut {NoStop}%
\bibitem [{\citenamefont {Espinosa}\ \emph {et~al.}(2018)\citenamefont {Espinosa}, \citenamefont {Racco},\ and\ \citenamefont {Riotto}}]{Espinosa:2018eve}%
  \BibitemOpen
  \bibfield  {author} {\bibinfo {author} {\bibfnamefont {Jos\'e~Ram\'on}\ \bibnamefont {Espinosa}}, \bibinfo {author} {\bibfnamefont {Davide}\ \bibnamefont {Racco}}, \ and\ \bibinfo {author} {\bibfnamefont {Antonio}\ \bibnamefont {Riotto}},\ }\bibfield  {title} {\enquote {\bibinfo {title} {{A Cosmological Signature of the SM Higgs Instability: Gravitational Waves}},}\ }\href {\doibase 10.1088/1475-7516/2018/09/012} {\bibfield  {journal} {\bibinfo  {journal} {JCAP}\ }\textbf {\bibinfo {volume} {09}},\ \bibinfo {pages} {012} (\bibinfo {year} {2018})},\ \Eprint {http://arxiv.org/abs/1804.07732} {arXiv:1804.07732 [hep-ph]} \BibitemShut {NoStop}%
\bibitem [{\citenamefont {Bartolo}\ \emph {et~al.}(2019)\citenamefont {Bartolo}, \citenamefont {De~Luca}, \citenamefont {Franciolini}, \citenamefont {Peloso}, \citenamefont {Racco},\ and\ \citenamefont {Riotto}}]{Bartolo:2018rku}%
  \BibitemOpen
  \bibfield  {author} {\bibinfo {author} {\bibfnamefont {N.}~\bibnamefont {Bartolo}}, \bibinfo {author} {\bibfnamefont {V.}~\bibnamefont {De~Luca}}, \bibinfo {author} {\bibfnamefont {G.}~\bibnamefont {Franciolini}}, \bibinfo {author} {\bibfnamefont {M.}~\bibnamefont {Peloso}}, \bibinfo {author} {\bibfnamefont {D.}~\bibnamefont {Racco}}, \ and\ \bibinfo {author} {\bibfnamefont {A.}~\bibnamefont {Riotto}},\ }\bibfield  {title} {\enquote {\bibinfo {title} {{Testing primordial black holes as dark matter with LISA}},}\ }\href {\doibase 10.1103/PhysRevD.99.103521} {\bibfield  {journal} {\bibinfo  {journal} {Phys. Rev. D}\ }\textbf {\bibinfo {volume} {99}},\ \bibinfo {pages} {103521} (\bibinfo {year} {2019})},\ \Eprint {http://arxiv.org/abs/1810.12224} {arXiv:1810.12224 [astro-ph.CO]} \BibitemShut {NoStop}%
\bibitem [{\citenamefont {Inomata}\ and\ \citenamefont {Nakama}(2019)}]{Inomata:2018epa}%
  \BibitemOpen
  \bibfield  {author} {\bibinfo {author} {\bibfnamefont {Keisuke}\ \bibnamefont {Inomata}}\ and\ \bibinfo {author} {\bibfnamefont {Tomohiro}\ \bibnamefont {Nakama}},\ }\bibfield  {title} {\enquote {\bibinfo {title} {{Gravitational waves induced by scalar perturbations as probes of the small-scale primordial spectrum}},}\ }\href {\doibase 10.1103/PhysRevD.99.043511} {\bibfield  {journal} {\bibinfo  {journal} {Phys. Rev. D}\ }\textbf {\bibinfo {volume} {99}},\ \bibinfo {pages} {043511} (\bibinfo {year} {2019})},\ \Eprint {http://arxiv.org/abs/1812.00674} {arXiv:1812.00674 [astro-ph.CO]} \BibitemShut {NoStop}%
\bibitem [{\citenamefont {Yuan}\ \emph {et~al.}(2019)\citenamefont {Yuan}, \citenamefont {Chen},\ and\ \citenamefont {Huang}}]{Yuan:2019udt}%
  \BibitemOpen
  \bibfield  {author} {\bibinfo {author} {\bibfnamefont {Chen}\ \bibnamefont {Yuan}}, \bibinfo {author} {\bibfnamefont {Zu-Cheng}\ \bibnamefont {Chen}}, \ and\ \bibinfo {author} {\bibfnamefont {Qing-Guo}\ \bibnamefont {Huang}},\ }\bibfield  {title} {\enquote {\bibinfo {title} {{Probing primordial\textendash{}black-hole dark matter with scalar induced gravitational waves}},}\ }\href {\doibase 10.1103/PhysRevD.100.081301} {\bibfield  {journal} {\bibinfo  {journal} {Phys. Rev. D}\ }\textbf {\bibinfo {volume} {100}},\ \bibinfo {pages} {081301} (\bibinfo {year} {2019})},\ \Eprint {http://arxiv.org/abs/1906.11549} {arXiv:1906.11549 [astro-ph.CO]} \BibitemShut {NoStop}%
\bibitem [{\citenamefont {Inomata}\ \emph {et~al.}(2019{\natexlab{a}})\citenamefont {Inomata}, \citenamefont {Kohri}, \citenamefont {Nakama},\ and\ \citenamefont {Terada}}]{Inomata:2019zqy}%
  \BibitemOpen
  \bibfield  {author} {\bibinfo {author} {\bibfnamefont {Keisuke}\ \bibnamefont {Inomata}}, \bibinfo {author} {\bibfnamefont {Kazunori}\ \bibnamefont {Kohri}}, \bibinfo {author} {\bibfnamefont {Tomohiro}\ \bibnamefont {Nakama}}, \ and\ \bibinfo {author} {\bibfnamefont {Takahiro}\ \bibnamefont {Terada}},\ }\bibfield  {title} {\enquote {\bibinfo {title} {{Gravitational Waves Induced by Scalar Perturbations during a Gradual Transition from an Early Matter Era to the Radiation Era}},}\ }\href {\doibase 10.1088/1475-7516/2019/10/071} {\bibfield  {journal} {\bibinfo  {journal} {JCAP}\ }\textbf {\bibinfo {volume} {10}},\ \bibinfo {pages} {071} (\bibinfo {year} {2019}{\natexlab{a}})},\ \bibinfo {note} {[Erratum: JCAP 08, E01 (2023)]},\ \Eprint {http://arxiv.org/abs/1904.12878} {arXiv:1904.12878 [astro-ph.CO]} \BibitemShut {NoStop}%
\bibitem [{\citenamefont {Inomata}\ \emph {et~al.}(2019{\natexlab{b}})\citenamefont {Inomata}, \citenamefont {Kohri}, \citenamefont {Nakama},\ and\ \citenamefont {Terada}}]{Inomata:2019ivs}%
  \BibitemOpen
  \bibfield  {author} {\bibinfo {author} {\bibfnamefont {Keisuke}\ \bibnamefont {Inomata}}, \bibinfo {author} {\bibfnamefont {Kazunori}\ \bibnamefont {Kohri}}, \bibinfo {author} {\bibfnamefont {Tomohiro}\ \bibnamefont {Nakama}}, \ and\ \bibinfo {author} {\bibfnamefont {Takahiro}\ \bibnamefont {Terada}},\ }\bibfield  {title} {\enquote {\bibinfo {title} {{Enhancement of Gravitational Waves Induced by Scalar Perturbations due to a Sudden Transition from an Early Matter Era to the Radiation Era}},}\ }\href {\doibase 10.1103/PhysRevD.108.049901} {\bibfield  {journal} {\bibinfo  {journal} {Phys. Rev. D}\ }\textbf {\bibinfo {volume} {100}},\ \bibinfo {pages} {043532} (\bibinfo {year} {2019}{\natexlab{b}})},\ \bibinfo {note} {[Erratum: Phys.Rev.D 108, 049901 (2023)]},\ \Eprint {http://arxiv.org/abs/1904.12879} {arXiv:1904.12879 [astro-ph.CO]} \BibitemShut {NoStop}%
\bibitem [{\citenamefont {Chen}\ \emph {et~al.}(2020)\citenamefont {Chen}, \citenamefont {Yuan},\ and\ \citenamefont {Huang}}]{Chen:2019xse}%
  \BibitemOpen
  \bibfield  {author} {\bibinfo {author} {\bibfnamefont {Zu-Cheng}\ \bibnamefont {Chen}}, \bibinfo {author} {\bibfnamefont {Chen}\ \bibnamefont {Yuan}}, \ and\ \bibinfo {author} {\bibfnamefont {Qing-Guo}\ \bibnamefont {Huang}},\ }\bibfield  {title} {\enquote {\bibinfo {title} {{Pulsar Timing Array Constraints on Primordial Black Holes with NANOGrav 11-Year Dataset}},}\ }\href {\doibase 10.1103/PhysRevLett.124.251101} {\bibfield  {journal} {\bibinfo  {journal} {Phys. Rev. Lett.}\ }\textbf {\bibinfo {volume} {124}},\ \bibinfo {pages} {251101} (\bibinfo {year} {2020})},\ \Eprint {http://arxiv.org/abs/1910.12239} {arXiv:1910.12239 [astro-ph.CO]} \BibitemShut {NoStop}%
\bibitem [{\citenamefont {Yuan}\ \emph {et~al.}(2020{\natexlab{a}})\citenamefont {Yuan}, \citenamefont {Chen},\ and\ \citenamefont {Huang}}]{Yuan:2019wwo}%
  \BibitemOpen
  \bibfield  {author} {\bibinfo {author} {\bibfnamefont {Chen}\ \bibnamefont {Yuan}}, \bibinfo {author} {\bibfnamefont {Zu-Cheng}\ \bibnamefont {Chen}}, \ and\ \bibinfo {author} {\bibfnamefont {Qing-Guo}\ \bibnamefont {Huang}},\ }\bibfield  {title} {\enquote {\bibinfo {title} {{Log-dependent slope of scalar induced gravitational waves in the infrared regions}},}\ }\href {\doibase 10.1103/PhysRevD.101.043019} {\bibfield  {journal} {\bibinfo  {journal} {Phys. Rev. D}\ }\textbf {\bibinfo {volume} {101}},\ \bibinfo {pages} {043019} (\bibinfo {year} {2020}{\natexlab{a}})},\ \Eprint {http://arxiv.org/abs/1910.09099} {arXiv:1910.09099 [astro-ph.CO]} \BibitemShut {NoStop}%
\bibitem [{\citenamefont {Dom\`enech}\ and\ \citenamefont {Sasaki}(2018)}]{Domenech:2017ems}%
  \BibitemOpen
  \bibfield  {author} {\bibinfo {author} {\bibfnamefont {Guillem}\ \bibnamefont {Dom\`enech}}\ and\ \bibinfo {author} {\bibfnamefont {Misao}\ \bibnamefont {Sasaki}},\ }\bibfield  {title} {\enquote {\bibinfo {title} {{Hamiltonian approach to second order gauge invariant cosmological perturbations}},}\ }\href {\doibase 10.1103/PhysRevD.97.023521} {\bibfield  {journal} {\bibinfo  {journal} {Phys. Rev. D}\ }\textbf {\bibinfo {volume} {97}},\ \bibinfo {pages} {023521} (\bibinfo {year} {2018})},\ \Eprint {http://arxiv.org/abs/1709.09804} {arXiv:1709.09804 [gr-qc]} \BibitemShut {NoStop}%
\bibitem [{\citenamefont {Dom\`enech}(2020)}]{Domenech:2019quo}%
  \BibitemOpen
  \bibfield  {author} {\bibinfo {author} {\bibfnamefont {Guillem}\ \bibnamefont {Dom\`enech}},\ }\bibfield  {title} {\enquote {\bibinfo {title} {{Induced gravitational waves in a general cosmological background}},}\ }\href {\doibase 10.1142/S0218271820500285} {\bibfield  {journal} {\bibinfo  {journal} {Int. J. Mod. Phys. D}\ }\textbf {\bibinfo {volume} {29}},\ \bibinfo {pages} {2050028} (\bibinfo {year} {2020})},\ \Eprint {http://arxiv.org/abs/1912.05583} {arXiv:1912.05583 [gr-qc]} \BibitemShut {NoStop}%
\bibitem [{\citenamefont {Ota}(2020)}]{Ota:2020vfn}%
  \BibitemOpen
  \bibfield  {author} {\bibinfo {author} {\bibfnamefont {Atsuhisa}\ \bibnamefont {Ota}},\ }\bibfield  {title} {\enquote {\bibinfo {title} {{Induced superhorizon tensor perturbations from anisotropic non-Gaussianity}},}\ }\href {\doibase 10.1103/PhysRevD.101.103511} {\bibfield  {journal} {\bibinfo  {journal} {Phys. Rev. D}\ }\textbf {\bibinfo {volume} {101}},\ \bibinfo {pages} {103511} (\bibinfo {year} {2020})},\ \Eprint {http://arxiv.org/abs/2001.00409} {arXiv:2001.00409 [astro-ph.CO]} \BibitemShut {NoStop}%
\bibitem [{\citenamefont {Cai}\ \emph {et~al.}(2019{\natexlab{b}})\citenamefont {Cai}, \citenamefont {Chen}, \citenamefont {Tong}, \citenamefont {Wang},\ and\ \citenamefont {Yan}}]{Cai:2019jah}%
  \BibitemOpen
  \bibfield  {author} {\bibinfo {author} {\bibfnamefont {Yi-Fu}\ \bibnamefont {Cai}}, \bibinfo {author} {\bibfnamefont {Chao}\ \bibnamefont {Chen}}, \bibinfo {author} {\bibfnamefont {Xi}~\bibnamefont {Tong}}, \bibinfo {author} {\bibfnamefont {Dong-Gang}\ \bibnamefont {Wang}}, \ and\ \bibinfo {author} {\bibfnamefont {Sheng-Feng}\ \bibnamefont {Yan}},\ }\bibfield  {title} {\enquote {\bibinfo {title} {{When Primordial Black Holes from Sound Speed Resonance Meet a Stochastic Background of Gravitational Waves}},}\ }\href {\doibase 10.1103/PhysRevD.100.043518} {\bibfield  {journal} {\bibinfo  {journal} {Phys. Rev. D}\ }\textbf {\bibinfo {volume} {100}},\ \bibinfo {pages} {043518} (\bibinfo {year} {2019}{\natexlab{b}})},\ \Eprint {http://arxiv.org/abs/1902.08187} {arXiv:1902.08187 [astro-ph.CO]} \BibitemShut {NoStop}%
\bibitem [{\citenamefont {Cai}\ \emph {et~al.}(2019{\natexlab{c}})\citenamefont {Cai}, \citenamefont {Pi}, \citenamefont {Wang},\ and\ \citenamefont {Yang}}]{Cai:2019elf}%
  \BibitemOpen
  \bibfield  {author} {\bibinfo {author} {\bibfnamefont {Rong-Gen}\ \bibnamefont {Cai}}, \bibinfo {author} {\bibfnamefont {Shi}\ \bibnamefont {Pi}}, \bibinfo {author} {\bibfnamefont {Shao-Jiang}\ \bibnamefont {Wang}}, \ and\ \bibinfo {author} {\bibfnamefont {Xing-Yu}\ \bibnamefont {Yang}},\ }\bibfield  {title} {\enquote {\bibinfo {title} {{Pulsar Timing Array Constraints on the Induced Gravitational Waves}},}\ }\href {\doibase 10.1088/1475-7516/2019/10/059} {\bibfield  {journal} {\bibinfo  {journal} {JCAP}\ }\textbf {\bibinfo {volume} {10}},\ \bibinfo {pages} {059} (\bibinfo {year} {2019}{\natexlab{c}})},\ \Eprint {http://arxiv.org/abs/1907.06372} {arXiv:1907.06372 [astro-ph.CO]} \BibitemShut {NoStop}%
\bibitem [{\citenamefont {Cai}\ \emph {et~al.}(2019{\natexlab{d}})\citenamefont {Cai}, \citenamefont {Pi}, \citenamefont {Wang},\ and\ \citenamefont {Yang}}]{Cai:2019amo}%
  \BibitemOpen
  \bibfield  {author} {\bibinfo {author} {\bibfnamefont {Rong-Gen}\ \bibnamefont {Cai}}, \bibinfo {author} {\bibfnamefont {Shi}\ \bibnamefont {Pi}}, \bibinfo {author} {\bibfnamefont {Shao-Jiang}\ \bibnamefont {Wang}}, \ and\ \bibinfo {author} {\bibfnamefont {Xing-Yu}\ \bibnamefont {Yang}},\ }\bibfield  {title} {\enquote {\bibinfo {title} {{Resonant multiple peaks in the induced gravitational waves}},}\ }\href {\doibase 10.1088/1475-7516/2019/05/013} {\bibfield  {journal} {\bibinfo  {journal} {JCAP}\ }\textbf {\bibinfo {volume} {05}},\ \bibinfo {pages} {013} (\bibinfo {year} {2019}{\natexlab{d}})},\ \Eprint {http://arxiv.org/abs/1901.10152} {arXiv:1901.10152 [astro-ph.CO]} \BibitemShut {NoStop}%
\bibitem [{\citenamefont {Bhattacharya}\ \emph {et~al.}(2020)\citenamefont {Bhattacharya}, \citenamefont {Mohanty},\ and\ \citenamefont {Parashari}}]{Bhattacharya:2019bvk}%
  \BibitemOpen
  \bibfield  {author} {\bibinfo {author} {\bibfnamefont {Sukannya}\ \bibnamefont {Bhattacharya}}, \bibinfo {author} {\bibfnamefont {Subhendra}\ \bibnamefont {Mohanty}}, \ and\ \bibinfo {author} {\bibfnamefont {Priyank}\ \bibnamefont {Parashari}},\ }\bibfield  {title} {\enquote {\bibinfo {title} {{Primordial black holes and gravitational waves in nonstandard cosmologies}},}\ }\href {\doibase 10.1103/PhysRevD.102.043522} {\bibfield  {journal} {\bibinfo  {journal} {Phys. Rev. D}\ }\textbf {\bibinfo {volume} {102}},\ \bibinfo {pages} {043522} (\bibinfo {year} {2020})},\ \Eprint {http://arxiv.org/abs/1912.01653} {arXiv:1912.01653 [astro-ph.CO]} \BibitemShut {NoStop}%
\bibitem [{\citenamefont {Pi}\ and\ \citenamefont {Sasaki}(2020)}]{Pi:2020otn}%
  \BibitemOpen
  \bibfield  {author} {\bibinfo {author} {\bibfnamefont {Shi}\ \bibnamefont {Pi}}\ and\ \bibinfo {author} {\bibfnamefont {Misao}\ \bibnamefont {Sasaki}},\ }\bibfield  {title} {\enquote {\bibinfo {title} {{Gravitational Waves Induced by Scalar Perturbations with a Lognormal Peak}},}\ }\href {\doibase 10.1088/1475-7516/2020/09/037} {\bibfield  {journal} {\bibinfo  {journal} {JCAP}\ }\textbf {\bibinfo {volume} {09}},\ \bibinfo {pages} {037} (\bibinfo {year} {2020})},\ \Eprint {http://arxiv.org/abs/2005.12306} {arXiv:2005.12306 [gr-qc]} \BibitemShut {NoStop}%
\bibitem [{\citenamefont {Choudhury}\ \emph {et~al.}(2024{\natexlab{f}})\citenamefont {Choudhury}, \citenamefont {Karde}, \citenamefont {Panda},\ and\ \citenamefont {Sami}}]{Choudhury:2023hfm}%
  \BibitemOpen
  \bibfield  {author} {\bibinfo {author} {\bibfnamefont {Sayantan}\ \bibnamefont {Choudhury}}, \bibinfo {author} {\bibfnamefont {Ahaskar}\ \bibnamefont {Karde}}, \bibinfo {author} {\bibfnamefont {Sudhakar}\ \bibnamefont {Panda}}, \ and\ \bibinfo {author} {\bibfnamefont {M.}~\bibnamefont {Sami}},\ }\bibfield  {title} {\enquote {\bibinfo {title} {{Scalar induced gravity waves from ultra slow-roll galileon inflation}},}\ }\href {\doibase 10.1016/j.nuclphysb.2024.116678} {\bibfield  {journal} {\bibinfo  {journal} {Nucl. Phys. B}\ }\textbf {\bibinfo {volume} {1007}},\ \bibinfo {pages} {116678} (\bibinfo {year} {2024}{\natexlab{f}})},\ \Eprint {http://arxiv.org/abs/2308.09273} {arXiv:2308.09273 [astro-ph.CO]} \BibitemShut {NoStop}%
\bibitem [{\citenamefont {Choudhury}(2024)}]{Choudhury:2023kam}%
  \BibitemOpen
  \bibfield  {author} {\bibinfo {author} {\bibfnamefont {Sayantan}\ \bibnamefont {Choudhury}},\ }\bibfield  {title} {\enquote {\bibinfo {title} {{Single field inflation in the light of Pulsar Timing Array Data: quintessential interpretation of blue tilted tensor spectrum through Non-Bunch Davies initial condition}},}\ }\href {\doibase 10.1140/epjc/s10052-024-12625-9} {\bibfield  {journal} {\bibinfo  {journal} {Eur. Phys. J. C}\ }\textbf {\bibinfo {volume} {84}},\ \bibinfo {pages} {278} (\bibinfo {year} {2024})},\ \Eprint {http://arxiv.org/abs/2307.03249} {arXiv:2307.03249 [astro-ph.CO]} \BibitemShut {NoStop}%
\bibitem [{\citenamefont {Bhattacharya}\ \emph {et~al.}(2024)\citenamefont {Bhattacharya}, \citenamefont {Choudhury}, \citenamefont {Dey}, \citenamefont {Ghosh}, \citenamefont {Karde},\ and\ \citenamefont {Mishra}}]{Bhattacharya:2023ysp}%
  \BibitemOpen
  \bibfield  {author} {\bibinfo {author} {\bibfnamefont {Gourab}\ \bibnamefont {Bhattacharya}}, \bibinfo {author} {\bibfnamefont {Sayantan}\ \bibnamefont {Choudhury}}, \bibinfo {author} {\bibfnamefont {Kritartha}\ \bibnamefont {Dey}}, \bibinfo {author} {\bibfnamefont {Saptarshi}\ \bibnamefont {Ghosh}}, \bibinfo {author} {\bibfnamefont {Ahaskar}\ \bibnamefont {Karde}}, \ and\ \bibinfo {author} {\bibfnamefont {Navneet~Suryaprakash}\ \bibnamefont {Mishra}},\ }\bibfield  {title} {\enquote {\bibinfo {title} {{Evading no-go for PBH formation and production of SIGWs using Multiple Sharp Transitions in EFT of single field inflation}},}\ }\href {\doibase 10.1016/j.dark.2024.101602} {\bibfield  {journal} {\bibinfo  {journal} {Phys. Dark Univ.}\ }\textbf {\bibinfo {volume} {46}},\ \bibinfo {pages} {101602} (\bibinfo {year} {2024})},\ \Eprint {http://arxiv.org/abs/2309.00973} {arXiv:2309.00973 [astro-ph.CO]} \BibitemShut {NoStop}%
\bibitem [{\citenamefont {Liu}\ \emph {et~al.}(2024{\natexlab{a}})\citenamefont {Liu}, \citenamefont {Chen},\ and\ \citenamefont {Huang}}]{Liu:2023ymk}%
  \BibitemOpen
  \bibfield  {author} {\bibinfo {author} {\bibfnamefont {Lang}\ \bibnamefont {Liu}}, \bibinfo {author} {\bibfnamefont {Zu-Cheng}\ \bibnamefont {Chen}}, \ and\ \bibinfo {author} {\bibfnamefont {Qing-Guo}\ \bibnamefont {Huang}},\ }\bibfield  {title} {\enquote {\bibinfo {title} {{Implications for the non-Gaussianity of curvature perturbation from pulsar timing arrays}},}\ }\href {\doibase 10.1103/PhysRevD.109.L061301} {\bibfield  {journal} {\bibinfo  {journal} {Phys. Rev. D}\ }\textbf {\bibinfo {volume} {109}},\ \bibinfo {pages} {L061301} (\bibinfo {year} {2024}{\natexlab{a}})},\ \Eprint {http://arxiv.org/abs/2307.01102} {arXiv:2307.01102 [astro-ph.CO]} \BibitemShut {NoStop}%
\bibitem [{\citenamefont {Jin}\ \emph {et~al.}(2023)\citenamefont {Jin}, \citenamefont {Chen}, \citenamefont {Yi}, \citenamefont {You}, \citenamefont {Liu},\ and\ \citenamefont {Wu}}]{Jin:2023wri}%
  \BibitemOpen
  \bibfield  {author} {\bibinfo {author} {\bibfnamefont {Jia-Heng}\ \bibnamefont {Jin}}, \bibinfo {author} {\bibfnamefont {Zu-Cheng}\ \bibnamefont {Chen}}, \bibinfo {author} {\bibfnamefont {Zhu}\ \bibnamefont {Yi}}, \bibinfo {author} {\bibfnamefont {Zhi-Qiang}\ \bibnamefont {You}}, \bibinfo {author} {\bibfnamefont {Lang}\ \bibnamefont {Liu}}, \ and\ \bibinfo {author} {\bibfnamefont {You}\ \bibnamefont {Wu}},\ }\bibfield  {title} {\enquote {\bibinfo {title} {{Confronting sound speed resonance with pulsar timing arrays}},}\ }\href {\doibase 10.1088/1475-7516/2023/09/016} {\bibfield  {journal} {\bibinfo  {journal} {JCAP}\ }\textbf {\bibinfo {volume} {09}},\ \bibinfo {pages} {016} (\bibinfo {year} {2023})},\ \Eprint {http://arxiv.org/abs/2307.08687} {arXiv:2307.08687 [astro-ph.CO]} \BibitemShut {NoStop}%
\bibitem [{\citenamefont {Liu}\ \emph {et~al.}(2023{\natexlab{c}})\citenamefont {Liu}, \citenamefont {Chen},\ and\ \citenamefont {Huang}}]{Liu:2023pau}%
  \BibitemOpen
  \bibfield  {author} {\bibinfo {author} {\bibfnamefont {Lang}\ \bibnamefont {Liu}}, \bibinfo {author} {\bibfnamefont {Zu-Cheng}\ \bibnamefont {Chen}}, \ and\ \bibinfo {author} {\bibfnamefont {Qing-Guo}\ \bibnamefont {Huang}},\ }\bibfield  {title} {\enquote {\bibinfo {title} {{Probing the equation of state of the early Universe with pulsar timing arrays}},}\ }\href {\doibase 10.1088/1475-7516/2023/11/071} {\bibfield  {journal} {\bibinfo  {journal} {JCAP}\ }\textbf {\bibinfo {volume} {11}},\ \bibinfo {pages} {071} (\bibinfo {year} {2023}{\natexlab{c}})},\ \Eprint {http://arxiv.org/abs/2307.14911} {arXiv:2307.14911 [astro-ph.CO]} \BibitemShut {NoStop}%
\bibitem [{\citenamefont {Yi}\ \emph {et~al.}(2024)\citenamefont {Yi}, \citenamefont {You}, \citenamefont {Wu}, \citenamefont {Chen},\ and\ \citenamefont {Liu}}]{Yi:2023npi}%
  \BibitemOpen
  \bibfield  {author} {\bibinfo {author} {\bibfnamefont {Zhu}\ \bibnamefont {Yi}}, \bibinfo {author} {\bibfnamefont {Zhi-Qiang}\ \bibnamefont {You}}, \bibinfo {author} {\bibfnamefont {You}\ \bibnamefont {Wu}}, \bibinfo {author} {\bibfnamefont {Zu-Cheng}\ \bibnamefont {Chen}}, \ and\ \bibinfo {author} {\bibfnamefont {Lang}\ \bibnamefont {Liu}},\ }\bibfield  {title} {\enquote {\bibinfo {title} {{Exploring the NANOGrav signal and planet-mass primordial black holes through Higgs inflation}},}\ }\href {\doibase 10.1088/1475-7516/2024/06/043} {\bibfield  {journal} {\bibinfo  {journal} {JCAP}\ }\textbf {\bibinfo {volume} {06}},\ \bibinfo {pages} {043} (\bibinfo {year} {2024})},\ \Eprint {http://arxiv.org/abs/2308.14688} {arXiv:2308.14688 [astro-ph.CO]} \BibitemShut {NoStop}%
\bibitem [{\citenamefont {Liu}\ \emph {et~al.}(2024{\natexlab{b}})\citenamefont {Liu}, \citenamefont {Wu},\ and\ \citenamefont {Chen}}]{Liu:2023hpw}%
  \BibitemOpen
  \bibfield  {author} {\bibinfo {author} {\bibfnamefont {Lang}\ \bibnamefont {Liu}}, \bibinfo {author} {\bibfnamefont {You}\ \bibnamefont {Wu}}, \ and\ \bibinfo {author} {\bibfnamefont {Zu-Cheng}\ \bibnamefont {Chen}},\ }\bibfield  {title} {\enquote {\bibinfo {title} {{Simultaneously probing the sound speed and equation of state of the early Universe with pulsar timing arrays}},}\ }\href {\doibase 10.1088/1475-7516/2024/04/011} {\bibfield  {journal} {\bibinfo  {journal} {JCAP}\ }\textbf {\bibinfo {volume} {04}},\ \bibinfo {pages} {011} (\bibinfo {year} {2024}{\natexlab{b}})},\ \Eprint {http://arxiv.org/abs/2310.16500} {arXiv:2310.16500 [astro-ph.CO]} \BibitemShut {NoStop}%
\bibitem [{\citenamefont {Choudhury}\ \emph {et~al.}(2024{\natexlab{g}})\citenamefont {Choudhury}, \citenamefont {Dey}, \citenamefont {Karde}, \citenamefont {Panda},\ and\ \citenamefont {Sami}}]{Choudhury:2023fwk}%
  \BibitemOpen
  \bibfield  {author} {\bibinfo {author} {\bibfnamefont {Sayantan}\ \bibnamefont {Choudhury}}, \bibinfo {author} {\bibfnamefont {Kritartha}\ \bibnamefont {Dey}}, \bibinfo {author} {\bibfnamefont {Ahaskar}\ \bibnamefont {Karde}}, \bibinfo {author} {\bibfnamefont {Sudhakar}\ \bibnamefont {Panda}}, \ and\ \bibinfo {author} {\bibfnamefont {M.}~\bibnamefont {Sami}},\ }\bibfield  {title} {\enquote {\bibinfo {title} {{Primordial non-Gaussianity as a saviour for PBH overproduction in SIGWs generated by pulsar timing arrays for Galileon inflation}},}\ }\href {\doibase 10.1016/j.physletb.2024.138925} {\bibfield  {journal} {\bibinfo  {journal} {Phys. Lett. B}\ }\textbf {\bibinfo {volume} {856}},\ \bibinfo {pages} {138925} (\bibinfo {year} {2024}{\natexlab{g}})},\ \Eprint {http://arxiv.org/abs/2310.11034} {arXiv:2310.11034 [astro-ph.CO]} \BibitemShut {NoStop}%
\bibitem [{\citenamefont {Choudhury}\ \emph {et~al.}(2023{\natexlab{d}})\citenamefont {Choudhury}, \citenamefont {Dey},\ and\ \citenamefont {Karde}}]{Choudhury:2023fjs}%
  \BibitemOpen
  \bibfield  {author} {\bibinfo {author} {\bibfnamefont {Sayantan}\ \bibnamefont {Choudhury}}, \bibinfo {author} {\bibfnamefont {Kritartha}\ \bibnamefont {Dey}}, \ and\ \bibinfo {author} {\bibfnamefont {Ahaskar}\ \bibnamefont {Karde}},\ }\bibfield  {title} {\enquote {\bibinfo {title} {{Untangling PBH overproduction in $w$-SIGWs generated by Pulsar Timing Arrays for MST-EFT of single field inflation}},}\ }\href@noop {} {\  (\bibinfo {year} {2023}{\natexlab{d}})},\ \Eprint {http://arxiv.org/abs/2311.15065} {arXiv:2311.15065 [astro-ph.CO]} \BibitemShut {NoStop}%
\bibitem [{\citenamefont {Chen}\ \emph {et~al.}(2024)\citenamefont {Chen}, \citenamefont {Li}, \citenamefont {Liu},\ and\ \citenamefont {Yi}}]{Chen:2024fir}%
  \BibitemOpen
  \bibfield  {author} {\bibinfo {author} {\bibfnamefont {Zu-Cheng}\ \bibnamefont {Chen}}, \bibinfo {author} {\bibfnamefont {Jun}\ \bibnamefont {Li}}, \bibinfo {author} {\bibfnamefont {Lang}\ \bibnamefont {Liu}}, \ and\ \bibinfo {author} {\bibfnamefont {Zhu}\ \bibnamefont {Yi}},\ }\bibfield  {title} {\enquote {\bibinfo {title} {{Probing the speed of scalar-induced gravitational waves with pulsar timing arrays}},}\ }\href {\doibase 10.1103/PhysRevD.109.L101302} {\bibfield  {journal} {\bibinfo  {journal} {Phys. Rev. D}\ }\textbf {\bibinfo {volume} {109}},\ \bibinfo {pages} {L101302} (\bibinfo {year} {2024})},\ \Eprint {http://arxiv.org/abs/2401.09818} {arXiv:2401.09818 [gr-qc]} \BibitemShut {NoStop}%
\bibitem [{\citenamefont {Choudhury}\ \emph {et~al.}(2024{\natexlab{h}})\citenamefont {Choudhury}, \citenamefont {Ganguly}, \citenamefont {Panda}, \citenamefont {SenGupta},\ and\ \citenamefont {Tiwari}}]{Choudhury:2024dzw}%
  \BibitemOpen
  \bibfield  {author} {\bibinfo {author} {\bibfnamefont {Sayantan}\ \bibnamefont {Choudhury}}, \bibinfo {author} {\bibfnamefont {Siddhant}\ \bibnamefont {Ganguly}}, \bibinfo {author} {\bibfnamefont {Sudhakar}\ \bibnamefont {Panda}}, \bibinfo {author} {\bibfnamefont {Soumitra}\ \bibnamefont {SenGupta}}, \ and\ \bibinfo {author} {\bibfnamefont {Pranjal}\ \bibnamefont {Tiwari}},\ }\bibfield  {title} {\enquote {\bibinfo {title} {{Obviating PBH overproduction for SIGWs generated by pulsar timing arrays in loop corrected EFT of bounce}},}\ }\href {\doibase 10.1088/1475-7516/2024/09/013} {\bibfield  {journal} {\bibinfo  {journal} {JCAP}\ }\textbf {\bibinfo {volume} {09}},\ \bibinfo {pages} {013} (\bibinfo {year} {2024}{\natexlab{h}})},\ \Eprint {http://arxiv.org/abs/2407.18976} {arXiv:2407.18976 [astro-ph.CO]} \BibitemShut {NoStop}%
\bibitem [{\citenamefont {Chen}\ and\ \citenamefont {Liu}(2024{\natexlab{b}})}]{Chen:2024twp}%
  \BibitemOpen
  \bibfield  {author} {\bibinfo {author} {\bibfnamefont {Zu-Cheng}\ \bibnamefont {Chen}}\ and\ \bibinfo {author} {\bibfnamefont {Lang}\ \bibnamefont {Liu}},\ }\bibfield  {title} {\enquote {\bibinfo {title} {{Can we distinguish the adiabatic fluctuations and isocurvature fluctuations with pulsar timing arrays?}}}\ }\href@noop {} {\  (\bibinfo {year} {2024}{\natexlab{b}})},\ \Eprint {http://arxiv.org/abs/2402.16781} {arXiv:2402.16781 [astro-ph.CO]} \BibitemShut {NoStop}%
\bibitem [{\citenamefont {Choudhury}\ \emph {et~al.}(2024{\natexlab{i}})\citenamefont {Choudhury}, \citenamefont {Karde}, \citenamefont {Panda},\ and\ \citenamefont {Sami}}]{Choudhury:2024one}%
  \BibitemOpen
  \bibfield  {author} {\bibinfo {author} {\bibfnamefont {Sayantan}\ \bibnamefont {Choudhury}}, \bibinfo {author} {\bibfnamefont {Ahaskar}\ \bibnamefont {Karde}}, \bibinfo {author} {\bibfnamefont {Sudhakar}\ \bibnamefont {Panda}}, \ and\ \bibinfo {author} {\bibfnamefont {M.}~\bibnamefont {Sami}},\ }\bibfield  {title} {\enquote {\bibinfo {title} {{Realisation of the ultra-slow roll phase in Galileon inflation and PBH overproduction}},}\ }\href {\doibase 10.1088/1475-7516/2024/07/034} {\bibfield  {journal} {\bibinfo  {journal} {JCAP}\ }\textbf {\bibinfo {volume} {07}},\ \bibinfo {pages} {034} (\bibinfo {year} {2024}{\natexlab{i}})},\ \Eprint {http://arxiv.org/abs/2401.10925} {arXiv:2401.10925 [astro-ph.CO]} \BibitemShut {NoStop}%
\bibitem [{\citenamefont {Noh}\ and\ \citenamefont {Hwang}(2003)}]{Noh:2003yg}%
  \BibitemOpen
  \bibfield  {author} {\bibinfo {author} {\bibfnamefont {Hyerim}\ \bibnamefont {Noh}}\ and\ \bibinfo {author} {\bibfnamefont {Jai-chan}\ \bibnamefont {Hwang}},\ }\bibfield  {title} {\enquote {\bibinfo {title} {{Second-order perturbations of the friedmann world model}},}\ }\href@noop {} {\  (\bibinfo {year} {2003})},\ \Eprint {http://arxiv.org/abs/astro-ph/0305123} {arXiv:astro-ph/0305123} \BibitemShut {NoStop}%
\bibitem [{\citenamefont {Hwang}\ \emph {et~al.}(2017)\citenamefont {Hwang}, \citenamefont {Jeong},\ and\ \citenamefont {Noh}}]{Hwang:2017oxa}%
  \BibitemOpen
  \bibfield  {author} {\bibinfo {author} {\bibfnamefont {Jai-Chan}\ \bibnamefont {Hwang}}, \bibinfo {author} {\bibfnamefont {Donghui}\ \bibnamefont {Jeong}}, \ and\ \bibinfo {author} {\bibfnamefont {Hyerim}\ \bibnamefont {Noh}},\ }\bibfield  {title} {\enquote {\bibinfo {title} {{Gauge dependence of gravitational waves generated from scalar perturbations}},}\ }\href {\doibase 10.3847/1538-4357/aa74be} {\bibfield  {journal} {\bibinfo  {journal} {Astrophys. J.}\ }\textbf {\bibinfo {volume} {842}},\ \bibinfo {pages} {46} (\bibinfo {year} {2017})},\ \Eprint {http://arxiv.org/abs/1704.03500} {arXiv:1704.03500 [astro-ph.CO]} \BibitemShut {NoStop}%
\bibitem [{\citenamefont {Gong}(2022)}]{Gong:2019mui}%
  \BibitemOpen
  \bibfield  {author} {\bibinfo {author} {\bibfnamefont {Jinn-Ouk}\ \bibnamefont {Gong}},\ }\bibfield  {title} {\enquote {\bibinfo {title} {{Analytic Integral Solutions for Induced Gravitational Waves}},}\ }\href {\doibase 10.3847/1538-4357/ac3a6c} {\bibfield  {journal} {\bibinfo  {journal} {Astrophys. J.}\ }\textbf {\bibinfo {volume} {925}},\ \bibinfo {pages} {102} (\bibinfo {year} {2022})},\ \Eprint {http://arxiv.org/abs/1909.12708} {arXiv:1909.12708 [gr-qc]} \BibitemShut {NoStop}%
\bibitem [{\citenamefont {Tomikawa}\ and\ \citenamefont {Kobayashi}(2020)}]{Tomikawa:2019tvi}%
  \BibitemOpen
  \bibfield  {author} {\bibinfo {author} {\bibfnamefont {Keitaro}\ \bibnamefont {Tomikawa}}\ and\ \bibinfo {author} {\bibfnamefont {Tsutomu}\ \bibnamefont {Kobayashi}},\ }\bibfield  {title} {\enquote {\bibinfo {title} {{Gauge dependence of gravitational waves generated at second order from scalar perturbations}},}\ }\href {\doibase 10.1103/PhysRevD.101.083529} {\bibfield  {journal} {\bibinfo  {journal} {Phys. Rev. D}\ }\textbf {\bibinfo {volume} {101}},\ \bibinfo {pages} {083529} (\bibinfo {year} {2020})},\ \Eprint {http://arxiv.org/abs/1910.01880} {arXiv:1910.01880 [gr-qc]} \BibitemShut {NoStop}%
\bibitem [{\citenamefont {Wang}\ and\ \citenamefont {Zhang}(2019)}]{Wang:2019zhj}%
  \BibitemOpen
  \bibfield  {author} {\bibinfo {author} {\bibfnamefont {Bo}~\bibnamefont {Wang}}\ and\ \bibinfo {author} {\bibfnamefont {Yang}\ \bibnamefont {Zhang}},\ }\bibfield  {title} {\enquote {\bibinfo {title} {{Second-order cosmological perturbations IV. Produced by scalar-tensor and tensor-tensor couplings during the radiation dominated stage}},}\ }\href {\doibase 10.1103/PhysRevD.99.123008} {\bibfield  {journal} {\bibinfo  {journal} {Phys. Rev. D}\ }\textbf {\bibinfo {volume} {99}},\ \bibinfo {pages} {123008} (\bibinfo {year} {2019})},\ \Eprint {http://arxiv.org/abs/1905.03272} {arXiv:1905.03272 [gr-qc]} \BibitemShut {NoStop}%
\bibitem [{\citenamefont {De~Luca}\ \emph {et~al.}(2020)\citenamefont {De~Luca}, \citenamefont {Franciolini}, \citenamefont {Kehagias},\ and\ \citenamefont {Riotto}}]{DeLuca:2019ufz}%
  \BibitemOpen
  \bibfield  {author} {\bibinfo {author} {\bibfnamefont {V.}~\bibnamefont {De~Luca}}, \bibinfo {author} {\bibfnamefont {G.}~\bibnamefont {Franciolini}}, \bibinfo {author} {\bibfnamefont {A.}~\bibnamefont {Kehagias}}, \ and\ \bibinfo {author} {\bibfnamefont {A.}~\bibnamefont {Riotto}},\ }\bibfield  {title} {\enquote {\bibinfo {title} {{On the Gauge Invariance of Cosmological Gravitational Waves}},}\ }\href {\doibase 10.1088/1475-7516/2020/03/014} {\bibfield  {journal} {\bibinfo  {journal} {JCAP}\ }\textbf {\bibinfo {volume} {03}},\ \bibinfo {pages} {014} (\bibinfo {year} {2020})},\ \Eprint {http://arxiv.org/abs/1911.09689} {arXiv:1911.09689 [gr-qc]} \BibitemShut {NoStop}%
\bibitem [{\citenamefont {Inomata}\ and\ \citenamefont {Terada}(2020)}]{Inomata:2019yww}%
  \BibitemOpen
  \bibfield  {author} {\bibinfo {author} {\bibfnamefont {Keisuke}\ \bibnamefont {Inomata}}\ and\ \bibinfo {author} {\bibfnamefont {Takahiro}\ \bibnamefont {Terada}},\ }\bibfield  {title} {\enquote {\bibinfo {title} {{Gauge Independence of Induced Gravitational Waves}},}\ }\href {\doibase 10.1103/PhysRevD.101.023523} {\bibfield  {journal} {\bibinfo  {journal} {Phys. Rev. D}\ }\textbf {\bibinfo {volume} {101}},\ \bibinfo {pages} {023523} (\bibinfo {year} {2020})},\ \Eprint {http://arxiv.org/abs/1912.00785} {arXiv:1912.00785 [gr-qc]} \BibitemShut {NoStop}%
\bibitem [{\citenamefont {Yuan}\ \emph {et~al.}(2020{\natexlab{b}})\citenamefont {Yuan}, \citenamefont {Chen},\ and\ \citenamefont {Huang}}]{Yuan:2019fwv}%
  \BibitemOpen
  \bibfield  {author} {\bibinfo {author} {\bibfnamefont {Chen}\ \bibnamefont {Yuan}}, \bibinfo {author} {\bibfnamefont {Zu-Cheng}\ \bibnamefont {Chen}}, \ and\ \bibinfo {author} {\bibfnamefont {Qing-Guo}\ \bibnamefont {Huang}},\ }\bibfield  {title} {\enquote {\bibinfo {title} {{Scalar induced gravitational waves in different gauges}},}\ }\href {\doibase 10.1103/PhysRevD.101.063018} {\bibfield  {journal} {\bibinfo  {journal} {Phys. Rev. D}\ }\textbf {\bibinfo {volume} {101}},\ \bibinfo {pages} {063018} (\bibinfo {year} {2020}{\natexlab{b}})},\ \Eprint {http://arxiv.org/abs/1912.00885} {arXiv:1912.00885 [astro-ph.CO]} \BibitemShut {NoStop}%
\bibitem [{\citenamefont {Nakamura}(2020)}]{Nakamura:2020pre}%
  \BibitemOpen
  \bibfield  {author} {\bibinfo {author} {\bibfnamefont {Kouji}\ \bibnamefont {Nakamura}},\ }\bibfield  {title} {\enquote {\bibinfo {title} {{Second-order Gauge-invariant Cosmological Perturbation Theory: Current Status updated in 2019}},}\ }\href {\doibase 10.9734/bpi/taps/v3} {\  (\bibinfo {year} {2020}),\ 10.9734/bpi/taps/v3},\ \Eprint {http://arxiv.org/abs/1912.12805} {arXiv:1912.12805 [gr-qc]} \BibitemShut {NoStop}%
\bibitem [{\citenamefont {Giovannini}(2020{\natexlab{a}})}]{Giovannini:2020qta}%
  \BibitemOpen
  \bibfield  {author} {\bibinfo {author} {\bibfnamefont {Massimo}\ \bibnamefont {Giovannini}},\ }\bibfield  {title} {\enquote {\bibinfo {title} {{Spurious gauge-invariance of higher-order contributions to the spectral energy density of the relic gravitons}},}\ }\href {\doibase 10.1142/S0217751X20501651} {\bibfield  {journal} {\bibinfo  {journal} {Int. J. Mod. Phys. A}\ }\textbf {\bibinfo {volume} {35}},\ \bibinfo {pages} {2050165} (\bibinfo {year} {2020}{\natexlab{a}})},\ \Eprint {http://arxiv.org/abs/2005.04962} {arXiv:2005.04962 [hep-th]} \BibitemShut {NoStop}%
\bibitem [{\citenamefont {Lu}\ \emph {et~al.}(2020)\citenamefont {Lu}, \citenamefont {Ali}, \citenamefont {Gong}, \citenamefont {Lin},\ and\ \citenamefont {Zhang}}]{Lu:2020diy}%
  \BibitemOpen
  \bibfield  {author} {\bibinfo {author} {\bibfnamefont {Yizhou}\ \bibnamefont {Lu}}, \bibinfo {author} {\bibfnamefont {Arshad}\ \bibnamefont {Ali}}, \bibinfo {author} {\bibfnamefont {Yungui}\ \bibnamefont {Gong}}, \bibinfo {author} {\bibfnamefont {Jiong}\ \bibnamefont {Lin}}, \ and\ \bibinfo {author} {\bibfnamefont {Fengge}\ \bibnamefont {Zhang}},\ }\bibfield  {title} {\enquote {\bibinfo {title} {{Gauge transformation of scalar induced gravitational waves}},}\ }\href {\doibase 10.1103/PhysRevD.102.083503(2020)} {\bibfield  {journal} {\bibinfo  {journal} {Phys. Rev. D}\ }\textbf {\bibinfo {volume} {102}},\ \bibinfo {pages} {083503} (\bibinfo {year} {2020})},\ \Eprint {http://arxiv.org/abs/2006.03450} {arXiv:2006.03450 [gr-qc]} \BibitemShut {NoStop}%
\bibitem [{\citenamefont {Chang}\ \emph {et~al.}(2021)\citenamefont {Chang}, \citenamefont {Wang},\ and\ \citenamefont {Zhu}}]{Chang:2020tji}%
  \BibitemOpen
  \bibfield  {author} {\bibinfo {author} {\bibfnamefont {Zhe}\ \bibnamefont {Chang}}, \bibinfo {author} {\bibfnamefont {Sai}\ \bibnamefont {Wang}}, \ and\ \bibinfo {author} {\bibfnamefont {Qing-Hua}\ \bibnamefont {Zhu}},\ }\bibfield  {title} {\enquote {\bibinfo {title} {{Note on gauge invariance of second order cosmological perturbations}},}\ }\href {\doibase 10.1088/1674-1137/ac0c74} {\bibfield  {journal} {\bibinfo  {journal} {Chin. Phys. C}\ }\textbf {\bibinfo {volume} {45}},\ \bibinfo {pages} {095101} (\bibinfo {year} {2021})},\ \Eprint {http://arxiv.org/abs/2009.11025} {arXiv:2009.11025 [astro-ph.CO]} \BibitemShut {NoStop}%
\bibitem [{\citenamefont {Ali}\ \emph {et~al.}(2021)\citenamefont {Ali}, \citenamefont {Gong},\ and\ \citenamefont {Lu}}]{Ali:2020sfw}%
  \BibitemOpen
  \bibfield  {author} {\bibinfo {author} {\bibfnamefont {Arshad}\ \bibnamefont {Ali}}, \bibinfo {author} {\bibfnamefont {Yungui}\ \bibnamefont {Gong}}, \ and\ \bibinfo {author} {\bibfnamefont {Yizhou}\ \bibnamefont {Lu}},\ }\bibfield  {title} {\enquote {\bibinfo {title} {{Gauge transformation of scalar induced tensor perturbation during matter domination}},}\ }\href {\doibase 10.1103/PhysRevD.103.043516} {\bibfield  {journal} {\bibinfo  {journal} {Phys. Rev. D}\ }\textbf {\bibinfo {volume} {103}},\ \bibinfo {pages} {043516} (\bibinfo {year} {2021})},\ \Eprint {http://arxiv.org/abs/2009.11081} {arXiv:2009.11081 [gr-qc]} \BibitemShut {NoStop}%
\bibitem [{\citenamefont {Chang}\ \emph {et~al.}(2020{\natexlab{a}})\citenamefont {Chang}, \citenamefont {Wang},\ and\ \citenamefont {Zhu}}]{Chang:2020iji}%
  \BibitemOpen
  \bibfield  {author} {\bibinfo {author} {\bibfnamefont {Zhe}\ \bibnamefont {Chang}}, \bibinfo {author} {\bibfnamefont {Sai}\ \bibnamefont {Wang}}, \ and\ \bibinfo {author} {\bibfnamefont {Qing-Hua}\ \bibnamefont {Zhu}},\ }\bibfield  {title} {\enquote {\bibinfo {title} {{Gauge Invariant Second Order Gravitational Waves}},}\ }\href@noop {} {\  (\bibinfo {year} {2020}{\natexlab{a}})},\ \Eprint {http://arxiv.org/abs/2009.11994} {arXiv:2009.11994 [gr-qc]} \BibitemShut {NoStop}%
\bibitem [{\citenamefont {Chang}\ \emph {et~al.}(2020{\natexlab{b}})\citenamefont {Chang}, \citenamefont {Wang},\ and\ \citenamefont {Zhu}}]{Chang:2020mky}%
  \BibitemOpen
  \bibfield  {author} {\bibinfo {author} {\bibfnamefont {Zhe}\ \bibnamefont {Chang}}, \bibinfo {author} {\bibfnamefont {Sai}\ \bibnamefont {Wang}}, \ and\ \bibinfo {author} {\bibfnamefont {Qing-Hua}\ \bibnamefont {Zhu}},\ }\bibfield  {title} {\enquote {\bibinfo {title} {{On the Gauge Invariance of Scalar Induced Gravitational Waves: Gauge Fixings Considered}},}\ }\href@noop {} {\  (\bibinfo {year} {2020}{\natexlab{b}})},\ \Eprint {http://arxiv.org/abs/2010.01487} {arXiv:2010.01487 [gr-qc]} \BibitemShut {NoStop}%
\bibitem [{\citenamefont {Giovannini}(2020{\natexlab{b}})}]{Giovannini:2020soq}%
  \BibitemOpen
  \bibfield  {author} {\bibinfo {author} {\bibfnamefont {Massimo}\ \bibnamefont {Giovannini}},\ }\bibfield  {title} {\enquote {\bibinfo {title} {{Effective anisotropic stresses of the relic gravitons}},}\ }\href {\doibase 10.1142/S0218271820501126} {\bibfield  {journal} {\bibinfo  {journal} {Int. J. Mod. Phys. D}\ }\textbf {\bibinfo {volume} {29}},\ \bibinfo {pages} {2050112} (\bibinfo {year} {2020}{\natexlab{b}})},\ \Eprint {http://arxiv.org/abs/2007.14956} {arXiv:2007.14956 [hep-th]} \BibitemShut {NoStop}%
\bibitem [{\citenamefont {Dom\`enech}\ and\ \citenamefont {Sasaki}(2021)}]{Domenech:2020xin}%
  \BibitemOpen
  \bibfield  {author} {\bibinfo {author} {\bibfnamefont {Guillem}\ \bibnamefont {Dom\`enech}}\ and\ \bibinfo {author} {\bibfnamefont {Misao}\ \bibnamefont {Sasaki}},\ }\bibfield  {title} {\enquote {\bibinfo {title} {{Approximate gauge independence of the induced gravitational wave spectrum}},}\ }\href {\doibase 10.1103/PhysRevD.103.063531} {\bibfield  {journal} {\bibinfo  {journal} {Phys. Rev. D}\ }\textbf {\bibinfo {volume} {103}},\ \bibinfo {pages} {063531} (\bibinfo {year} {2021})},\ \Eprint {http://arxiv.org/abs/2012.14016} {arXiv:2012.14016 [gr-qc]} \BibitemShut {NoStop}%
\bibitem [{\citenamefont {Cai}\ \emph {et~al.}(2021{\natexlab{b}})\citenamefont {Cai}, \citenamefont {Yang},\ and\ \citenamefont {Zhao}}]{Cai:2021ndu}%
  \BibitemOpen
  \bibfield  {author} {\bibinfo {author} {\bibfnamefont {Rong-Gen}\ \bibnamefont {Cai}}, \bibinfo {author} {\bibfnamefont {Xing-Yu}\ \bibnamefont {Yang}}, \ and\ \bibinfo {author} {\bibfnamefont {Long}\ \bibnamefont {Zhao}},\ }\bibfield  {title} {\enquote {\bibinfo {title} {{Energy spectrum of gravitational waves}},}\ }\href@noop {} {\  (\bibinfo {year} {2021}{\natexlab{b}})},\ \Eprint {http://arxiv.org/abs/2109.06865} {arXiv:2109.06865 [astro-ph.CO]} \BibitemShut {NoStop}%
\bibitem [{\citenamefont {Cai}\ \emph {et~al.}(2022)\citenamefont {Cai}, \citenamefont {Yang},\ and\ \citenamefont {Zhao}}]{Cai:2021jbi}%
  \BibitemOpen
  \bibfield  {author} {\bibinfo {author} {\bibfnamefont {Rong-Gen}\ \bibnamefont {Cai}}, \bibinfo {author} {\bibfnamefont {Xing-Yu}\ \bibnamefont {Yang}}, \ and\ \bibinfo {author} {\bibfnamefont {Long}\ \bibnamefont {Zhao}},\ }\bibfield  {title} {\enquote {\bibinfo {title} {{On the energy of gravitational waves}},}\ }\href {\doibase 10.1007/s10714-022-02972-x} {\bibfield  {journal} {\bibinfo  {journal} {Gen. Rel. Grav.}\ }\textbf {\bibinfo {volume} {54}},\ \bibinfo {pages} {89} (\bibinfo {year} {2022})},\ \Eprint {http://arxiv.org/abs/2109.06864} {arXiv:2109.06864 [gr-qc]} \BibitemShut {NoStop}%
\bibitem [{\citenamefont {Tomita}(1967)}]{Tomita:1967wkp}%
  \BibitemOpen
  \bibfield  {author} {\bibinfo {author} {\bibfnamefont {Kenji}\ \bibnamefont {Tomita}},\ }\bibfield  {title} {\enquote {\bibinfo {title} {{Non-Linear Theory of Gravitational Instability in the Expanding Universe}},}\ }\href {\doibase 10.1143/PTP.37.831} {\bibfield  {journal} {\bibinfo  {journal} {Prog. Theor. Phys.}\ }\textbf {\bibinfo {volume} {37}},\ \bibinfo {pages} {831--846} (\bibinfo {year} {1967})}\BibitemShut {NoStop}%
\bibitem [{\citenamefont {Matarrese}\ \emph {et~al.}(1993)\citenamefont {Matarrese}, \citenamefont {Pantano},\ and\ \citenamefont {Saez}}]{Matarrese:1992rp}%
  \BibitemOpen
  \bibfield  {author} {\bibinfo {author} {\bibfnamefont {Sabino}\ \bibnamefont {Matarrese}}, \bibinfo {author} {\bibfnamefont {Ornella}\ \bibnamefont {Pantano}}, \ and\ \bibinfo {author} {\bibfnamefont {Diego}\ \bibnamefont {Saez}},\ }\bibfield  {title} {\enquote {\bibinfo {title} {{A General relativistic approach to the nonlinear evolution of collisionless matter}},}\ }\href {\doibase 10.1103/PhysRevD.47.1311} {\bibfield  {journal} {\bibinfo  {journal} {Phys. Rev. D}\ }\textbf {\bibinfo {volume} {47}},\ \bibinfo {pages} {1311--1323} (\bibinfo {year} {1993})}\BibitemShut {NoStop}%
\bibitem [{\citenamefont {Matarrese}\ \emph {et~al.}(1994)\citenamefont {Matarrese}, \citenamefont {Pantano},\ and\ \citenamefont {Saez}}]{Matarrese:1993zf}%
  \BibitemOpen
  \bibfield  {author} {\bibinfo {author} {\bibfnamefont {Sabino}\ \bibnamefont {Matarrese}}, \bibinfo {author} {\bibfnamefont {Ornella}\ \bibnamefont {Pantano}}, \ and\ \bibinfo {author} {\bibfnamefont {Diego}\ \bibnamefont {Saez}},\ }\bibfield  {title} {\enquote {\bibinfo {title} {{General relativistic dynamics of irrotational dust: Cosmological implications}},}\ }\href {\doibase 10.1103/PhysRevLett.72.320} {\bibfield  {journal} {\bibinfo  {journal} {Phys. Rev. Lett.}\ }\textbf {\bibinfo {volume} {72}},\ \bibinfo {pages} {320--323} (\bibinfo {year} {1994})},\ \Eprint {http://arxiv.org/abs/astro-ph/9310036} {arXiv:astro-ph/9310036} \BibitemShut {NoStop}%
\bibitem [{\citenamefont {Matarrese}\ \emph {et~al.}(1998)\citenamefont {Matarrese}, \citenamefont {Mollerach},\ and\ \citenamefont {Bruni}}]{Matarrese:1997ay}%
  \BibitemOpen
  \bibfield  {author} {\bibinfo {author} {\bibfnamefont {Sabino}\ \bibnamefont {Matarrese}}, \bibinfo {author} {\bibfnamefont {Silvia}\ \bibnamefont {Mollerach}}, \ and\ \bibinfo {author} {\bibfnamefont {Marco}\ \bibnamefont {Bruni}},\ }\bibfield  {title} {\enquote {\bibinfo {title} {{Second order perturbations of the Einstein-de Sitter universe}},}\ }\href {\doibase 10.1103/PhysRevD.58.043504} {\bibfield  {journal} {\bibinfo  {journal} {Phys. Rev. D}\ }\textbf {\bibinfo {volume} {58}},\ \bibinfo {pages} {043504} (\bibinfo {year} {1998})},\ \Eprint {http://arxiv.org/abs/astro-ph/9707278} {arXiv:astro-ph/9707278} \BibitemShut {NoStop}%
\bibitem [{\citenamefont {Noh}\ and\ \citenamefont {Hwang}(2004)}]{Noh:2004bc}%
  \BibitemOpen
  \bibfield  {author} {\bibinfo {author} {\bibfnamefont {Hyerim}\ \bibnamefont {Noh}}\ and\ \bibinfo {author} {\bibfnamefont {Jai-chan}\ \bibnamefont {Hwang}},\ }\bibfield  {title} {\enquote {\bibinfo {title} {{Second-order perturbations of the Friedmann world model}},}\ }\href {\doibase 10.1103/PhysRevD.69.104011} {\bibfield  {journal} {\bibinfo  {journal} {Phys. Rev. D}\ }\textbf {\bibinfo {volume} {69}},\ \bibinfo {pages} {104011} (\bibinfo {year} {2004})}\BibitemShut {NoStop}%
\bibitem [{\citenamefont {Carbone}\ and\ \citenamefont {Matarrese}(2005)}]{Carbone:2004iv}%
  \BibitemOpen
  \bibfield  {author} {\bibinfo {author} {\bibfnamefont {Carmelita}\ \bibnamefont {Carbone}}\ and\ \bibinfo {author} {\bibfnamefont {Sabino}\ \bibnamefont {Matarrese}},\ }\bibfield  {title} {\enquote {\bibinfo {title} {{A Unified treatment of cosmological perturbations from super-horizon to small scales}},}\ }\href {\doibase 10.1103/PhysRevD.71.043508} {\bibfield  {journal} {\bibinfo  {journal} {Phys. Rev. D}\ }\textbf {\bibinfo {volume} {71}},\ \bibinfo {pages} {043508} (\bibinfo {year} {2005})},\ \Eprint {http://arxiv.org/abs/astro-ph/0407611} {arXiv:astro-ph/0407611} \BibitemShut {NoStop}%
\bibitem [{\citenamefont {Nakamura}(2007)}]{Nakamura:2004rm}%
  \BibitemOpen
  \bibfield  {author} {\bibinfo {author} {\bibfnamefont {Kouji}\ \bibnamefont {Nakamura}},\ }\bibfield  {title} {\enquote {\bibinfo {title} {{Second-order gauge invariant cosmological perturbation theory: Einstein equations in terms of gauge invariant variables}},}\ }\href {\doibase 10.1143/PTP.117.17} {\bibfield  {journal} {\bibinfo  {journal} {Prog. Theor. Phys.}\ }\textbf {\bibinfo {volume} {117}},\ \bibinfo {pages} {17--74} (\bibinfo {year} {2007})},\ \Eprint {http://arxiv.org/abs/gr-qc/0605108} {arXiv:gr-qc/0605108} \BibitemShut {NoStop}%
\bibitem [{\citenamefont {Dom\`enech}(2024)}]{Domenech:2023jve}%
  \BibitemOpen
  \bibfield  {author} {\bibinfo {author} {\bibfnamefont {Guillem}\ \bibnamefont {Dom\`enech}},\ }\bibfield  {title} {\enquote {\bibinfo {title} {{Cosmological gravitational waves from isocurvature fluctuations}},}\ }\href {\doibase 10.1007/s43673-023-00109-z} {\bibfield  {journal} {\bibinfo  {journal} {AAPPS Bull.}\ }\textbf {\bibinfo {volume} {34}},\ \bibinfo {pages} {4} (\bibinfo {year} {2024})},\ \Eprint {http://arxiv.org/abs/2311.02065} {arXiv:2311.02065 [gr-qc]} \BibitemShut {NoStop}%
\bibitem [{\citenamefont {Dom\`enech}(2021)}]{Domenech:2021ztg}%
  \BibitemOpen
  \bibfield  {author} {\bibinfo {author} {\bibfnamefont {Guillem}\ \bibnamefont {Dom\`enech}},\ }\bibfield  {title} {\enquote {\bibinfo {title} {{Scalar Induced Gravitational Waves Review}},}\ }\href {\doibase 10.3390/universe7110398} {\bibfield  {journal} {\bibinfo  {journal} {Universe}\ }\textbf {\bibinfo {volume} {7}},\ \bibinfo {pages} {398} (\bibinfo {year} {2021})},\ \Eprint {http://arxiv.org/abs/2109.01398} {arXiv:2109.01398 [gr-qc]} \BibitemShut {NoStop}%
\bibitem [{\citenamefont {Dom\`enech}\ \emph {et~al.}(2022)\citenamefont {Dom\`enech}, \citenamefont {Passaglia},\ and\ \citenamefont {Renaux-Petel}}]{Domenech:2021and}%
  \BibitemOpen
  \bibfield  {author} {\bibinfo {author} {\bibfnamefont {Guillem}\ \bibnamefont {Dom\`enech}}, \bibinfo {author} {\bibfnamefont {Samuel}\ \bibnamefont {Passaglia}}, \ and\ \bibinfo {author} {\bibfnamefont {S\'ebastien}\ \bibnamefont {Renaux-Petel}},\ }\bibfield  {title} {\enquote {\bibinfo {title} {{Gravitational waves from dark matter isocurvature}},}\ }\href {\doibase 10.1088/1475-7516/2022/03/023} {\bibfield  {journal} {\bibinfo  {journal} {JCAP}\ }\textbf {\bibinfo {volume} {03}},\ \bibinfo {pages} {023} (\bibinfo {year} {2022})},\ \Eprint {http://arxiv.org/abs/2112.10163} {arXiv:2112.10163 [astro-ph.CO]} \BibitemShut {NoStop}%
\bibitem [{\citenamefont {Dom\`enech}\ and\ \citenamefont {Tr\"ankle}(2025)}]{Domenech:2024wao}%
  \BibitemOpen
  \bibfield  {author} {\bibinfo {author} {\bibfnamefont {Guillem}\ \bibnamefont {Dom\`enech}}\ and\ \bibinfo {author} {\bibfnamefont {Jan}\ \bibnamefont {Tr\"ankle}},\ }\bibfield  {title} {\enquote {\bibinfo {title} {{From formation to evaporation: Induced gravitational wave probes of the primordial black hole reheating scenario}},}\ }\href {\doibase 10.1103/PhysRevD.111.063528} {\bibfield  {journal} {\bibinfo  {journal} {Phys. Rev. D}\ }\textbf {\bibinfo {volume} {111}},\ \bibinfo {pages} {063528} (\bibinfo {year} {2025})},\ \Eprint {http://arxiv.org/abs/2409.12125} {arXiv:2409.12125 [gr-qc]} \BibitemShut {NoStop}%
\bibitem [{\citenamefont {Pitrou}\ \emph {et~al.}(2013)\citenamefont {Pitrou}, \citenamefont {Roy},\ and\ \citenamefont {Umeh}}]{Pitrou:2013hga}%
  \BibitemOpen
  \bibfield  {author} {\bibinfo {author} {\bibfnamefont {Cyril}\ \bibnamefont {Pitrou}}, \bibinfo {author} {\bibfnamefont {Xavier}\ \bibnamefont {Roy}}, \ and\ \bibinfo {author} {\bibfnamefont {Obinna}\ \bibnamefont {Umeh}},\ }\bibfield  {title} {\enquote {\bibinfo {title} {{xPand: An algorithm for perturbing homogeneous cosmologies}},}\ }\href {\doibase 10.1088/0264-9381/30/16/165002} {\bibfield  {journal} {\bibinfo  {journal} {Class. Quant. Grav.}\ }\textbf {\bibinfo {volume} {30}},\ \bibinfo {pages} {165002} (\bibinfo {year} {2013})},\ \Eprint {http://arxiv.org/abs/1302.6174} {arXiv:1302.6174 [astro-ph.CO]} \BibitemShut {NoStop}%
\end{thebibliography}%

\clearpage
\appendix

\section{Metric perturbations up to first order}\label{appendix}
At leading order, the Einstein equation and energy conservation yield
\begin{equation}
\begin{aligned}
    &3\mH^2=8\pi a^2 (\rho_m+\rho_r), \\
    & \mH^2+2\mH'=-\frac{8\pi}{3} a^2 \rho_r,\\
    &\rho_m'+3\mH \rho_m=0, \\
    &\rho_r'+4\mH\rho_r =0 .
\end{aligned}
\end{equation}
The solution for the scale factor is given by
\begin{equation}
    {a(\eta)\over a_{\mathrm{eq}}}=2\left(\eta \over \eta_*\right) + \left(\eta \over \eta_*\right)^2,
\end{equation}
where $\eta_*=\eta_{\mathrm{eq}}/(\sqrt{2}-1)$. From the above equations, we can find a solution such that
\begin{equation}
    \rho_m (\eta)={1
    \over 2}\rho_{\mathrm{eq}}\left(a \over a_{\mathrm{eq}}\right)^{-3},\qquad\qquad 
    \rho_r (\eta)= {1
    \over 2}\rho_{\mathrm{eq}}\left(a \over a_{\mathrm{eq}}\right)^{-4},
\end{equation}
where $\rho_{\mathrm{eq}}=\rho_m(\eta_{\mathrm{eq}})+\rho_r(\eta_{\mathrm{eq}})$  is the total energy density at matter-radiation equality.

At first order, energy conservation gives
\begin{equation}
\begin{aligned}
    &\delta\rho_m'+3\mH\delta\rho_m+\rho_m\left( -3\psi' + \Delta E'+\Delta v_m \right)=0,\\
    &\delta\rho_r'+4\mH\delta\rho_r+{4\over 3}\rho_r\left( -3\psi' + \Delta E'+\Delta v_r \right)=0,\\
    &v_m'+\mH v_m+(\phi+\mH B+B') = 0, \\
    &v_r'+{1\over 4}\frac{\delta\rho_r}{\rho_r} +(\phi+B') = 0 .
\end{aligned}
\end{equation}
The first-order Einstein equations for the $00$ and $0i$ components are given by
\begin{equation}
\begin{aligned}
    &3\mH^2\phi+\mH(3\psi'+\Delta B-\Delta E')-\Delta\psi = -4\pi a^2 (\delta\rho_m+\delta\rho_r),\\
    &\mH^2 B-\mH'B+\mH\phi+\psi'=-4\pi a^2\left(\rho_m v_m +{4\over 3}\rho_r v_r\right).
\end{aligned}
\end{equation}

\end{document}